\documentclass[useAMS,usenatbib,a4paper,fleqn]{mn2e}

\usepackage{amsfonts,amssymb,amsmath}
\usepackage{aas_macros}
\usepackage{times,varioref,multirow,textcomp,comment,xtab,units}
\usepackage[usenames,dvipsnames,svgnames,hyperref]{xcolor}
\usepackage[pdftex]{graphicx}
\usepackage{comment}
\usepackage[
    pdftex,
    a4paper=true,
    plainpages=true,
    pdfpagelabels,
    breaklinks=true,
    bookmarks=true,
    bookmarksopen=false,
    bookmarksopenlevel=2
    bookmarksnumbered=true,
    bookmarkstype=toc,
    colorlinks=true,
    citecolor=RoyalBlue,
    linkcolor=ForestGreen,
    menucolor=Teal,
    urlcolor=DarkOrange,
]{hyperref}

\providecommand{\adsurl}[1]{\href{#1}{ADS}}

\def \Msunh{\ h^{-1}{\rm M_\odot}}

\def \Mpch{\ h^{-1}{\rm Mpc}}
\def \kpch{\ h^{-1}{\rm kpc}}
\def \LCDM{$\Lambda$CDM}

\newcommand{\Eqref}[1]{Eq.~(\ref{#1})}
\newcommand{\Figref}[1]{Fig.~\ref{#1}}
\newcommand{\Secref}[1]{\S\ref{#1}}  
\newcommand{\Tableref}[1]{Table~\ref{#1}}

\usepackage[roman]{parnotes}

\def \dmsim{DM}
\def \nrsim{NR}
\def \radsim{FP}
\def \vmax{V_{\rm max}}
\def \nifty{nIFTy}

\def \gadget{{\sc Gadget}}

\def \arepo{{\sc Arepo}}

\def \gadgetxart{{\sc G3-X-Art}}

\def \magneticum{{\sc G3-Magneticum}}
\def \pesph{{\sc G3-PESPH}}
\def \music{{\sc G3-Music}}
\def \musicpi{{\sc G3-MusicPi}}
\def \owls{{\sc G3-OWLS}}
\def \gadgettwox{{\sc G2-X}}
\def \ramses{{\sc ramses}}

\def \rvmax{R_{V_{\rm max}}}
\def \vmax{V_{\rm max}}

\def \MstarMh{M_*/M_{\rm h}}

\newcommand{\sigdec}[1]{{\bf\color{red}#1}}
\newcommand{\siginc}[1]{{\bf\color{blue}#1}}

\defcitealias{springel2003}{SH03}
\defcitealias{schaye2008}{SDV08}
\hypersetup{
  pdfauthor = {Pascal J. Elahi},
  pdfkeywords = {nIFTY Galaxies},
  pdftitle = {nIFTY III: The Similarity and Diversity of Galaxies and Subhaloes}
}

\begin{document}
\title[nIFTy III: Galaxies \& Subhaloes]{nIFTY galaxy cluster simulations III: The Similarity \& Diversity of Galaxies \& Subhaloes}
\author[Elahi {\it et al}.]{Pascal J. Elahi$^{1}$\thanks{\href{mailto:pelahi@physics.usyd.edu.au}{pelahi@physics.usyd.edu.au}},
Alexander Knebe$^{2,3}$, 
Frazer R. Pearce$^{4}$,
Chris Power$^{5,6}$,
Gustavo Yepes$^{2}$,
\newauthor
Weiguang Cui$^{5,6}$,
Daniel Cunnama$^{7}$,
Scott T. Kay$^{8}$,
Federico Sembolini$^{2,9}$,
\newauthor
Alexander M. Beck$^{10}$,
Romeel Dav\'e$^{11,12,13,14}$,
Sean February$^{15}$,
\newauthor
Shuiyao Huang$^{16}$,
Neal Katz$^{16}$,
Ian G. McCarthy$^{17}$,
Giuseppe Murante$^{18}$,
Valentin Perret$^{19}$,
\newauthor
Ewald Puchwein$^{20}$,
Alexandro Saro$^{21}$,
Romain Teyssier$^{22}$\\
$^{1}$Sydney Institute for Astronomy, A28, School of Physics, The University of Sydney, NSW 2006, Australia\\
$^{2}$Departamento de F\'isica Te\'{o}rica, M\'{o}dulo 8, Facultad de Ciencias, Universidad Aut\'{o}noma de Madrid, 28049 Madrid, Spain\\
$^{3}$Astro-UAM, UAM, Unidad Asociada CSIC\\
$^{4}$School of Physics \& Astronomy, University of Nottingham, Nottingham NG7 2RD, UK\\
$^{5}$International Centre for Radio Astronomy Research, University of Western Australia, 35 Stirling Highway, Crawley, Western Australia 6009, Australia\\
$^{6}$ARC Centre of Excellence for All-Sky Astrophysics (CAASTRO)\\
$^{7}$Physics Department, University of the Western Cape, Cape Town 7535, South Africa\\
$^{9}$Jodrell Bank Centre for Astrophysics, School of Physics and Astronomy, The University of Manchester, Manchester M13 9PL, UK\\
$^{9}$Dipartimento di Fisica, Sapienza Universit$\grave{a}$  di Roma, Piazzale Aldo Moro 5, I-00185 Roma, Italy\\
$^{10}$University Observatory Munich, Scheinerstr. 1, D-81679 Munich, Germany\\
$^{11}$Physics Department, University of Western Cape, Bellville, Cape Town 7535, South Africa\\
$^{12}$South African Astronomical Observatory, PO Box 9, Observatory, Cape Town 7935, South Africa\\
$^{13}$African Institute of Mathematical Sciences, Muizenberg, Cape Town 7945, South Africa\\
$^{14}$Astronomy Department and CCAPP, Ohio State University, Columbus, OH 43210, USA\\
$^{15}$Center for High Performance Computing, CSIR Campus, 15 Lower Hope Street, Rosebank, Cape Town 7701, South Africa\\
$^{16}$Astronomy Department, University of Massachusetts, Amherst, MA 01003, USA\\
$^{17}$Astrophysics Research Institute, Liverpool John Moores University, 146 Brownlow Hill, Liverpool L3 5RF, UK\\
$^{18}$INAF, Osservatorio Astronomico di Trieste, via G.B. Tiepolo 11, I-34143 Trieste, Italy\\
$^{18}$Institute for Computational Science, ETH Zurich, Wolfgang-Pauli-Strasse 16, CH-8093, Zurich, Switzerland\\
$^{20}$Institute of Astronomy and Kavli Institute for Cosmology, University of Cambridge, Madingley Road, Cambridge CB3 0HA, UK\\
$^{21}$Department of Physics, Ludwig-Maximilians-Universitat, Scheinerstr. 1, 81679 Munchen, Germany\\
$^{22}$Institute of Theoretical Physics, Universit\"at Z\"urich, Winterthurerstrasse 190, CH-8057 Z\"urich, Switzerland\\
}

\maketitle
\clearpage
\pdfbookmark[1]{Abstract}{sec:abstract}
\begin{abstract}
We examine subhaloes and galaxies residing in a simulated \LCDM\ galaxy cluster ($M^{\rm crit}_{200}=1.1\times10^{15}h^{-1}M_\odot$) produced by hydrodynamical codes ranging from classic Smooth Particle Hydrodynamics (SPH), newer SPH codes, adaptive and moving mesh codes. These codes use subgrid models to capture galaxy formation physics. We compare how well these codes reproduce the same subhaloes/galaxies in gravity-only, non-radiative hydrodynamics and {\em full feedback physics} runs by looking at the overall subhalo/galaxy distribution and on an individual objects basis. We find the subhalo population is reproduced to within $\lesssim10\%$ for both dark matter only and non-radiative runs, with individual objects showing code-to-code scatter of $\lesssim0.1$~dex, although the gas in non-radiative simulations shows significant scatter. Including feedback physics significantly increases the diversity. Subhalo mass and $\vmax$ distributions vary by $\approx20\%$. The galaxy populations also show striking code-to-code variations. Although the Tully-Fisher relation is similar in almost all codes, the number of galaxies with $10^{9}\Msunh\lesssim~M_{*}\lesssim10^{12}\Msunh$ can differ by a factor of 4. Individual galaxies show code-to-code scatter of $\sim0.5$~dex in stellar mass. Moreover, systematic differences exist, with some codes producing galaxies $70\%$ smaller than others. The diversity partially arises from the inclusion/absence of AGN feedback. Our results combined with our companion papers demonstrate that subgrid physics is not just subject to fine-tuning, but the complexity of building galaxies {\em in all environments} remains a challenge. We argue even basic galaxy properties, such as stellar mass to halo mass, should be treated with errors bars of $\sim0.2-0.4$~dex.
\end{abstract}
\maketitle

\section{Introduction}\label{sec:intro}
The complex environment of galaxy clusters provides a challenging and unique astrophysical laboratory with which to test our theories of cosmic structure formation and the processes that govern galaxy formation. The progenitors of these massive structures collapsed at high redshift, and so their present day properties probe cosmic structure formation over a large fraction of the Universe's lifetime. A cluster's galaxy population is comprised of both those that have orbited within the dense, violent environment for several dynamical times and newly accreted field galaxies. Modelling these systems has been a great challenge given the enormous range in both spatial and temporal scales probed: from the local cooling of gas; conversion of gas to stars; and injection of energy into the surrounding galactic medium from supernovae; to merger driven star bursts and the powerful AGN outflows from massive galaxies that affect the large-scale intra-cluster medium. 

\par
Hydrodynamical simulations traditionally used either Lagrangian Smoothed-Particle-Hydrodynamics (SPH) techniques (e.g.~\citealp{gingold1977,lucy1977,monaghan1992,katz1996}; and see \citealp{springel2010b} for a review) or Eulerian grid-based solvers sometimes aided by Adaptive Mesh Refinement (AMR) techniques \cite[e.g.][]{cen1992,bryan1995,art}. Ideally, synthetic galaxies should be similar regardless of code or technique used. However, early comparisons of hydrodynamical N-body codes showed worrying differences between numerical approaches and even codes. The classic Santa Barbara Cluster Comparison Project, \cite{frenk1999a}, compared the properties of a galaxy cluster formed in a non-radiative cosmological simulation using 12 then state-of-the-art mesh- and particle-based codes and found a large scatter in almost all bulk properties. The key difference confirmed in many other studies was the presence of a core in the radial entropy profile in mesh based codes that was absent in SPH codes \cite[e.g.][]{dolag2005a,voit2005,mitchell2009}.

\par
Some of these differences can be attributed to the underlying technique used, whether SPH or mesh based. By its very nature of SPH can smooth out shocks, dampen subsonic turbulence, and suppress fluid instabilities, at least for vanilla SPH \cite[e.g.][]{okamoto2003,agertz2007,tasker2008}. Mesh codes by construction are not Galilean invariant, consequently results are sensitive to the presence of bulk velocities and significant advection errors can occur when fluids with sharp gradients move across cells in a manner un-aligned with the grid, generating entropy spuriously through artificially enhanced mixing \cite[e.g.][]{wadsley2008,tasker2008}. AMR codes, which use flexible but necessarily ad-hoc refinement criteria, have artefacts arising from the loss of accuracy at refinement boundaries. When coupled to gravity, this loss of accuracy leads to suppression of low amplitude gravitational instabilities, which are seeds for cosmological structure formation, and violate energy and momentum conservation in the long-range forces whenever cells are refined or de-refined \cite[e.g.][]{oshea2005,heitmann2008}. Consequently, even for some simple non-radiative problems, classic Lagrangian and Eulerian codes will not converge to the same solution \cite[e.g.][]{tasker2008,hubber2013}. Modern codes have attempted to address some of the inherent issues with each method by the inclusion of higher order dissipative switches \cite[e.g.][]{read2010}, new SPH kernels, different SPH formulations \cite[e.g.][]{hopkins2013}, sub-grid physics in mesh codes \cite[e.g.][]{maier2009}, and hybrid methods \cite[e.g.][]{arepo,gizmo}. 

\par 
Comparisons are further complicated by the inclusion of uncertain baryonic physics governing galaxy formation. Though most codes attempt to reproduce the observed galaxy population, implementations of feedback physics vary and typically increases the code-to-code scatter. For instance, \cite{scannapieco2012} found that different star formation and stellar feedback implementations lead to significant differences in the morphology, angular momentum and stellar mass of an isolated individual galaxy. Some of the differences are a simple result of different subgrid physics. Several studies have investigated tuning parameters using in subgrid models, clearly showing the need for some tuning \cite[e.g.][]{haas2013a,haas2013b,lebrun2014a,crain2015a}, although typically these models focus on varying parameters and not necessarily changing the subgrid implementation. Using the same SPH code, \cite{duffy2010} showed different subgrid models produced different baryonic distributions. However, different models need not necessarily produce different galaxy populations. \cite{durier2012} showed that two significantly different implementations of supernova feedback in SPH codes, thermal and kinetic, do converge. In \cite{scannapieco2012}, the resulting disc galaxy was typically too concentrated but recent developments have shown that there are codes capable of producing more realistic disc galaxies \cite[e.g.][]{vogelsberger2014a,feldmann2015a,schaye2015a,wang2015a,murante2015a}, motivating new comparison projects using individual galaxies such as the ongoing AGORA project \cite[][]{kim2014a}.

\par
The appearance of numerous modern SPH and mesh methods and significant developments in modelling the processes governing galaxy formation warrants a second look at synthetic clusters. Hence, sixteen years later, the nIFTy comparison project aims to revisit the Santa Barbara comparison with new state-of-the-art hydrodynamical codes. The first paper in this series of comparisons, \cite{nifty1}, studied the bulk properties of the cluster environment using a single well-resolved cluster with twelve modern codes in pure N-body and adiabatic runs. This comparison clearly demonstrated that:
\begin{enumerate}
    \item The dark matter distribution in pure Dark Matter (DM) only simulations show $\lesssim20\%$ variation in the dark matter density profile.
    \item In non-radiative runs, the variation in the dark matter density profile remains at $\lesssim20\%$, but the gas distribution shows variations of up to $\sim100\%$.
    \item Newer SPH codes that use higher order kernels and more complex methods for modelling dissipative physics are in close agreement with mesh codes, with variations of $\lesssim10\%$, and more significantly these codes {\em reproduce the entropy core seen in numerous mesh codes}. 
\end{enumerate}
Clearly, the latest SPH codes have removed the long standing problem of falling entropy profiles seen in \cite{frenk1999a}.

\par 
In paper II, \cite{nifty2}, we examined the bulk properties of this same cluster in full physics runs. The inclusion of cooling, star formation and feedback significantly increases the scatter between codes, with baryon and stellar fractions varying by $30\%$. Furthermore, full physics removes between classic and modern SPH codes in regards to entropy profiles, i.e., full physics + classic SPH can produce entropy cores. Intriguingly, the dividing line in properties like the temperature profile between codes is not the inclusion/absence of AGN, although AGN play an important role in limiting the effect of overcooling.

\par
The next question, which we examine here, is whether codes reproduce not just the same overall cluster environment but also individual subhaloes \& galaxies residing in the cluster. Here we examine multiple subhaloes/galaxies, and the change in the differences between codes with the inclusion of more complex physics, going from pure dark matter simulations to full feedback physics simulations. The goal is to identify the origins of any differences and determine relative ``error'' bars for predictions from hydrodynamical simulations. This paper is organised as follows: we briefly describe the numerical methods in \Secref{sec:methods}, highlighting the differences between the codes in \Secref{sec:codes}. Our findings are presented in sections \ref{sec:subhalopop}-\ref{sec:crosscomp}, where we compare the subhalo/galaxy population as a whole and compare individual objects respectively. We end with discussion in \Secref{sec:discussion}.

\section{Numerical Methods}\label{sec:methods}
\subsection{Codes}\label{sec:codes}
The initial \nifty\ comparison project, as presented in \citet{nifty1}, included 13 codes -- the {\small{CART}} variant of {\sc{ART}}, \ramses, \arepo, {\sc{Hydra}} and 9 variants of the \gadget\ code. In this study as in \cite{nifty2}, we consider the subset of these codes in which full subgrid physics has been included: one Adaptive Mesh Refinement (AMR) code, \ramses, the moving mesh code, \arepo and 9 variants of the SPH \gadget\ code. The subgrid physics included span the range from codes only including Cooling and Star Formation (CSF) to those that also include supermassive black hole formation and associated Active Galactic Nuclei (AGN). Two codes, \arepo\ \& \music, have been run with variant subgrid physics. The salient features of each code are summarised in \Tableref{tab:codes:summary}. A comprehensive summary of the approach taken to solving the hydrodynamic equations in each of these codes can be found in \cite{nifty1} and description of subgrid models in \cite{nifty2} (and Appendix \ref{sec:app:codes}).

\begin{table*}
\setlength\tabcolsep{3pt}
\centering\footnotesize
\caption{A brief summary of the codes.}
\label{tab:codes:summary}
\begin{tabular}{@{\extracolsep{\fill}}p{0.08\textwidth}lcc p{0.675\textwidth}}
\hline
    Type & Code & SN & AGN & Comments \\
\hline
    Mesh 
    & \ramses        &  & \checkmark 
        & Salpeter IMF; {\bf No} SN feedback; thermal AGN; average metallicity.
        \\& & & & For more details see \cite{ramses,teyssier2011} \& appendix \ref{code:ramses}.
        \\
    \hline
    Moving Mesh 
    & \arepo        & \checkmark & \checkmark 
        & Chabrier IMF; \cite{springel2003} (hereafter SH03) SF; kinetic SN; thermal AGN; tracks 9 individual elements. 
        \\& & & & {\bf Variant}:\arepo-SH that uses subgrid physics of \music\ (no AGN, \citetalias{springel2003} SF, kinetic SN). 
        \\& & & & For more details see \cite{vogelsberger2013,vogelsberger2014a} \& appendix \ref{code:arepo}.\\
    \hline
    Classic SPH
    & \music        & \checkmark &
        & Salpeter IMF; \citetalias{springel2003} SF; thermal \& kinetic SN. 
        \\& & & & {\bf Variant}: \musicpi\ that uses modified kinetic feedback, metal dependent cooling.
        \\& & & & For more details see \cite{sembolini2013a} \& appendix \ref{code:music}. \\

    & \owls         & \checkmark & \checkmark 
        & Chabrier IMF; \cite{schaye2008} (hereafter SDV08) SF ; kinetic SN; thermal AGN; CLOUDY \cite[][]{cloudy} (element-by-element) cooling; tracks 11 individual elements.
        \\& & & & For more details see \cite{schaye2010} \& appendix \ref{code:owls}.\\

    & \gadgettwox   & \checkmark & \checkmark 
        & Salpeter IMF; \citetalias{schaye2008} SF; thermal SN; thermal AGN, 
        \\& & & & For more details see \cite{pike2014} \& appendix \ref{code:gadget2x}. \\
    \hline
    Modern SPH
    & \gadgetxart      & \checkmark & \checkmark 
        & Chabrier IMF; \citetalias{springel2003} SF; kinetic SN; thermal ``quasar'' \& ``radio'' AGN; tracks individual 16 elements;  C4 Wendland kernel; artificial conduction to promote mixing; and time-dependent artificial viscosity.
        \\& & & & For more details see \cite{beck2016a} \& appendix \ref{code:gadget3x}. \\
    & \pesph        & \checkmark &
        & Chabrier IMF; \citetalias{springel2003} SF based scheme with additional quenching in massive galaxies based on \cite{rafieferantsoa2015}; probabilistic kinetic SN driven wind scheme; tracks 4 individual elements; pressure-entropy formulation of SPH of \cite{hopkins2013}; HOCTS(n=5) kernel with 128 neighbours.
        \\& & & & For more details see Huang et al, in prep  \& appendix \ref{code:pesph}.\\
    & G3-Magneticum & \checkmark & \checkmark
        & Chabrier IMF; \citetalias{springel2003} SF; thermal \& kinetic SN feedback; thermal ``quasar'' \& ``radio'' AGN;  CLOUDY \cite[][]{cloudy} (element-by-element) cooling; tracks 11 individual elements; C6 Wendland kernel with 295 neighbours. 
        \\& & & & For more details see \cite{hirschmann2014a} \& appendix \ref{code:magneticum}. \\
    \hline
\end{tabular}
\label{tab:codes}
\end{table*}

\par
We note that there are several unique combinations of subgrid physics modules: \ramses\ has AGN feedback but {\em NO supernova feedback}; \pesph\ does not explicitly include AGN feedback but does have additional quenching for massive galaxies. Some codes also have full physics variants, most notably \arepo, which has a model {\em without AGN physics}. 

\subsection{Data}\label{sec:data}
The cluster we have used for the \nifty\ comparison was drawn from the \href{http://www.music.ft.uam.es}{\music-2 cluster catalogue}\footnote{\url{http://music.ft.uam.es}} \cite[][]{sembolini2013a,sembolini2014a,biffi2014a}, which consists of a mass limited sample of re-simulated haloes selected from the MultiDark cosmological simulation \cite[][]{riebe2013}. The MultiDark run simulated a 1~Gpc$/h$ volume with $2048^3$ dark matter particles in a $(h,\Omega_m,\Omega_b,\Omega_\Lambda,\sigma_8,n_s)=(0.7,0.27,0.0469,0.73,0.82,0.95)$ cosmology based on the best-fit parameters to WMAP7+BAO+SNI data \cite[][]{jarosik2011} using {\sc ART} \cite[][]{art} and the data is accessible online via the \href{http://www.MultiDark.org}{\it MultiDark Database}\footnote{\url{http://www.MultiDark.org}}.

\par
The \music-2 cluster catalogue was constructed by selecting all the objects with masses $>10^{15}~\Msunh$ at $z=0$. These objects were then resimulated with 8 times better mass resolution using the  zooming technique described in \cite{klypin2001}. We focus on one cluster in particular, a moderately unrelaxed object with a mass of $\approx1.1\times10^{15}~\Msunh$. The mass resolution of the \nifty\ cluster in the pure dark matter simulations is $m_{\rm DM}=1.09\times10^{9}~\Msunh$, and in the gas physics runs, $m_{\rm DM}=9.01\times10^{8}~\Msunh$ \& $m_{\rm gas}=1.9\times10^{8}~\Msunh$. Several sets of these simulations were produced by each code. Here we focus on the so-called aligned runs, which is the set of simulations that result in approximately the same gravitational accuracy\footnote{For more details on how these simulations were aligned see \cite{nifty1}} for those codes that have produced full physics runs, i.e., subgrid physics modelling the formation of stars (and possibly black holes).

\subsection{Analysis}
The output produced by the codes was all analysed using a unified pipeline. Haloes and subhaloes were identified and their properties calculated using {\sc VELOCIraptor} \cite[aka {\sc stf}][freely available \href{https://github.com/pelahi/VELOCIraptor-STF.git}{\url{https://github.com/pelahi/VELOCIraptor-STF.git}}]{elahi2011}. This code first identifies haloes using a 3DFOF algorithm \cite[3D Friends-of-Friends in configuration space, see][]{fof} and then identifies substructures using a phase-space FOF algorithm on particles that appear to be dynamically distinct from the mean halo background, i.e. particles which have a local velocity distribution that differs significantly from the mean, i.e. smooth background halo. Since this approach is capable of not only finding subhaloes, but also tidal streams surrounding subhaloes as well as tidal streams from completely disrupted subhaloes \cite[][]{elahi2013a}, for this analysis we also ensure that a group is self-bound. Bound baryonic content of dark matter subhaloes is determined by associating gas and star particles with the closest dark matter particle in phase space belonging to a (sub)halo \cite[see ][for a study on identifying synthetic galaxies]{knebe2013a}. The internal self-energy of the gas is take into account when determining whether these particles are bound. If we were interested in identifying gas outflows from galaxies, we could relax this condition but for the purposes of this study, we require particles to be strictly bound. Galaxies are defined as any self-bound structure that contains 10 or more star particles, although for the purposes of this study we are generally interested in galaxies containing more than 100 star particles. We have not searched for self-bound star particle groups containing no dark matter, which are generally not produced by any of the codes, nor have we decomposed the stellar structures to search for bulges and discs.

\par
To match (sub)haloes across codes, we used the halo merger tree code which is part of the {\sc VELOCIraptor} package \cite[see][for more details]{srisawat2013}. This code is a particle correlator and relies on particle IDs being continuous across the simulations and time. As continuity of particle IDs is only guaranteed for dark matter N-body particles, we limit our cross-matching to {\em only} these particles. This means that in principle it is possible to have a gas or stellar ``galaxy'', whose dark matter halo has been mostly stripped away, i.e., baryon dominated, appear to have no analogue in another catalogue. However, the likelihood of such a circumstance for a well resolved self-bound object is negligible. The cross-matching between catalogue $A$ \& $B$ is done by identifying for each object in catalogue A the object in catalogue B that maximises the merit function:
\begin{equation}
    \mathcal{M}_{A_{i}B_{j}} = N_{A_{i}\bigcap B_{j}}^2/(N_{A_{i}}N_{B_{j}}),\label{eqn:merit}
\end{equation}
where $N_{A_{i}\bigcap B_{j}}$ is the number of particles shared between objects $i$ and $j$ and $N_{A_{i}}$ and $N_{B_{j}}$ are the total number of particles in the corresponding object in catalogues $A$ and $B$, respectively. Here we use $\mathcal{M}\geq0.2$, which has been shown to be a reasonable threshold in previous studies \cite[e.g.][]{libeskind2011}.  We arbitrarily use \music\ as our reference catalogue.

\section{The subhalo/galaxy population}\label{sec:subhalopop}
\begin{table}
\centering
\caption{Number of dark matter subhaloes and, for the full physics simulation, number of galaxies at $z=0$ with dark matter mass $M_{\rm S}\geq 2\times10^{10}\Msunh$ within $2\Mpch$ of the cluster centre. We define galaxies as objects that contain $\geq10$ star particles, that is stellar masses of $M_*\geq 1.9\times10^{9}\Msunh$, assuming one generation of star particles is produced by a gas particle. We have highlighted values which significantly \siginc{increase} or \sigdec{decrease} (by $\gtrsim25\%$) going from \dmsim$\rightarrow$\nrsim$\rightarrow$\radsim.}
\begin{tabular}{l|cccc}
\hline
    Code & \multicolumn{4}{c}{Number of subhaloes} \\
    & \dmsim & \nrsim & \radsim & Galaxies\\
\hline
    \music              & 378 & 303 & \siginc{428} & 325 \\
    \musicpi\           & $\shortparallel$ & $\shortparallel$ & \siginc{435} & 324 \\
    \ramses             & 290 & \sigdec{174} & 182 & 16 \\
    \arepo              & 360 & \sigdec{243} & 294 & 76 \\
    \arepo-SH           & $\shortparallel$ & $\shortparallel$ & \siginc{341} & 220 \\
    \gadgetxart & 381  & 356 & 388 & 262 \\
    \owls               & 383 & 327 & \siginc{440} & 307 \\
    \pesph              & 371 & 328 & \siginc{425} & 273\\
    \gadgettwox         & 399 & \sigdec{294} & 319 & 186 \\
    \magneticum         & 380 & 341 & 330 & 176 \\
\end{tabular}
\label{tab:nsubs}
\parnotes
\end{table}
We begin with the simplest comparison, the total number of (sub)haloes/galaxies within $2~\Mpch$ of the cluster's centre is listed in \Tableref{tab:nsubs} for each type of simulation, dark matter only (\dmsim), non-radiative (\nrsim) and full physics (\radsim) runs. When comparing the number of subhaloes, we could of course use the virial radius, $R_{200}^c$, which is $\sim2~\Mpch$ for all the simulations (\music\ has $R_{200}^c=1.69\Mpch$). However, since this radius does change from one simulation to the next by a few percent, for simplicity we fix the radial cut to $2~\Mpch$. 

\par
We see that for the \dmsim\ run, most codes have similar number of subhaloes to within Poisson errors\footnote{Despite the fact that all codes use the same initial conditions, a object in one code may lie just outside the radial cut used whereas in another code the object lies just within as a result of differences in the gravitational integration (see appendix in \cite{nifty1} for related discussion on aligning codes). Moreover, the same object will experience slightly different tidal forces in each code and a subhalo that lies above the resolution threshold used in one code may have been stripped enough to lie below it in another.}. This pattern is also observed in the non-radiative simulations. \arepo, the moving mesh code, is a moderate outlier. The main outlier is the sole adaptive mesh code, \ramses, which has $20\%$ fewer dark matter subhaloes in the \dmsim\ run. This number drops by $\sim40\%\ (30\%)$ going from \dmsim$\rightarrow$\nrsim\ for \ramses\ (\arepo), whereas in most SPH codes it decreases by only $\sim10-20\%$. The SPH outlier is \gadgettwox, a classic SPH code, where the number of subhaloes decreases by $25\%$. 

\par
The picture as always is more complex with the addition of feedback physics. Recall that certain codes, \music, \arepo, and \gadgettwox\ have more than one flavour of full physics runs. In almost all cases, going from \nrsim$\rightarrow$\radsim, i.e., including cooling and feedback processes, increases the total number of subhaloes. Most SPH codes have even more in the \radsim\ runs than in the \dmsim, the notable exception being \magneticum\ and \gadgettwox, which behave similarly to \arepo\ and \ramses. Some of this increase is due to the resolution limit imposed: subhaloes must be composed of 20 or more particles, be they star particles, gas particles or dark matter particles. Thus in the \radsim\ runs, subhaloes with lower dark matter masses are counted if they also contain baryons. However most of the increase occurs at masses above the resolution threshold imposed and is a result of the influence of baryons on dark matter.

\par 
The diversity in the number of subhaloes in the full physics runs is mirrored by the galaxy population. Most codes result in the cluster containing on the order of 200 galaxies, though this number ranges from 16 to 325. As our synthetic cluster is of similar to the Virgo cluster one would expect $\sim60$ massive galaxies (stellar masses $M_*\gtrsim10^{9.5}$), although the total number of cluster members is $\sim1000$ \cite[][]{boselli2014a}. Caution should be used when directly comparing numbers is given the likely differences in merger histories between Virgo and our synthetic cluster and the complexity of estimating stellar masses from observations but we should expect similar numbers of galaxies. Typically most codes produce more galaxies than one might naively expect. The two codes that stand out are the mesh codes, which have far fewer galaxies (and subhaloes) than the SPH codes. \ramses\ has the fewest subhaloes and startlingly few galaxies, by far the lowest of any of the codes. \arepo\ (Illustris physics) also has few galaxies, similar to that observed in Virgo, and a low fraction of subhaloes hosting galaxies. Its variant, \arepo-SH has numbers similar to the SPH codes. Amongst the SPH codes, \magneticum\ has the smallest galaxy population and a low galaxy occupation fraction of $\sim50\%$. Other codes, like \music\ \& \owls\ have occupation fractions of $\sim80\%$ and between $\sim270-320$ galaxies. 

\par 
Perhaps the most relevant change to note is that due to different flavours of subgrid physics. The \arepo-SH simulation, which has the same subgrid physics as \music, has a moderate change in the number of subhaloes but an enormous change in the galaxy population compared to \arepo. The \music\ variant shows little change in the number. 

\subsection{Subhaloes}
\label{sec:subhalopop:subhaloes}
\begin{figure*}
    \centering
    \includegraphics[height=0.33\textheight,trim=0.cm 0.9cm 2.5cm 1.5cm, clip=true]{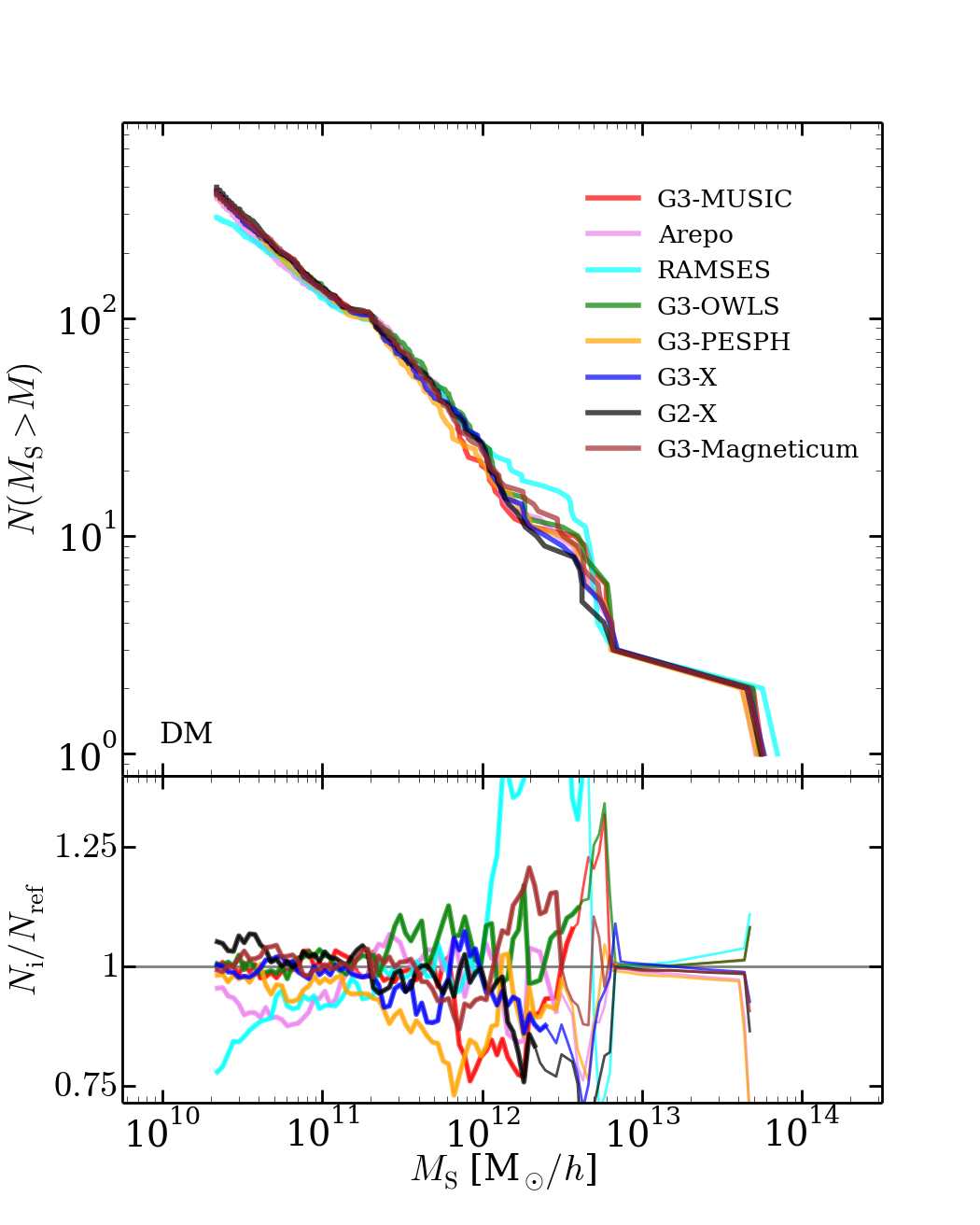}\hspace{-2.75pt}
    \includegraphics[height=0.33\textheight,trim=3.3cm 0.9cm 2.5cm 1.5cm, clip=true]{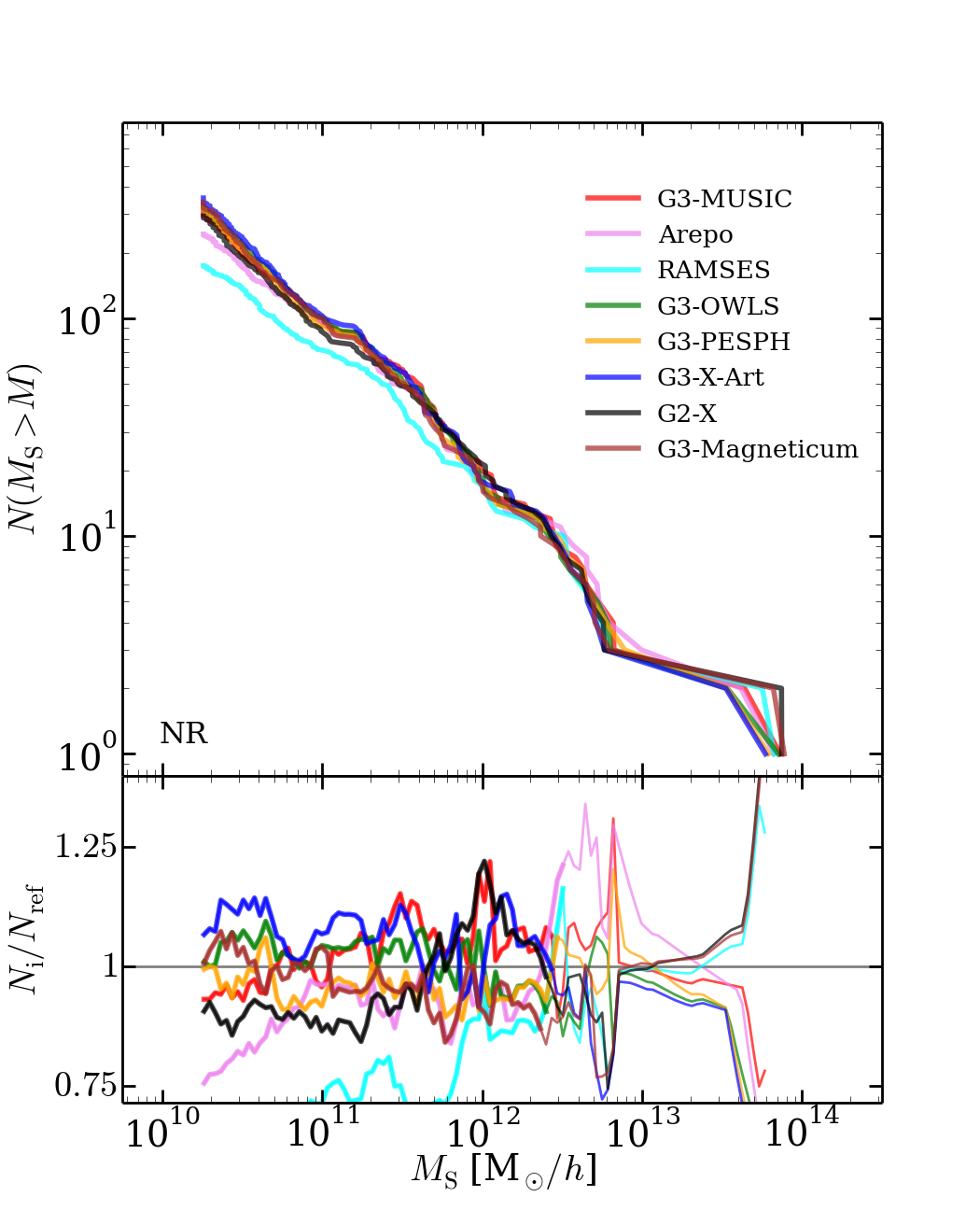}\hspace{-2.75pt}
    \includegraphics[height=0.33\textheight,trim=3.3cm 0.9cm 2.5cm 1.5cm, clip=true]{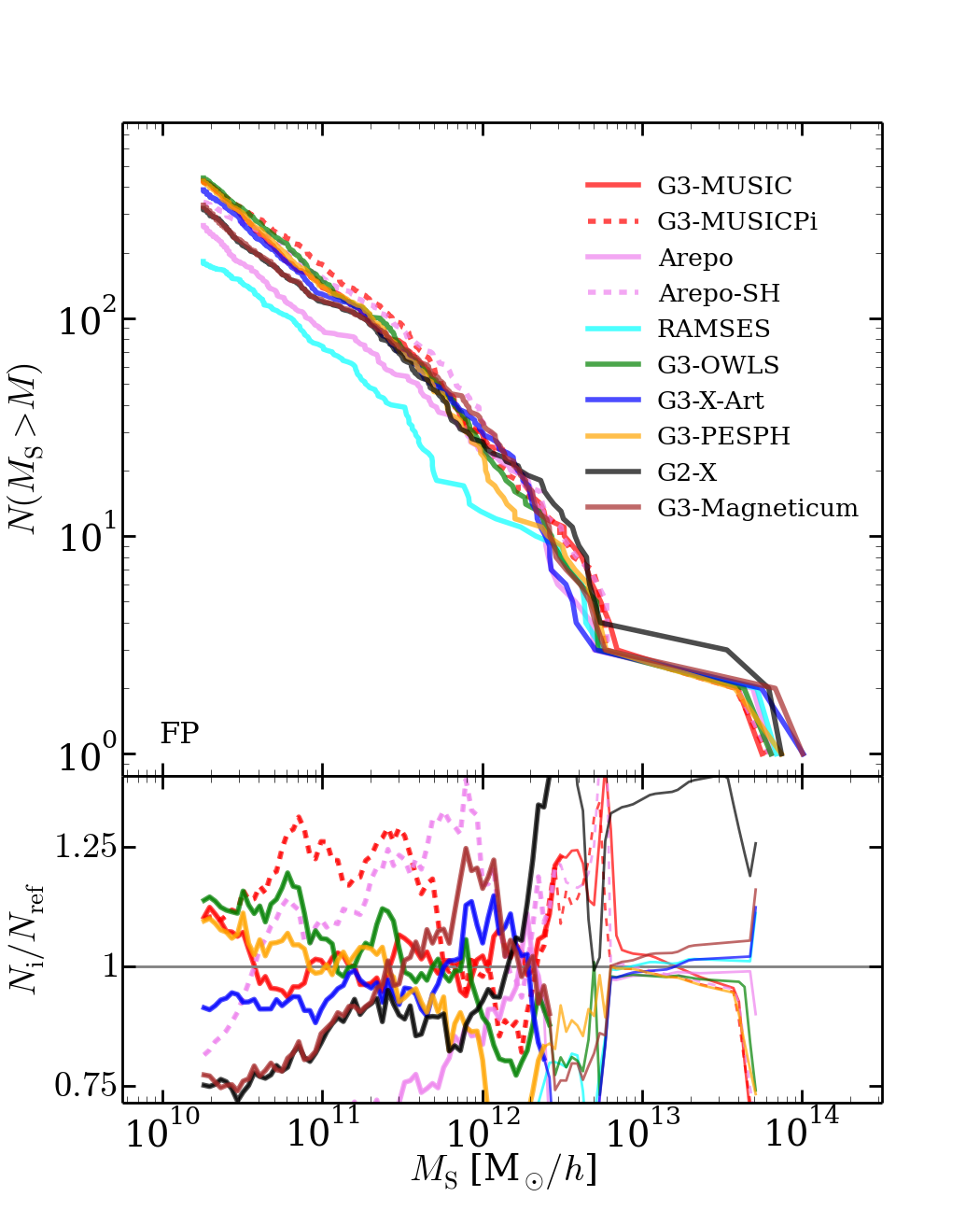}\hspace{-2.75pt}
    \caption{The subhalo cumulative mass function (top) and the difference between a given catalogue and the median calculated using all other catalogues (bottom). The thin lines in the residual plots correspond to where the catalogue's cumulative distribution has fewer than 10 subhaloes, ie: the region where the statistical error is $\gtrsim10\%$. Three types of simulations are shown: \dmsim\ (left), \nrsim\ (middle), and \radsim\ (right) respectively.}
    \label{fig:massdistrib}
\end{figure*}

\begin{figure*}
    \centering
    \includegraphics[height=0.33\textheight,trim=0.cm 0.9cm 2.5cm 1.5cm, clip=true]{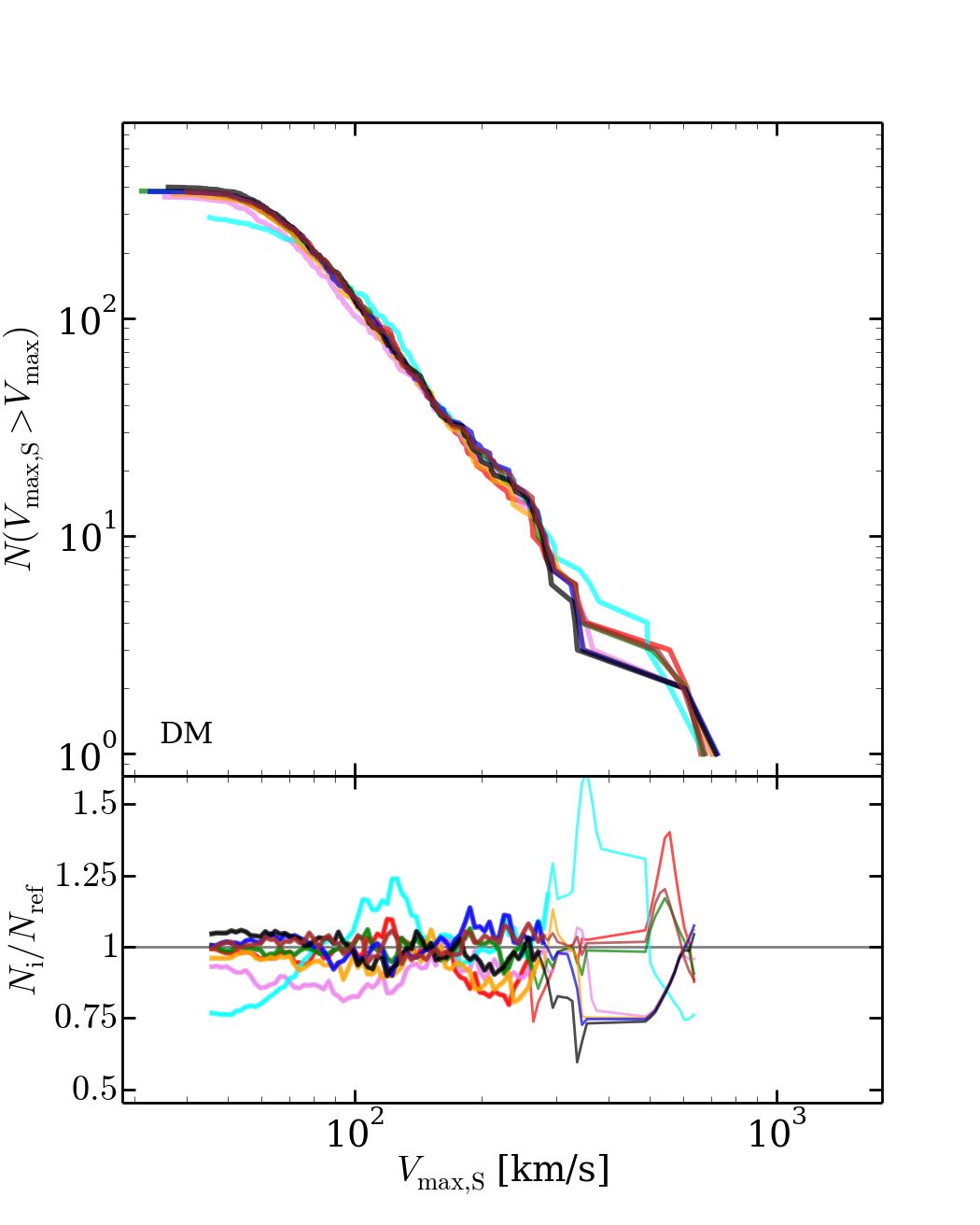}\hspace{-2.75pt}
    \includegraphics[height=0.33\textheight,trim=3.3cm 0.9cm 2.5cm 1.5cm, clip=true]{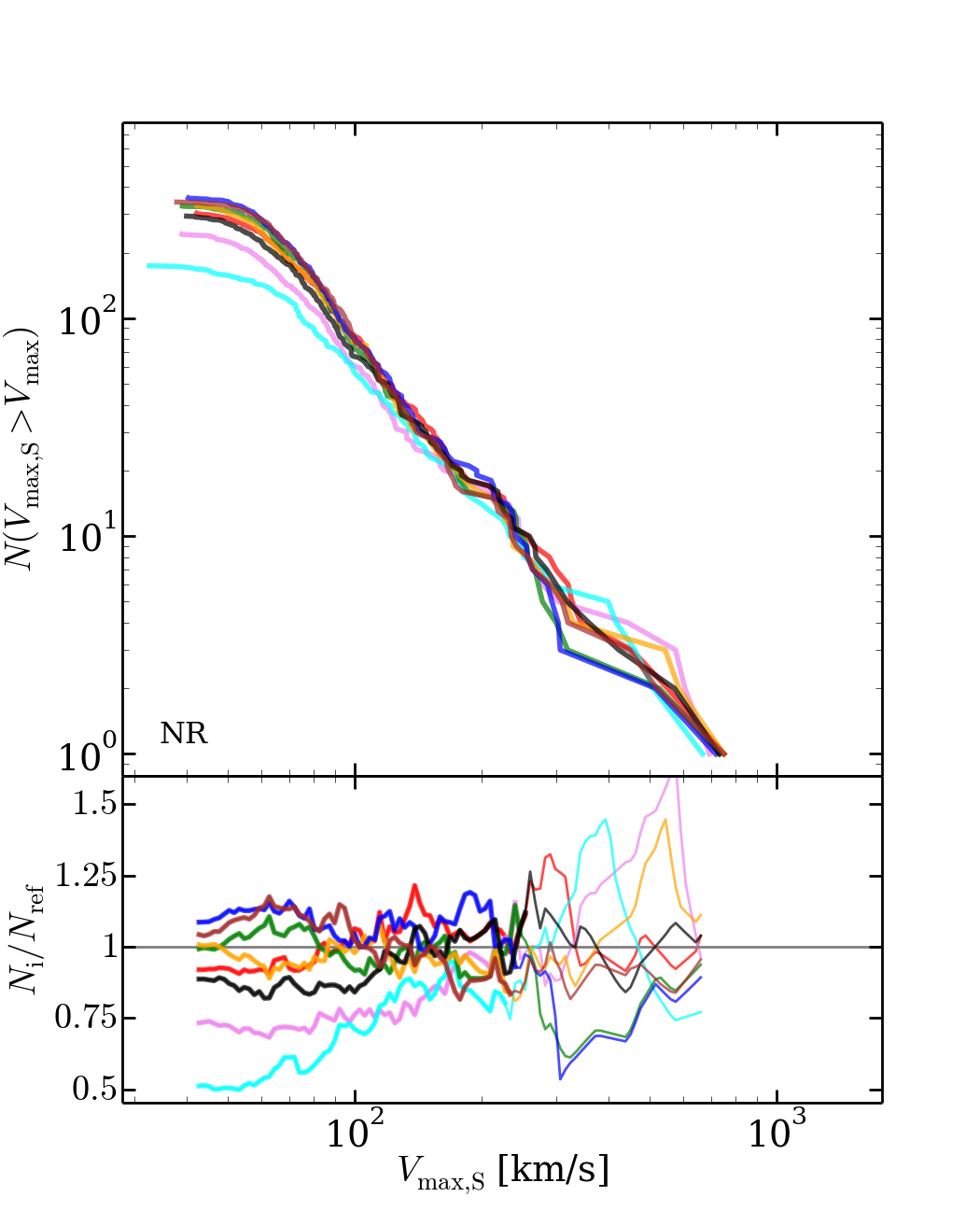}\hspace{-2.75pt}
    \includegraphics[height=0.33\textheight,trim=3.3cm 0.9cm 2.5cm 1.5cm, clip=true]{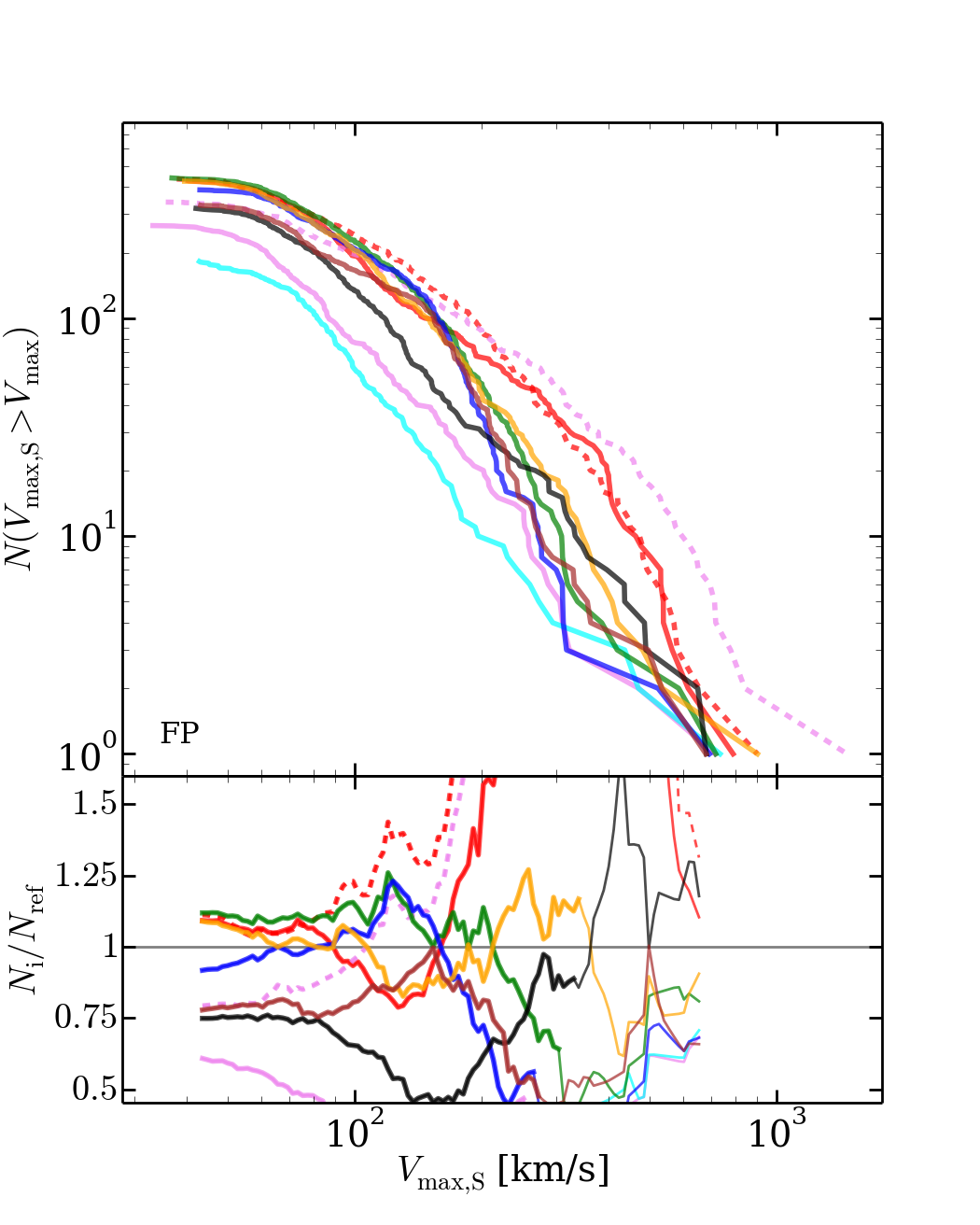}\hspace{-2.75pt}
    \caption{Similar to \Figref{fig:massdistrib} but for the subhalo cumulative maximum circular velocity distribution. For legend see \Figref{fig:massdistrib}.}
    \label{fig:vmaxdistrib}
\end{figure*}

\begin{figure*}
    \centering
    \includegraphics[height=0.33\textheight,trim=0.cm 0.9cm 2.5cm 1.5cm, clip=true]{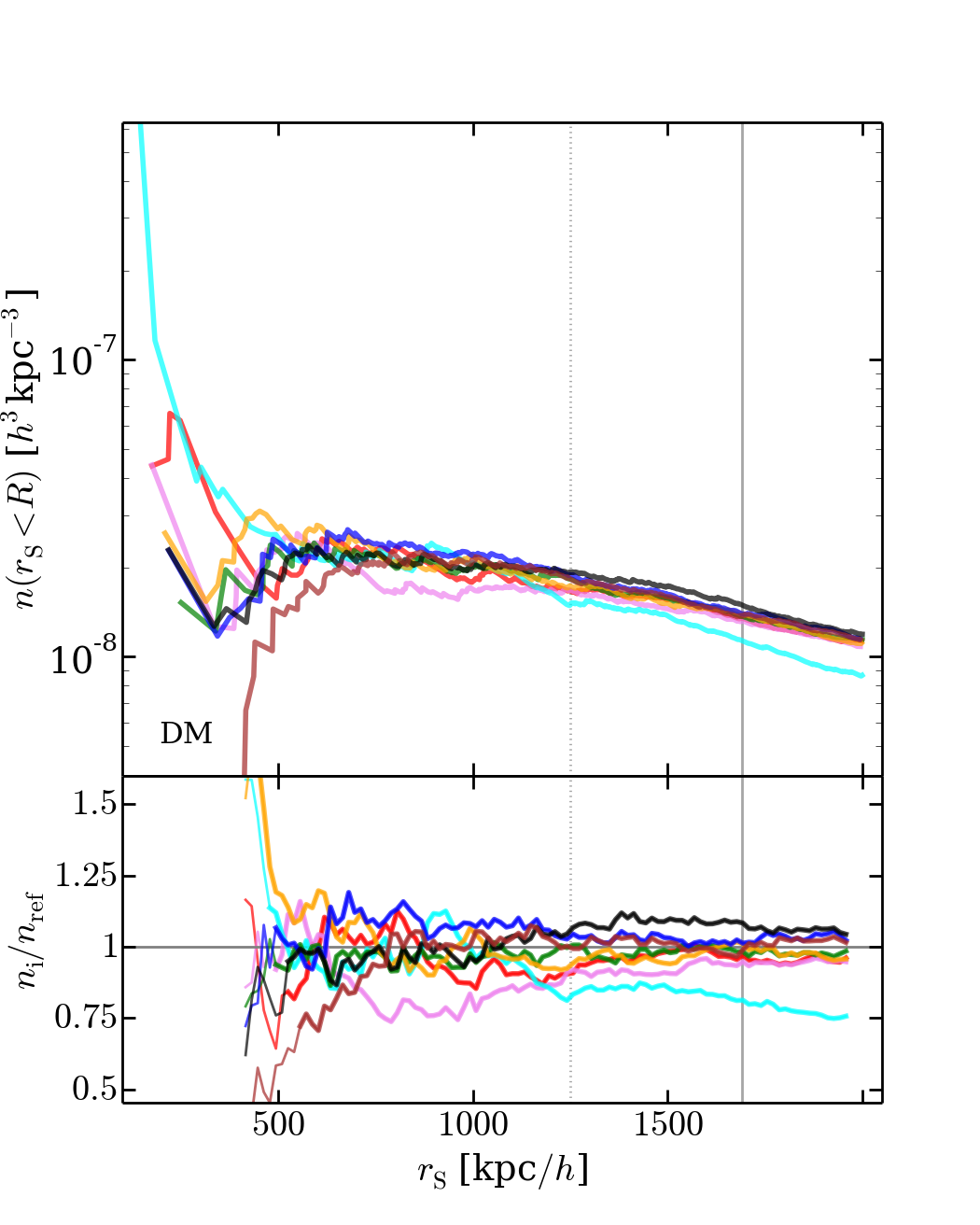}\hspace{-2.75pt}
    \includegraphics[height=0.33\textheight,trim=3.3cm 0.9cm 2.5cm 1.5cm, clip=true]{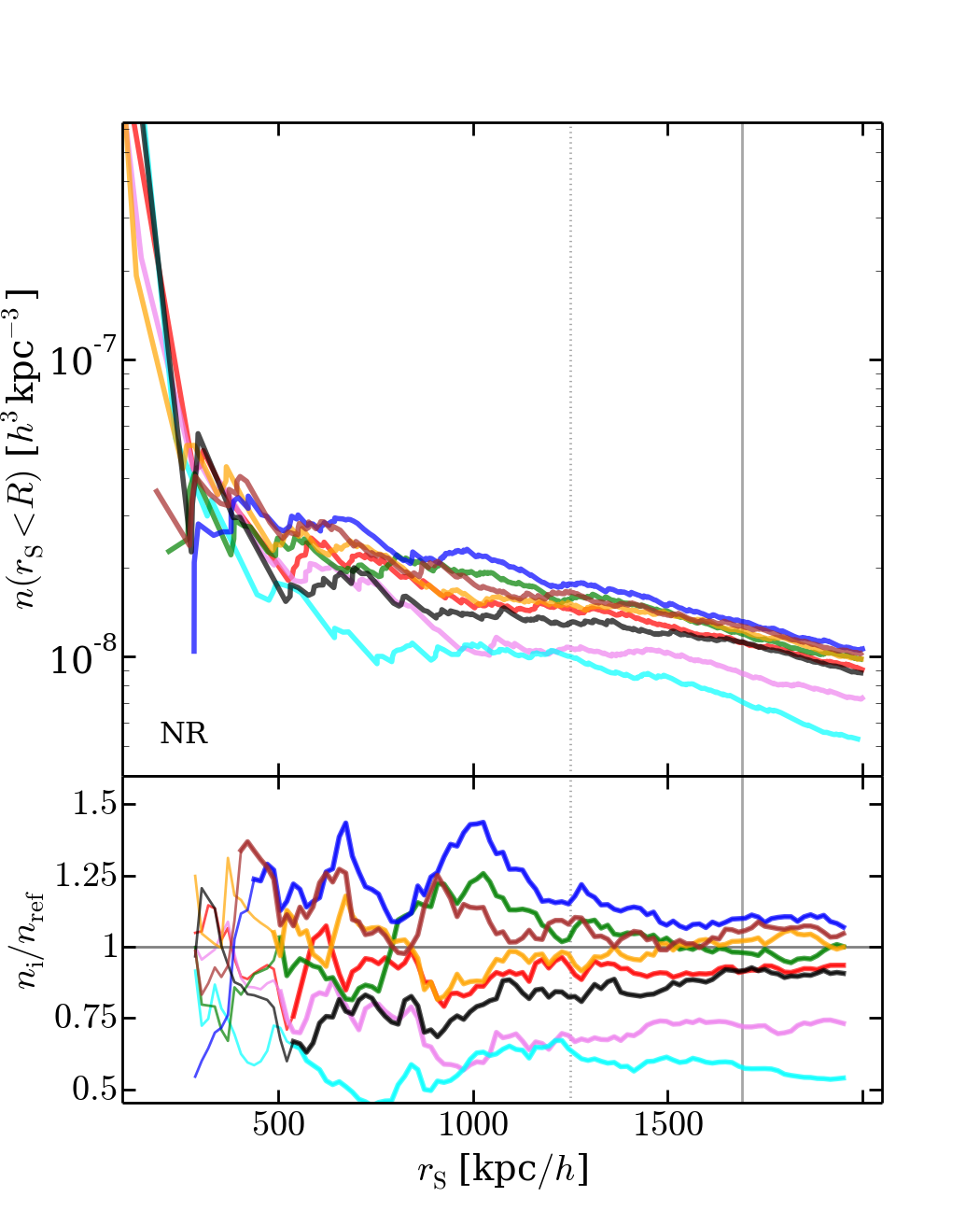}\hspace{-2.75pt}
    \includegraphics[height=0.33\textheight,trim=3.3cm 0.9cm 2.5cm 1.5cm, clip=true]{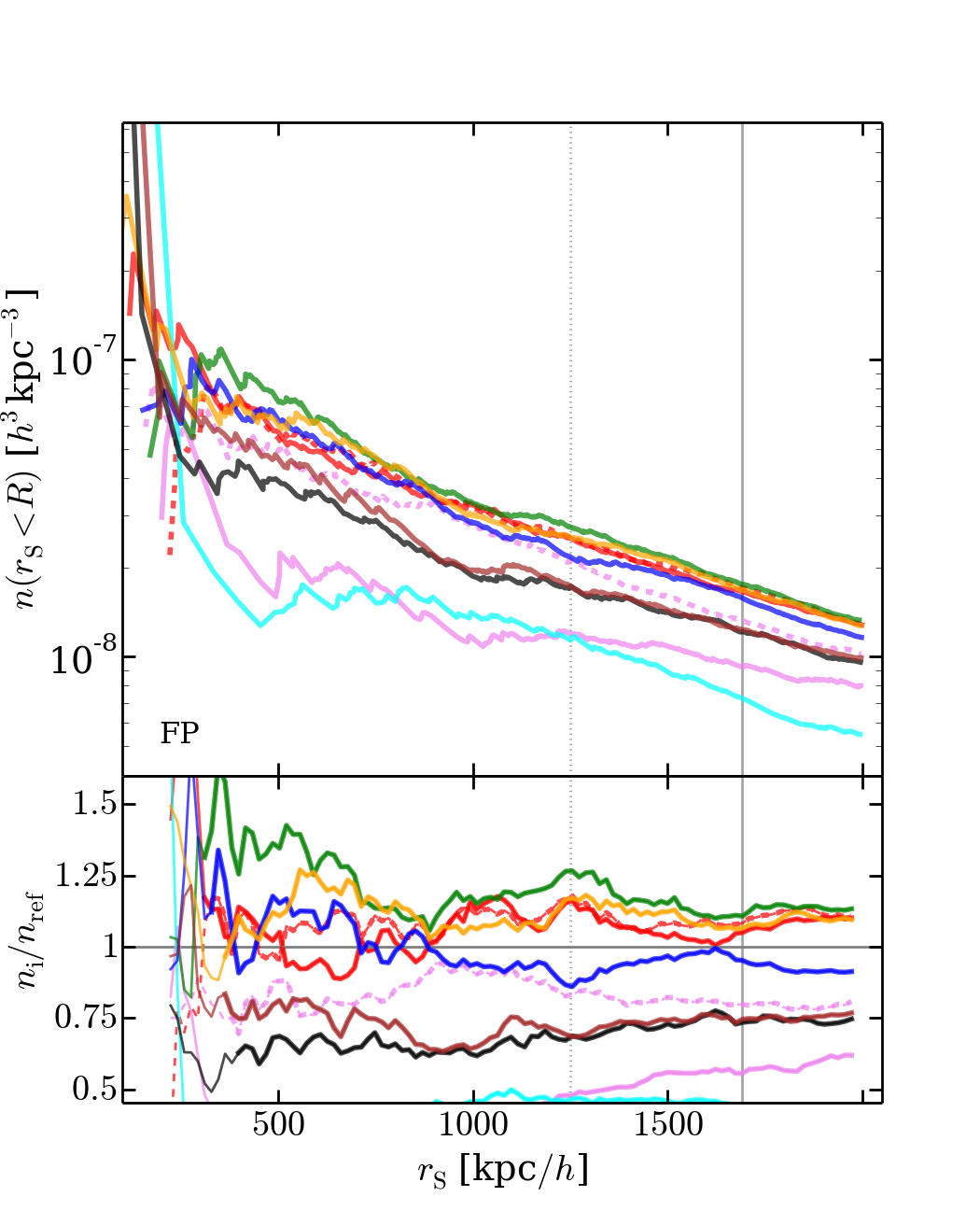}\hspace{-2.75pt}
    \caption{Similar to \Figref{fig:massdistrib} but for the mean enclosed subhalo number density. We plot a solid (dashed) gray vertical line at $R_{200}^{c}$ ($R_{500}^{c}$) of the \music\ cluster for reference. For legend see \Figref{fig:massdistrib}.}
    \label{fig:raddistrib}
\end{figure*}
We next examine the cumulative mass and maximum circular velocity distributions shown in \Figref{fig:massdistrib} \& \ref{fig:vmaxdistrib}, with the lower panels showing the ratio of the distribution from one code relative to the median calculated using all other codes. The mass distribution shows all codes produce similar DM results. The only noticeable feature is the lack of small subhaloes in \ramses, which matches the other codes reasonably well for subhaloes composed of $\gtrsim50$ particles. The overall scatter for all codes using the \gadget\ Tree-PM gravity solver is $\lesssim10\%$. The excess of $M_{\rm S}\sim10^{12}\Msunh$ subhaloes in \ramses\ is a result of these subhaloes residing outside our radial cut in most other simulations and a few subhaloes having slightly higher masses in \ramses. 

\par 
The general picture becomes worse with the inclusion of gas, although not significantly so despite the variety of approaches modelling gas. The scatter is typically $\lesssim15\%$. The key feature is that mesh codes, particularly \ramses, have fewer objects. However, these codes do not exhibit the precisely the same behaviour: \arepo\ has fewer low mass, poorly resolved objects, whereas \ramses\ actually has a slightly flatter mass distribution across a wide range of masses. This lack of subhaloes in \ramses\ is likely related to the known issue of lower small-scale power found in \ramses\ compared to \gadget\ at $z>1$ where these objects should form \cite[See figure 1 of][]{schneider2015a}.

\par
Including feedback physics increases the scatter to $\gtrsim25\%$ for masses $\lesssim10^{13}~\Msunh$, although the form of the mass function is generally unchanged. The codes that bracket the overall variation are the \music\ variant and \ramses. 

\par
The maximum circular velocity is less affected by tidal forces and differences in the position of a given subhalo relative to the host but is sensitive to changes in the central concentration of subhaloes \cite[e.g.][]{onions2013a}. Therefore, we might expect less scatter arising from differences in the position of a subhalo and see biases in the central concentration that would not be evident in the mass distribution for well resolved haloes where $\vmax$ can be accurately measured. Like the mass distribution, the DM simulations agree with one another if one accounts for the difference in normalisation, i.e., the residuals are flat. The non-radiative simulations also have little code-to-code scatter, with two exceptions. Both \ramses\ \& \arepo\ have fewer subhaloes low $\vmax$ subhaloes and \ramses\ has also flatter slopes (the residuals have a slight tilt). 

\par
Feedback physics causes the $\vmax$ variation to be more pronounced than that seen in the mass. This variation is a result of appreciable  amounts of baryons being moved around by the different cooling and feedback physics included by each code. \music\ subhaloes have higher circular velocities, whereas most other codes have steeper slopes with more low $\vmax$ subhaloes. It is worth noting that \arepo-SH, the variant not including AGN feedback (dashed lines), not only contains many more galaxies (see \Tableref{tab:nsubs}) but also contains subhaloes with high $\vmax$. 

\par
We examine the radial distribution via the enclosed number density $n(r_{\rm S}<R)$ in \Figref{fig:raddistrib}, where $r_{\rm S}$ is the radial distance a subhalo is from the cluster centre and $R$ is the radial distance cut. For all simulation types, almost all codes produce the same overall shape, i.e., the residuals are flat though the normalisation varies. Only in the very outskirts are significant differences apparent, which is not unexpected given that the DM profiles of the overall cluster agree to within $\sim10\%$. The \dmsim\ simulations show the smallest amount of scatter and the \radsim\ simulations the most. 
The outlier in all simulation types is \ramses, which drops faster than other codes. Note that the major jumps in the residuals seen in the core region are a result of differences in the positions of the few subhaloes identified deep within the host.

\subsection{Baryons}
\label{sec:subhalopop:baryons}
Here we focus on the baryonic component and the changes in the subhalo population resulting from the inclusion of adiabatic and full physics. Gas cooling can contract the core of a field dark matter halo \cite[e.g.][]{gnedin2004}, though the effect on a subhalo in the hot cluster environment is not as clear cut. Stripping of cold, low entropy gas contained in a subhalo as it falls into the cluster environment can counter adiabatic contraction. Codes treat mixing of low entropy gas differently, and consequently, the concentration of subhaloes should differ. 

\par
\begin{figure}
    \centering
    \includegraphics[width=0.425\textwidth,trim=0.cm 3.0cm 1.25cm 1.15cm, clip=true]{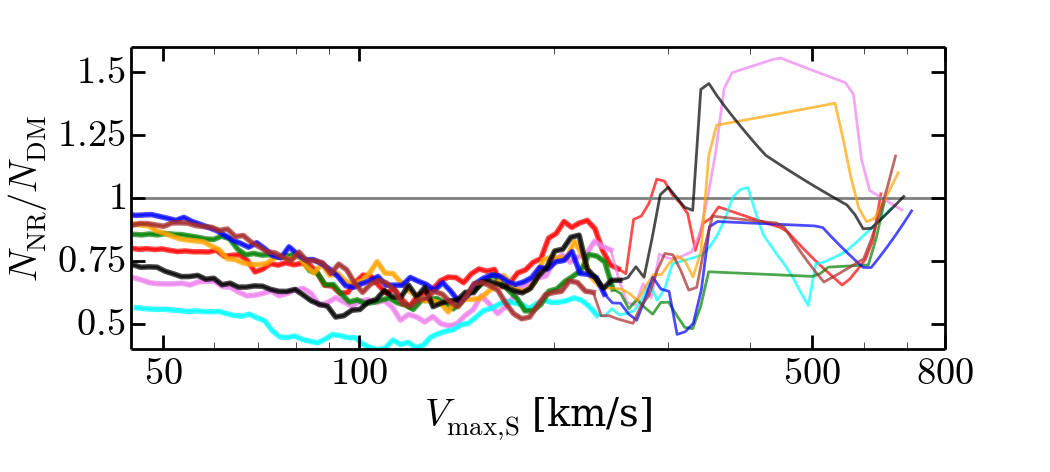}
    \\\vspace{-1.75pt}
    \includegraphics[width=0.425\textwidth,trim=0.cm 3.0cm 1.25cm 1.2cm, clip=true]{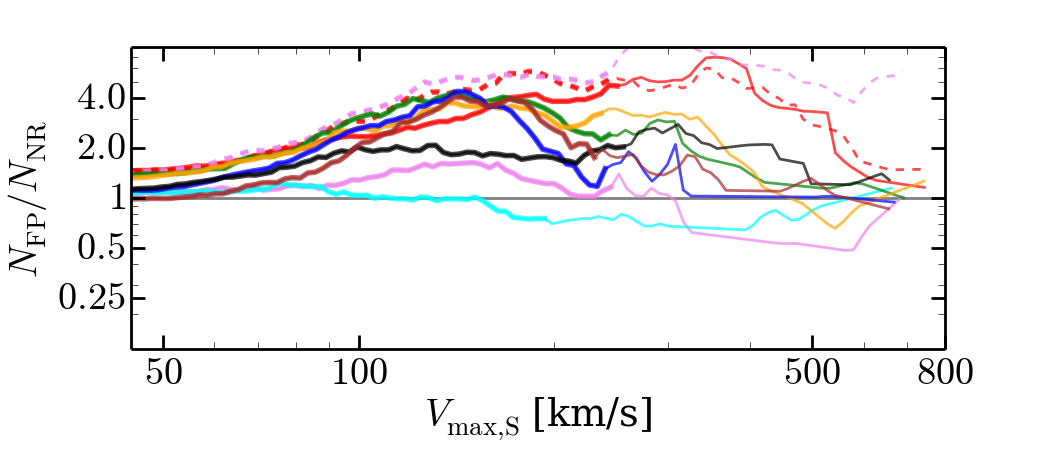}
    \\\vspace{-1.75pt}
    \includegraphics[width=0.425\textwidth,trim=0.cm 0.5cm 1.25cm 1.2cm, clip=true]{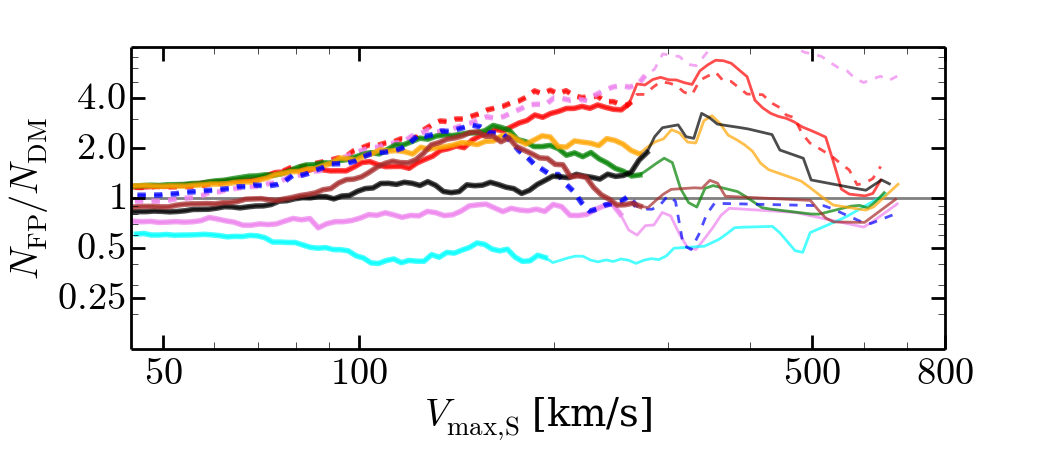}
    \caption{Difference in the cumulative $\vmax$ distribution between the \nrsim\ and \dmsim\ simulation (top), \radsim\ and \nrsim (middle), and \radsim\ and \dmsim (bottom). Line colours styles are the same as in \Figref{fig:massdistrib}. For legend see \Figref{fig:massdistrib}.}
     \label{fig:vmaxcompdistrib}
\end{figure}

 
We highlight the differences in the $\vmax$ distribution between the runs in \Figref{fig:vmaxcompdistrib}. The ratio between \nrsim\ \& \dmsim\ has a noticeable tilt for haloes with $\vmax\lesssim200$~km/s for the SPH simulations, with the two mesh codes, \ramses \& \arepo, having less pronounced tilts. Adiabatic physics results in fewer subhaloes, less centrally concentrated subhaloes, increasing the number of low $\vmax$ subhaloes over high $\vmax$ subhaloes due to the efficient stripping of gas from small subhaloes.

\par
In full physics runs gas can cool and contract, centrally concentrating material and forming galaxies, although this can be completely counteracted by feedback physics \cite[e.g.][]{abadi2010,dicintio2011}. The middle and bottom panels of \Figref{fig:vmaxcompdistrib}, show that, for SPH codes, cooling and feedback physics has counteracted the expansion of subhaloes arising in the adiabatic runs, with \magneticum\ and both \music\ variants experiencing the largest change. Interestingly, the residual for \ramses\ \& \arepo\ with AGN feedback are flat, i.e., feedback processes have balanced the contraction due to cooling, leaving haloes relatively unchanged from how they appear in pure DM simulations. Without AGN feedback, \arepo-SH, produces more high-$\vmax$ subhaloes in line with the changes seen in \music. 

\par
\begin{figure}
    \centering
    \includegraphics[width=0.425\textwidth,trim=0.cm 3.0cm 1.25cm 1.15cm, clip=true]{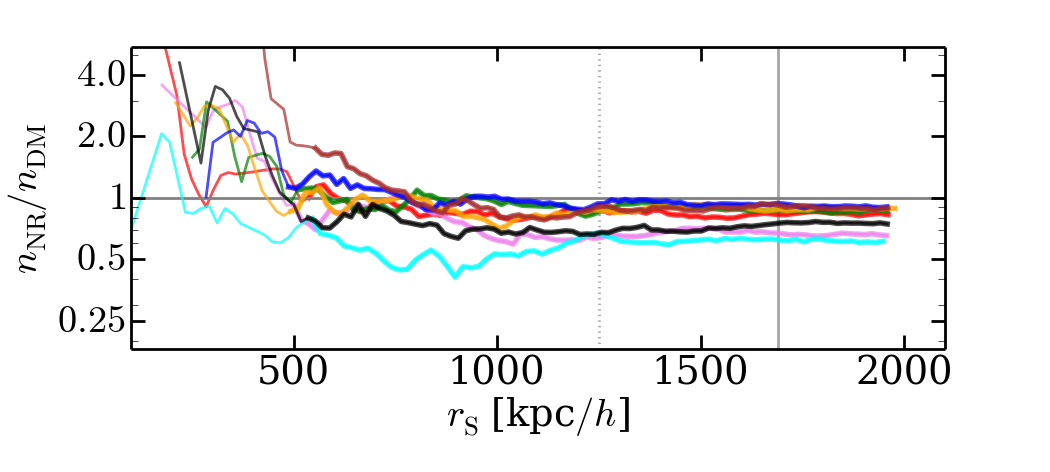}
    \\\vspace{-1.75pt}
    \includegraphics[width=0.425\textwidth,trim=0.cm 3.0cm 1.25cm 1.2cm, clip=true]{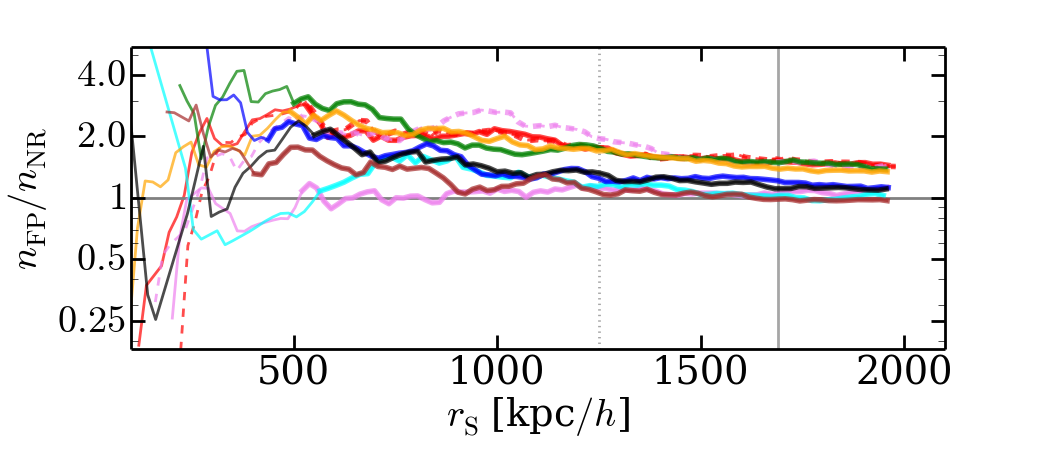}
    \\\vspace{-1.75pt}
    \includegraphics[width=0.425\textwidth,trim=0.cm 0.5cm 1.25cm 1.2cm, clip=true]{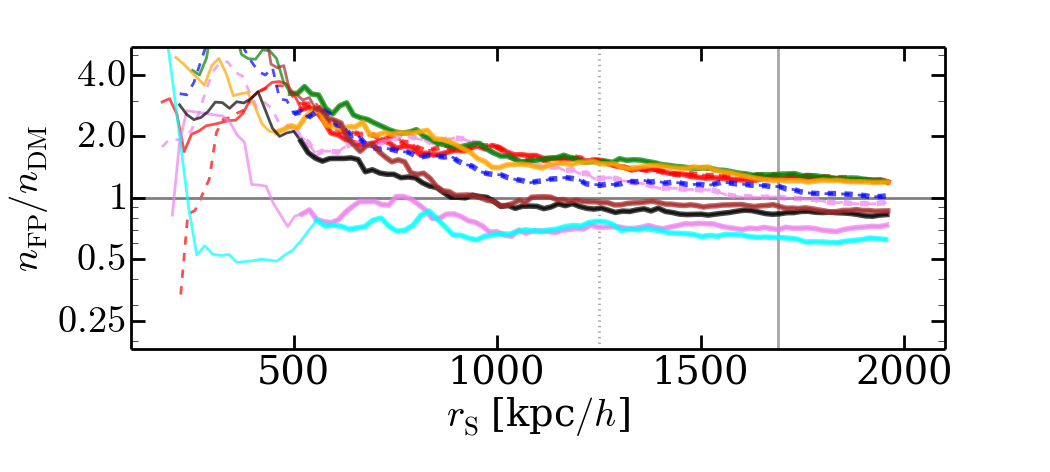}
    \caption{Difference in the average enclosed number density of subhaloes similar to \Figref{fig:vmaxcompdistrib}. For legend see \Figref{fig:massdistrib}.}
     \label{fig:radcompdistrib}
\end{figure}

 
The radial distribution shown in \Figref{fig:radcompdistrib} is not significantly affected by additional physics except in the very centre. The inclusion of gas increases the number density within $500~\kpch$. This very central concentration is removed going from \nrsim\ to \radsim, although full physics runs are still centrally biased compared to pure \dmsim\ simulations, in agreement with \cite{libeskind2010}. Only \ramses\ appears to have flat residuals. 

\par
\begin{figure}
    \centering
    \includegraphics[width=0.38\textwidth,trim=0.cm 2.5cm 2.5cm 2.6cm, clip=true]{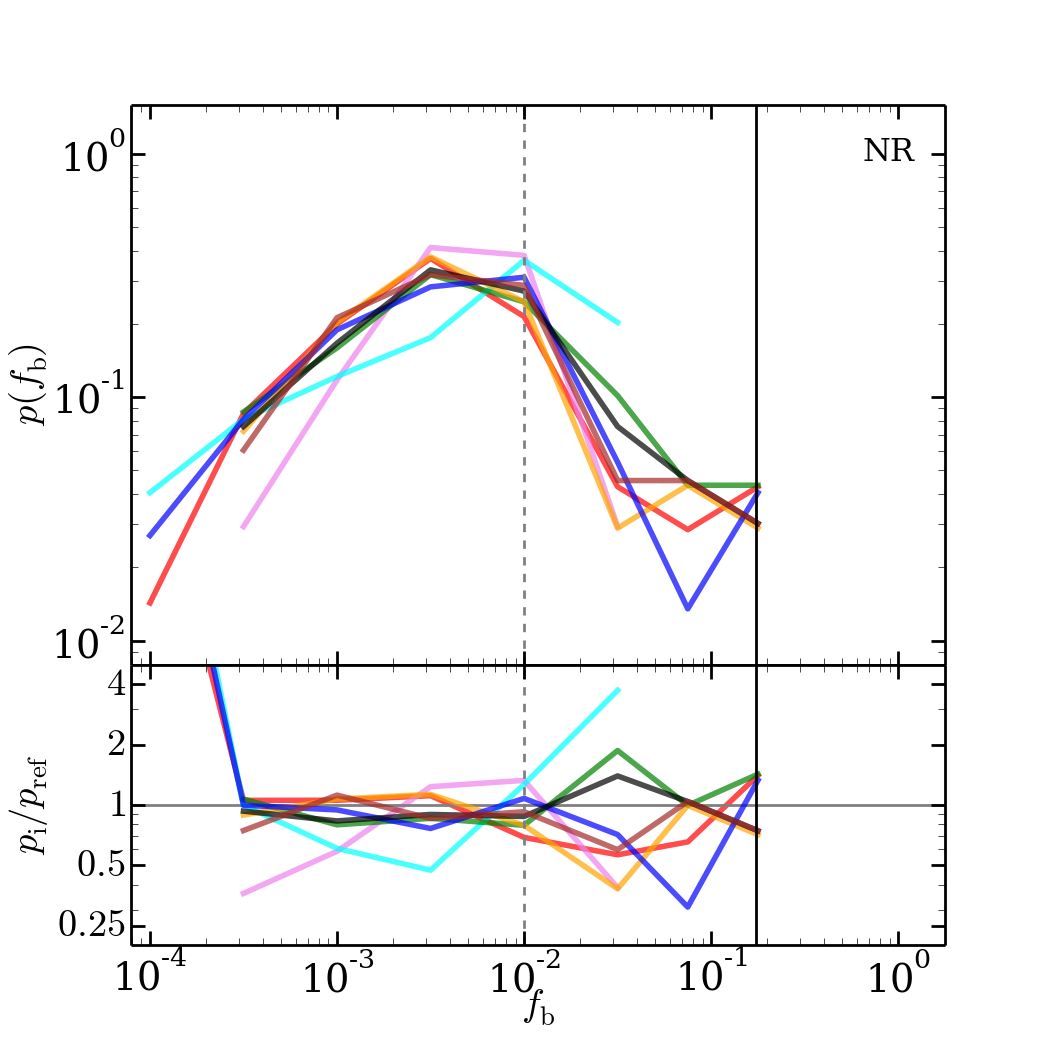}
    \\\vspace{-1.75pt}
    \includegraphics[width=0.38\textwidth,trim=0.cm 0.cm 2.5cm 2.7cm, clip=true]{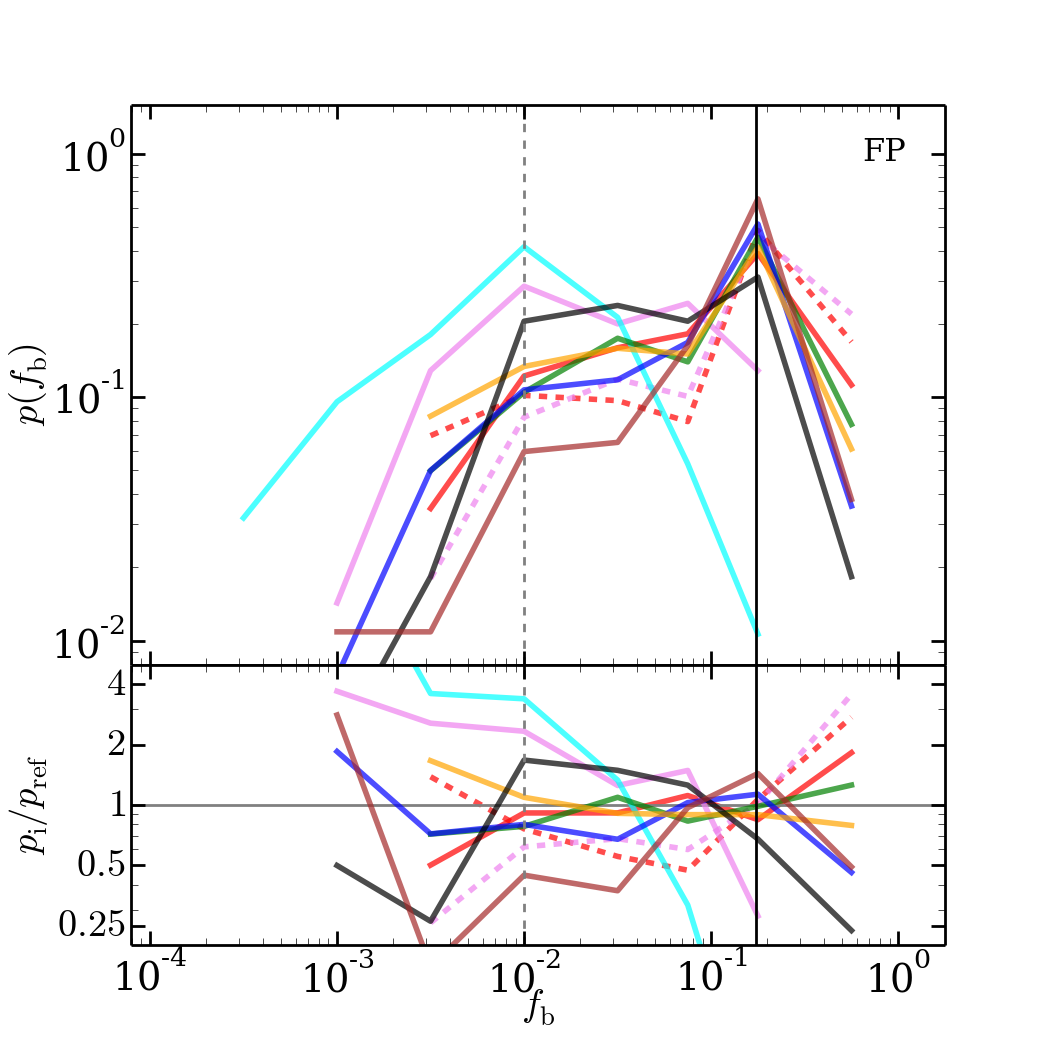}
    \caption{Baryon fraction distribution (top) and the residuals relative to the median calculated in the same fashion as \Figref{fig:massdistrib} (bottom subpanel). We show the \nrsim\ and \radsim\ simulations in the top and bottom major panels respectively. We indicate the cosmic baryon fraction $\Omega_b/\Omega_m$ by a solid vertical line. Colour, line types are the same as in \Figref{fig:massdistrib}.}
    \label{fig:baryonfracdistrib}
\end{figure}

We next examine baryon fractions in \Figref{fig:baryonfracdistrib}, where we show $f_{\rm b}$ of all objects containing some amount of bound gas or stars. Focusing on the non-radiative simulations, the first notable feature is that the peak of the $f_{\rm b}$ distribution is significantly less that $\Omega_b/\Omega_m$, the cosmic baryon fraction (solid vertical line). The hot cluster environment efficiently strips baryons away from subhaloes. Most codes have the same overall shape, a lognormal centred on $f_{\rm b}\sim3\times10^{-3}$. \ramses\ may be an exception as it is not as strongly peaked as the other codes. There is also the suggestion of a second peak around the cosmic baryon fraction. These two distributions arise from galaxies that have resided in the cluster environment and newly infalling galaxies that have yet to be stripped, of which there are few within $2\Mpch$. 

\par
The bottom two panels of \Figref{fig:baryonfracdistrib} shows that gas cooling and star formation allows subhaloes to retain significantly higher baryon fractions in the cluster environment. There are even a few subhaloes with $f_{\rm b}>\Omega_b/\Omega_m$. These are typically undergoing some tidal disruption, which has momentarily increased $f_{\rm b}$. Key is the increase in the code-to-code scatter. \arepo\ peaks and plateaus at $f_{\rm b}\gtrsim10^{-2}$, whereas most SPH codes have peaks at $\Omega_b/\Omega_m$, indicating \arepo's feedback processes are stronger and/or more efficient in moving material out of host subhalo. \ramses\ is even more extreme, containing no subhaloes with $f_{\rm b}$ close to the cosmic baryon fraction. Interestingly, \gadgettwox\ has a broad baryon fraction distribution, with a less noticeable peak at $\Omega_b/\Omega_m$.

\begin{figure}
    \centering
    \includegraphics[width=0.38\textwidth,trim=0.cm 30.5cm 2.5cm 7cm, clip=true]{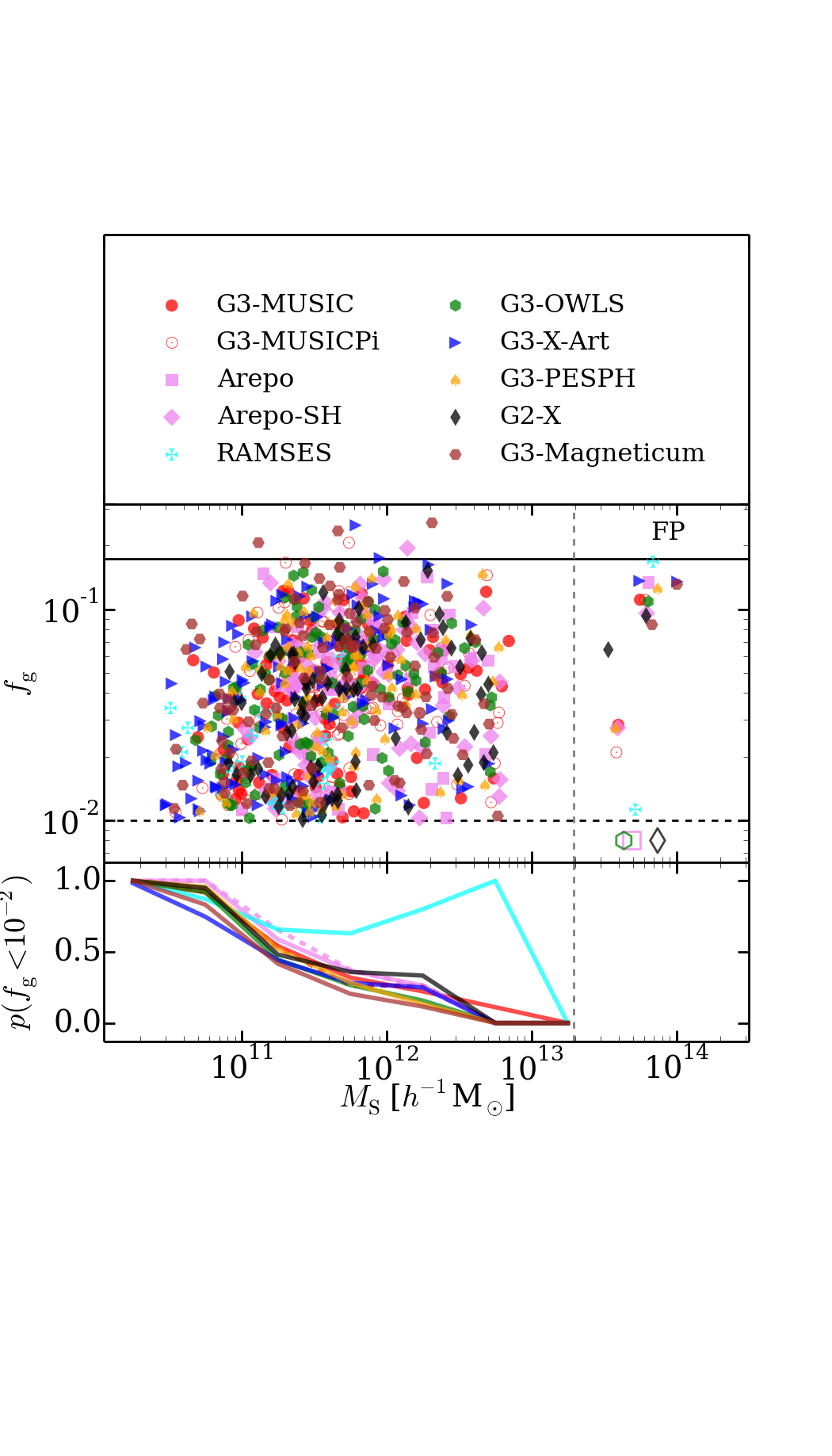}
    \\\vspace{-1.75pt}
    \includegraphics[width=0.38\textwidth,trim=0.cm 13.2cm 2.5cm 16.15cm, clip=true]{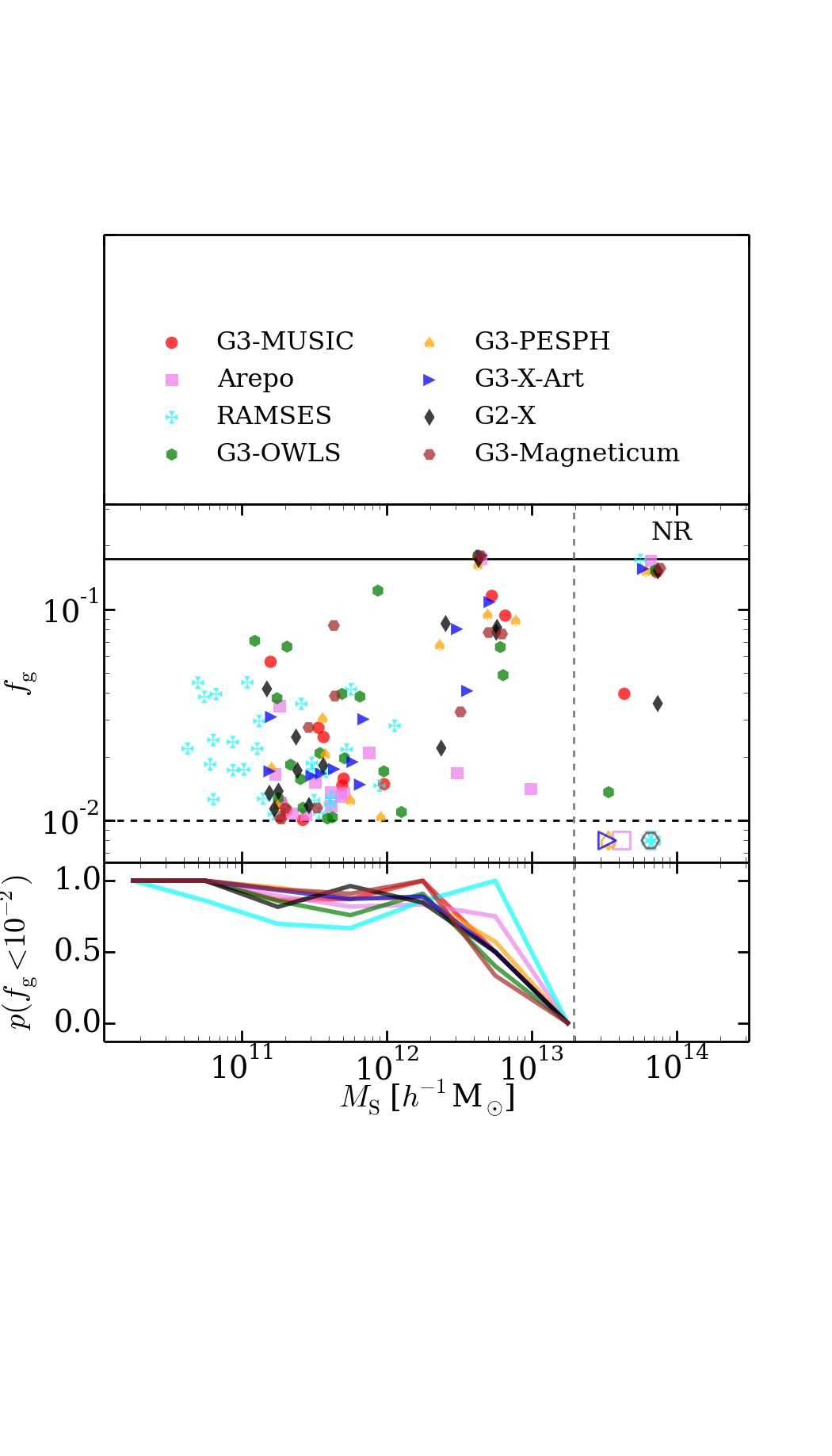}
    \\\vspace{-1.75pt}
    \includegraphics[width=0.38\textwidth,trim=0.cm 10.75cm 2.5cm 16.15cm, clip=true]{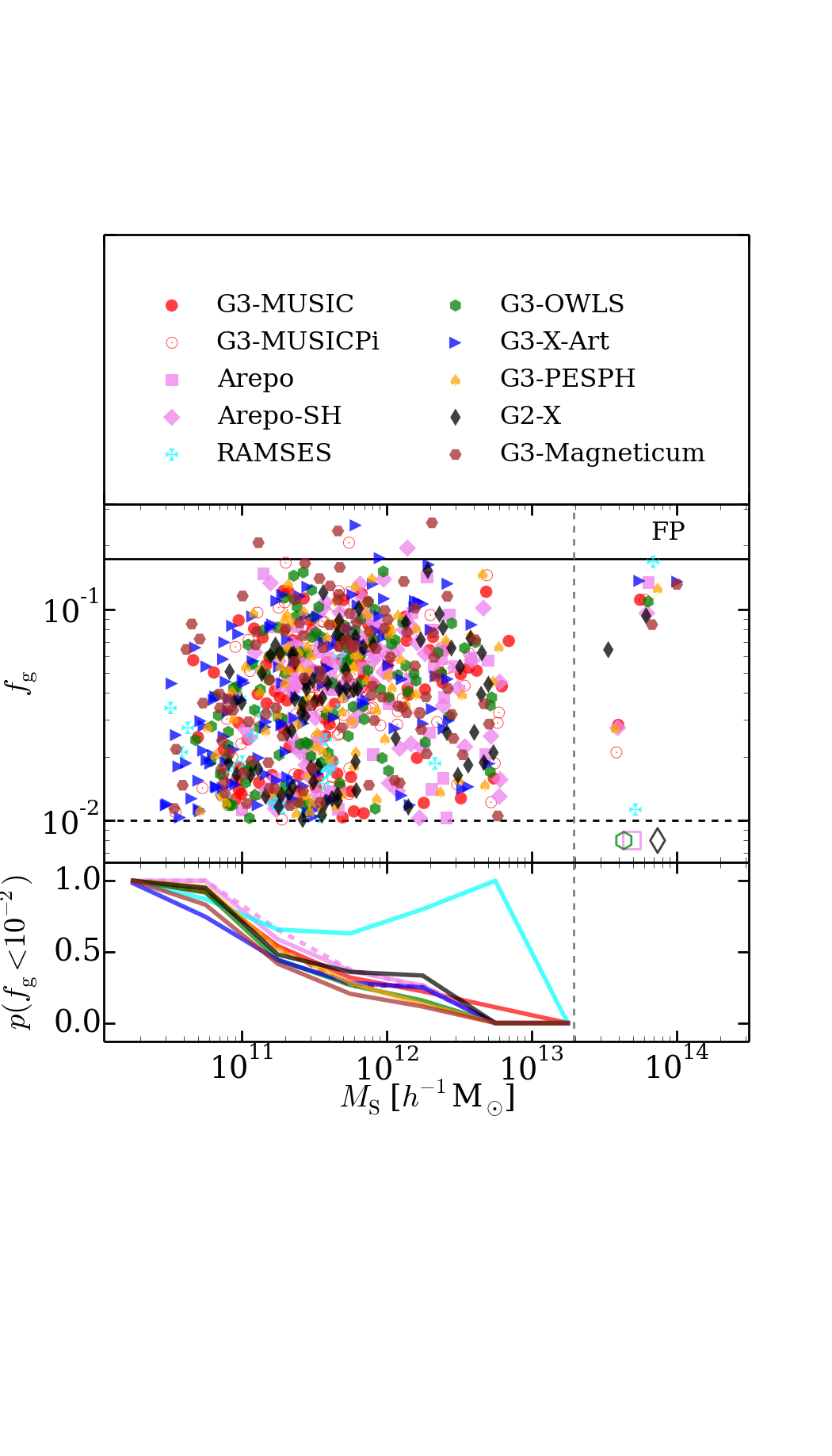}
    \caption{Gas fraction versus mass for \nrsim\ (top two subpanels) and \radsim\ (bottom two subpanels) simulations. The cosmic baryon fraction is shown by a solid black horizontal line. In the bottom subpanels we show the probability of a galaxy being stripped of all its gas, here assumed to be at $f_{\rm g}\leq10^{-2}$, in a given mass bin for all objects of $M\leq2\times10^{13}~\Msunh$ (dashed vertical line), above which there are very few objects. For those larger objects with $f_{\rm b}\leq10^{-2}$, we plot them as open markers in the top panel. Colour, line types are the same as in \Figref{fig:massdistrib}, markers are indicated in the legend. For line legend see \Figref{fig:massdistrib}.}
     \label{fig:gasfracmass}
\end{figure}

\par
Figure \ref{fig:baryonfracdistrib} showed that in non-radiative simulations, regardless of code, few subhaloes retain their gas (baryons) in the cluster environment. In \Figref{fig:gasfracmass} we show the gas fraction, $f_{\rm g}$, versus subhalo mass of all objects with non-negligible amounts in the upper subpanel and in the lower subpanel the probability that a subhalo of a given mass retains negligible amounts of gas (here we use $f_{\rm b}<10^{-2}$ based on \Figref{fig:baryonfracdistrib}). Reassuringly, most \nrsim\ simulations produce similar distributions in the mass of subhaloes which are unable to retain significant gas fractions. Only subhaloes with $M\gtrsim2\times10^{12}~\Msunh$ or $\gtrsim10^{-3}$ times the host cluster mass retain some gas. Note that both mesh codes are more likely to have massive gas poor subhaloes than SPH codes with \ramses\ again being an outlier.

\par
Code variations are also seen on an individual object basis. The most massive subhalo shown here has recently entered the cluster environment and consequently has $f_{\rm b}\approx \Omega_b/\Omega_m$ in all \nrsim\ runs. However, the second largest subhalo, which lies closer to the cluster centre, has been completely stripped in the mesh or new SPH codes (open points), but retains some gas in the more classic SPH codes. This hint of bimodality between codes is not too surprising considering studies of mesh and SPH codes using the blob test show that SPH codes increase the survival time of dense gas clumps exposed to a shock front or hot environment as a result of the artificially suppressed mixing present in the classic SPH formalism \cite[e.g.][]{tasker2008}. Generally \ramses\ \& \arepo\ have smaller $f_{\rm b}$ than classic SPH codes, which is consistent with the quicker gas depletion of substructures found in \arepo\ compared to \gadget\ \cite[][]{sijacki2012a}.

\par
We should note that \ramses\ has a few very low mass subhaloes with non-negligible gas fractions. These subhaloes reside at radii of $\gtrsim1500\kpch$ outside the hot cluster environment. The reason for this population is partially due to \ramses's adaptive mesh, which is able to follow much smaller parcels of gas. 

\par
The lower two panels of \Figref{fig:gasfracmass} show that feedback physics changes the picture. Across all codes, only objects with $M\lesssim10^{11}~\Msunh$ are now devoid of gas, stripped by the combination of the cluster environment and internal feedback processes. The notable exception is \ramses, which has a peak at much higher masses. This peak is partially a statistical fluke, there are only three subhaloes in this mass range and they have all been stripped of gas.  In general, the probability of being gas poor monotonically decreases with increasing mass, with \arepo\ and particularly \ramses\ having higher likelihoods than the other codes and \magneticum\ and \gadgetxart\ having the lowest. 

\subsection{Galaxies}
\label{sec:subhalopop:stars}
\begin{figure}
    \centering
    \includegraphics[width=0.45\textwidth,trim=0.cm 0.5cm 2.5cm 2.4cm, clip=true]{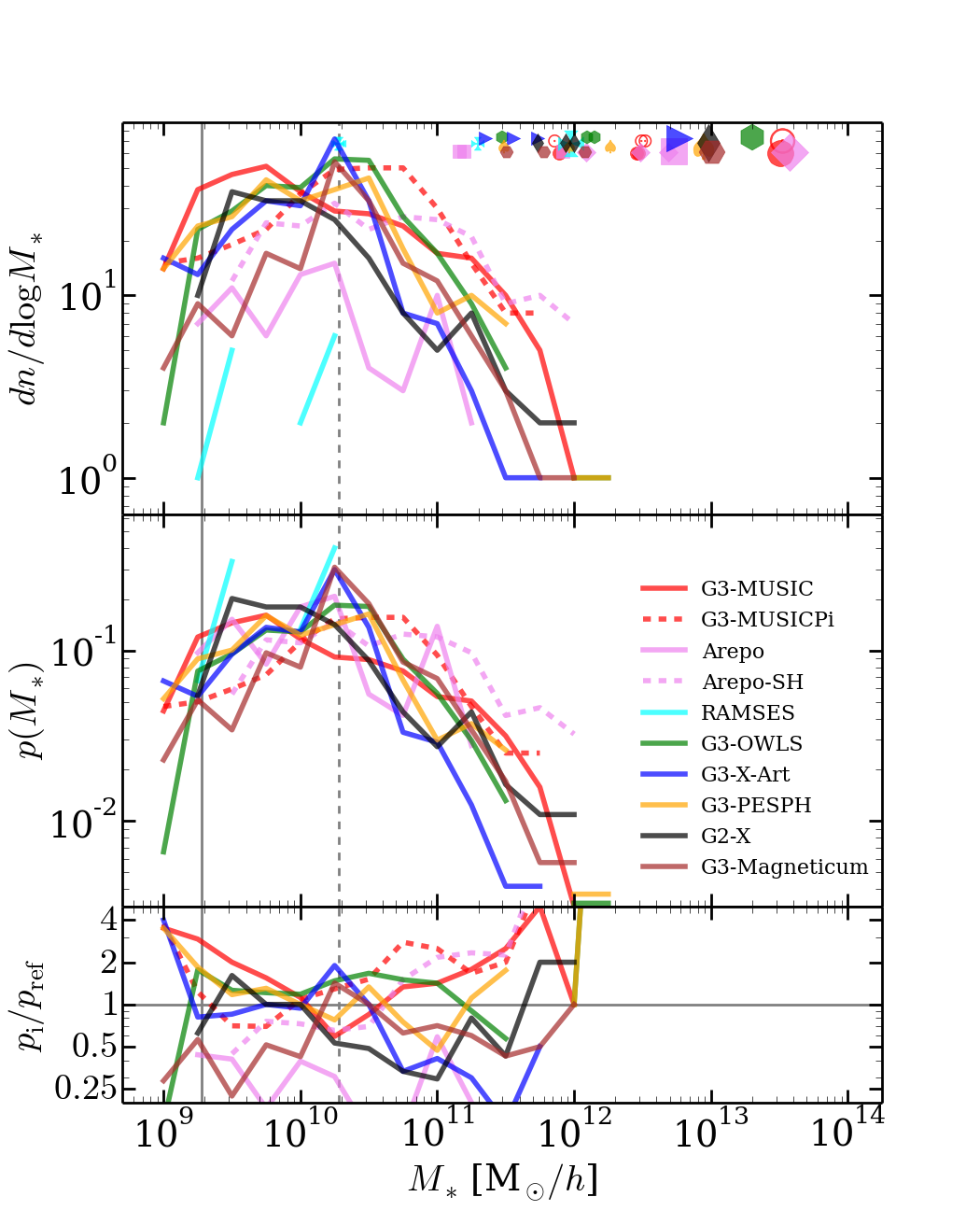}
    \caption{Galaxy Stellar Mass Function (GSMF) (top), normalised GSMF, i.e., probability of a galaxy having a mass within 0.25 dex of some mass $M_*$ (middle), and the residuals of the probability relative to the median calculated in the same fashion as \Figref{fig:massdistrib} (bottom). We also plot the stellar mass of the BCG (includes the inter-cluster stars), and the three largest galaxies as large markers in the top panel (y ordinate is arbitrary value). Vertical solid line is at a mass of $10m_{\rm gas}$ (resolution limit for codes which have one generation of stars produced by a gas particle) and we also show a dashed line at $100m_{\rm gas}$. Colour, line types, and markers are the same as in \Figref{fig:gasfracmass}. For  marker legend see \Figref{fig:gasfracmass}.}
    \label{fig:stellarmassfunction}
\end{figure}

Hydrodynamical codes typically seek to reproduce the observed galaxy population, hence the first comparison to be made is the resulting stellar mass function of galaxies. However, as is evident from \Tableref{tab:nsubs}, different codes result in significantly different number of galaxies. Therefore we examine both the galaxy stellar mass function (GSMF) and the normalised one, i.e.,~the probability of a galaxy having a stellar mass within a specific range, in \Figref{fig:stellarmassfunction}. Note that we do not compare our GSMF to observations as there appears to be significant cluster-to-cluster variation \cite[see Fig 12 of][for instance]{boselli2011a}.

\par
The stellar mass function shows large code-to-code scatter both for small and large galaxies, even when the differences in normalisation are removed. Even the brightest central galaxy (BCG, including the intracluster light) differs by a factor of $\gtrsim4$. The two mesh codes {\em with} AGN feedback, \ramses\ \& \arepo, produce the smallest BCG and \arepo-SH\ {\em without AGN feedback} produces the largest BCG. In fact, \ramses\ severely stunts the growth to $10^{12}\Msunh$, a factor of 10 times less massive than the next smallest BCG. Amongst the SPH codes, \music\ and \owls\ produce the largest, \gadgetxart\ and \pesph\ the smallest, differing by a factor of $\sim5$. 

\par 
This diversity is not simply due to different formulations of SPH or mesh codes evolving gas, the building blocks of stars, differently. For instance, the probability and number of low mass galaxies in \gadgettwox\ \& \music\ differ by $\gtrsim2$ for $M_*\lesssim5\times10^{9}~\Msunh$, with \gadgettwox\ having more. \music\ has much larger galaxies than \gadgettwox. This is in spite of the fact that both use standard SPH; the differences lie in the subgrid physics. That is not to say that all codes disagree. \music\ and \gadgettwox\ have monotonically decreasing stellar mass functions above masses of $\sim2\times10^{9}~\Msunh$. Other codes typically produce stellar mass functions that are strongly suppressed for $M_*\lesssim10^{10}~\Msunh$, with \magneticum\ showing the strongest suppression. However, this turn over likely arises partially due to resolution effects and not solely due to SN feedback, as indicated by the fact that it occurs for masses corresponding to less than 100 star particles.

\par 
We can see the effects of different subgrid physics by looking at \arepo\ and \arepo-SH, that is galaxies produced including/ignoring AGN physics. With the modified subgrid physics (specifically the lack of AGN feedback), \arepo-SH is able to reproduce the BCG seen in \music\ and also has similar numbers of massive galaxies. In fact, it is more biased towards massive galaxies than \music. AGN physics is, however, not a precise dividing line between codes. \pesph, which does not included AGN feedback but has a modified SN feedback, has a BCG similar to \gadgettwox\ and GSMF similar to \owls, AGN SPH codes. 
\begin{figure}
    \centering
    \includegraphics[width=0.45\textwidth,trim=0.cm 0.0cm 2.5cm 2.4cm, clip=true]{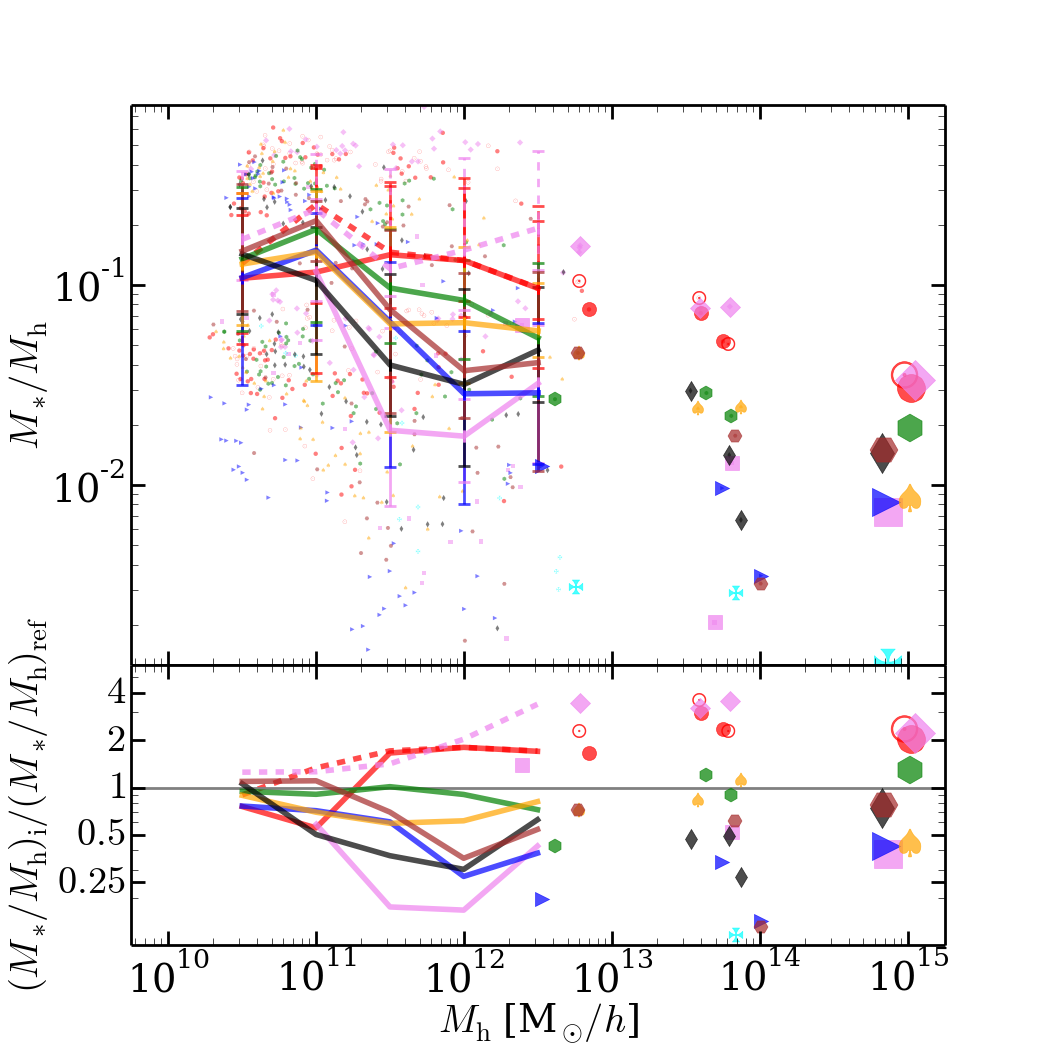}
    \caption{Stellar mass to host (sub)halo relation (top) and the residuals relative to the median calculated in the same fashion as \Figref{fig:massdistrib} (bottom). In the top panel, we bin the data in host mass and plot median and 0.16,0.84 quantiles along with the data lying outside this range as small filled points. Similar to \Figref{fig:stellarmassfunction} we also plot the BCG and next three largest galaxies as points. Colour, line types, and markers are the same as in \Figref{fig:gasfracmass}. For legends see \Figref{fig:gasfracmass} \& \ref{fig:stellarmassfunction}.}
    \label{fig:stellarhalorelation}
\end{figure}

\par 
The interplay between gas cooling and feedback is what transforms the (sub)halo mass function (\Figref{fig:massdistrib}) to the GSMF (\Figref{fig:stellarmassfunction}). In small haloes, supernova (SN) feedback should blow out gas from small haloes, whereas star formation is suppressed in larger haloes by the energy injected into the surroundings by the supermassive black hole (AGN) residing in the (sub)halo centre. Despite the fact that subgrid physics in each code attempts to model these processes, the stellar mass to host halo mass relation, seen in \Figref{fig:stellarhalorelation}, has large code-to-code scatter. Most codes have the same overall shape: $\MstarMh$ decreases with increasing halo mass, with plateaus for $M_{\rm h}\lesssim10^{11}~\Msunh$ and $M_{\rm h}\gtrsim10^{12}$. However, \music\ (and \musicpi\ \& \arepo-SH) has an almost constant average $\MstarMh$ relation and the efficiency of star formation only seems to gradually decrease for much larger host haloes\footnote{Although the downturn in $\MstarMh$ with decreasing $M_{\rm h}$ due to SN feedback at small halo masses is hinted at here, typically, this effect would most noticeable at host (sub)halo masses of $\lesssim10^{10}~\Msunh$, below the mass resolution used here.}. Even for codes with similar $\MstarMh$ shapes, the actual average $\MstarMh$ for a given halo mass can vary by a factor of $\sim4$. For instance, galaxies in \gadgettwox\ are far more dark matter dominated that those in \owls. 

\par
\begin{figure}
    \centering
    \includegraphics[width=0.45\textwidth,trim=0.cm 0.0cm 2.5cm 2.4cm, clip=true]{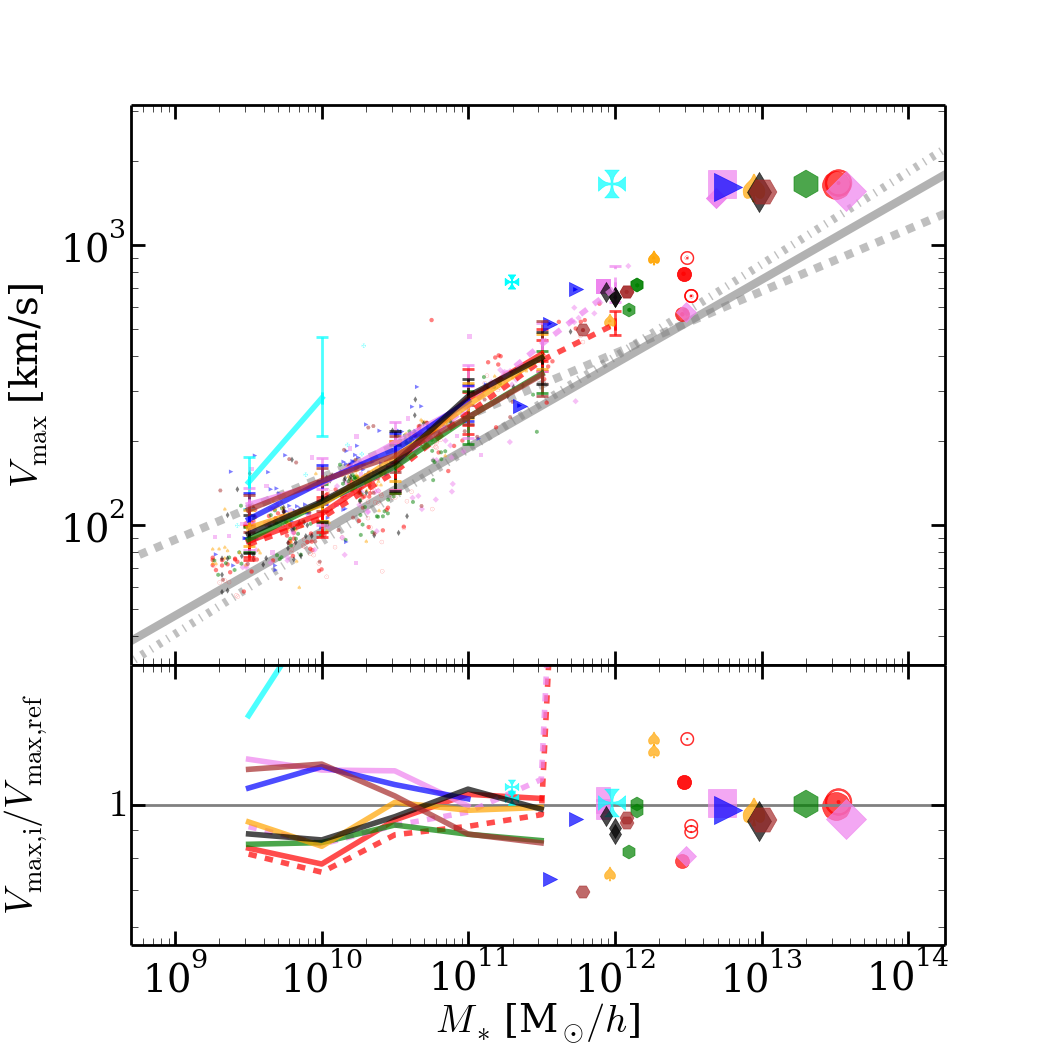}
    \caption{Relation between circular velocity and stellar mass (top) and the residuals relative to the median calculated in the same fashion as \Figref{fig:massdistrib} (bottom), similar to \Figref{fig:stellarhalorelation}. For reference we also plot observational fits of Kassin et al. (2007) (thick solid gray line), Dutton et al. (2011) (thick dashed gray line), \& Cortese et al. (2014) (thick dash-dotted line) to field galaxies. For legends see \Figref{fig:gasfracmass} \& \ref{fig:stellarmassfunction}.}
    \label{fig:tullyfisher}
\end{figure}

Finally we examine how well a common observational relation is reproduced, the Tully-Fisher/Faber-Jackson relations in \Figref{fig:tullyfisher}. Here we limit ourselves to a simple comparison, the maximum circular velocity as a function of stellar mass. We find that in contrast to the diversity seen in the other relations, there is little code-to-code scatter in the average relation. A galaxy with a mass $M_*$ will on average reside in a similar potential well $\phi\propto GM/R$ regardless of code but the total mass associated with and size of that potential will vary from code-to-code. The only code that truly not follow the same slope and amplitude is \ramses, where galaxies reside in far more massive subhaloes. \magneticum, \gadgetxart\ and \arepo\ also have low mass galaxies residing in larger hosts than other codes, although to a lesser extent than \ramses.
\par 
The relation produced by hydrodynamical codes differs from the observed relations shown in this figure, however, this is not completely unexpected given the environment probed here, a high density cluster. Observations typically stack galaxies from a wide variety of environments. Furthermore, we have not split our galaxies into morphological types and estimated rotational or dispersive velocities from line-of-sight measurements within some radius, necessary if we wish to compare directly to observations. Overall, codes only differ from the observed relation significantly at high stellar masses. Only the BCGs lie off the observed trend, however, as these galaxies lie at the centre of the cluster, $\vmax$ no longer probes the central galaxy but the overall cluster potential. 

\section{One-to-one comparisons}
\label{sec:crosscomp}
Section \ref{sec:subhalopop} shows most codes reproduce the same bulk distribution when running DM or NR simulations. The question is whether this agreement masks a variation in an individual object's properties. The properties of an individual object that has experienced the same mass accretion history and dynamical environment should be the same. A quick glance at the previous section shows that all the codes reproduce the same large subhalo in terms of total mass. This subhalo is somewhat unique in that is has only been recently accreted around $z\approx0.2$ and lies at the outer edge of the cluster environment. The second most massive subhalo, which has resided for a longer period of time in the cluster environment, shows larger code-to-code variation. Here we expand this line of comparison and search for counterparts between the subhalo catalogues produced by different simulation codes and compare their properties. 

\par 
When comparing properties we could cross correlate all catalogues with one another and compare codes relative to a virtual median object. However not all objects are found in all catalogues. Moreover, using a median (or mean) implies a median model. What is this median model? If most codes were similar than that medial model is easily understood and variations about this median give rise to differences in properties, i.e., scatter. However, this is not the case as we have codes that have attempted to incorporate different feedback physics. As we are not only interested in the scatter between codes but how different subgrid implementations effect galaxies, i.e. systematic differences, we use the \music\ catalogue as our reference, though any one could be used.

\par
\begin{figure*}
    \centering
    \includegraphics[height=0.34\textheight,trim=0.cm 1.cm 4.12cm 2.cm, clip=true]{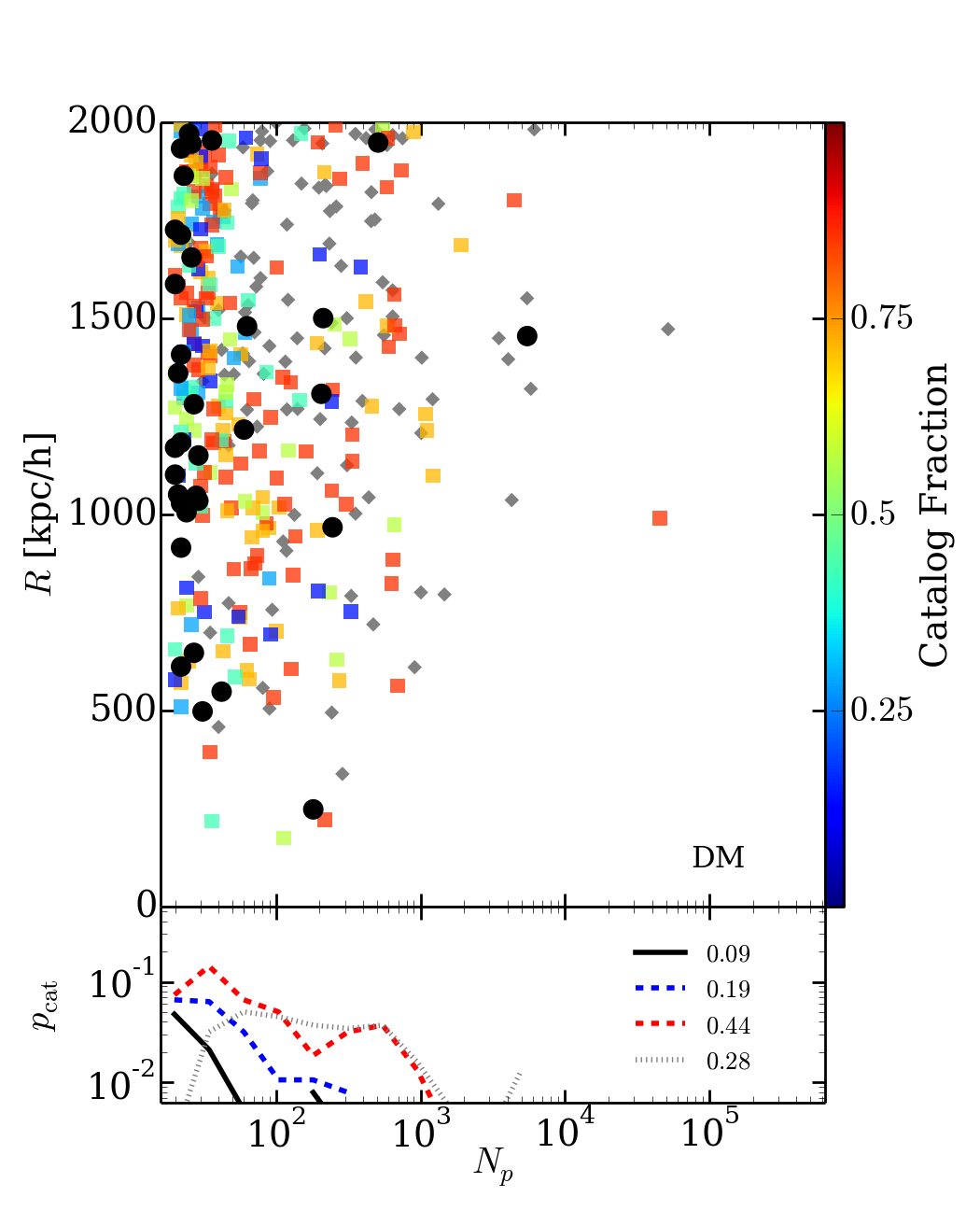}\hspace{-2.6pt}
    \includegraphics[height=0.34\textheight,trim=4.34cm 1.cm 4.12cm 2.cm, clip=true]{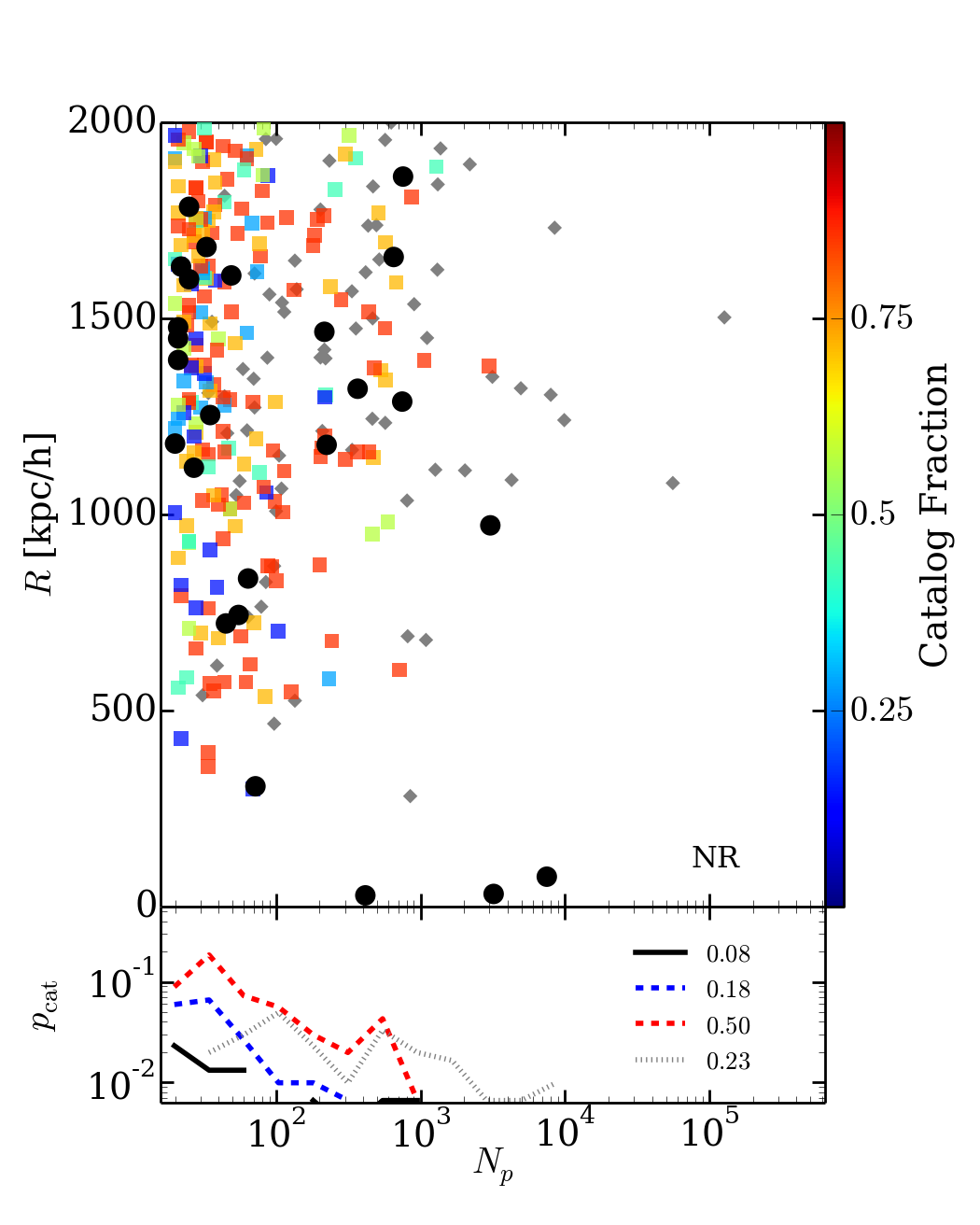}\hspace{-2.6pt}
    \includegraphics[height=0.34\textheight,trim=4.34cm 1.cm 0.cm 2.cm, clip=true]{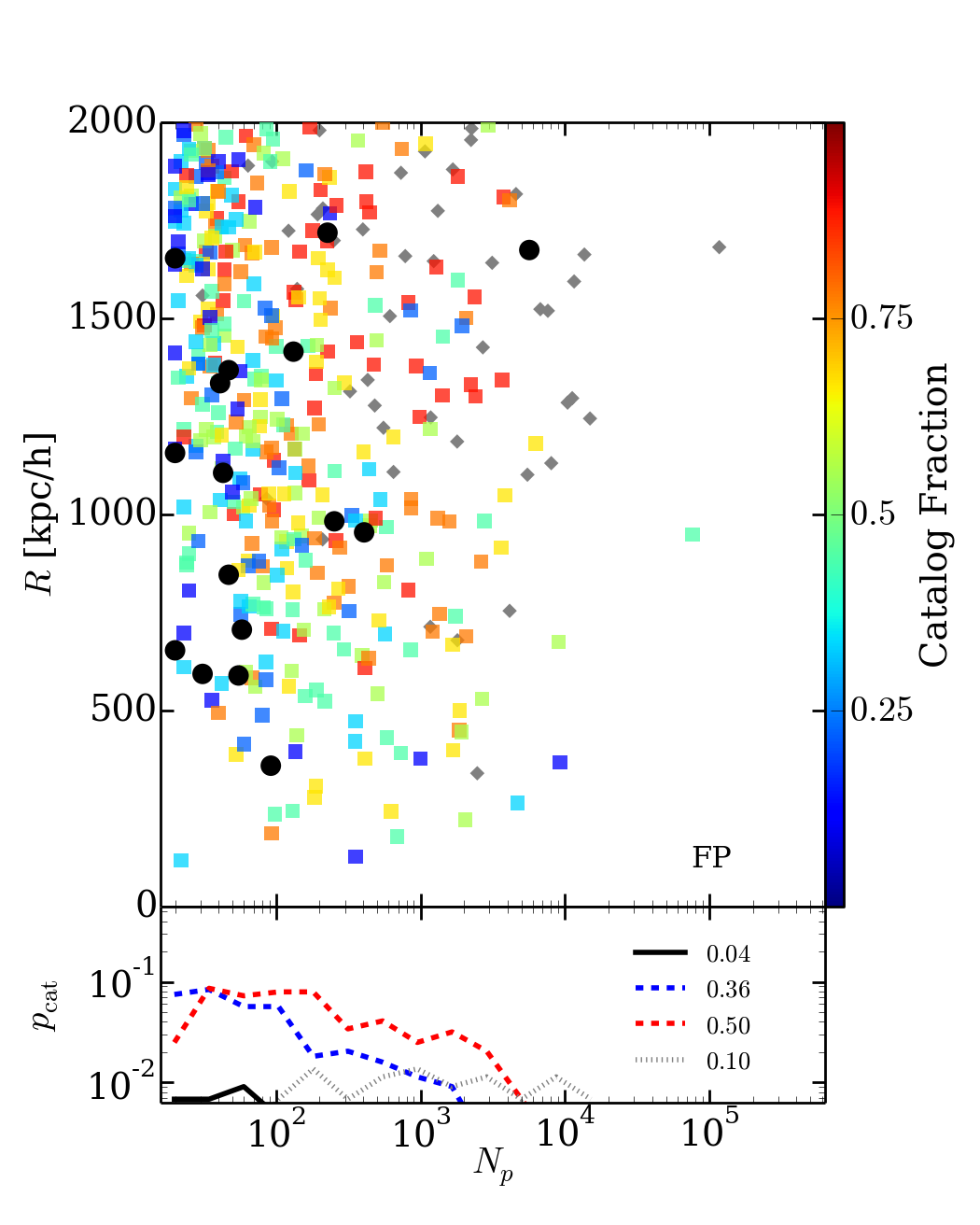}
    \caption{The distribution of ``missing'' subhaloes, specifically the number of particles in the dark matter subhalo and its radial position from the centre of the host halo (top panel). We use the \music\ catalogue as our reference. Subhaloes that are missing in {\em all} other simulations are plotted as large black circles. Subhaloes that are missing in one or more catalogues but not all of them are plotted as solid squares, with the colour showing the fraction of catalogues it is missing in. Subhaloes identified in {\em all} catalogues are plotted as gray diamonds. In the lower panel we show the probability distribution of missing subhaloes (solid black line), subhaloes found in fewer than $50\%$ of the catalogues (dashed blue line), subhaloes found in more than $50\%$ of the catalogues (dashed red line), and subhaloes found in all catalogues (dotted gray line), along with the total fraction of the catalogue in each of these subcategories. The three types of simulations are shown: \dmsim\ (left), \nrsim\ (middle), and \radsim\ (right) respectively.}
     \label{fig:crosscat}
\end{figure*}

Before we compare properties, it is important to check whether this is a viable exercise by identifying subhaloes for which no counterpart is found. Recall that a counterpart is one which satisfies \Eqref{eqn:merit} with a merit of 0.2. If there are numerous missing subhaloes, then comparing individual objects is not informative as the codes have produced clusters with wildly different internal structures. We compare catalogues in \Figref{fig:crosscat}, where we plot for every subhalo identified in the \music\ catalogue, the number of particles in a subhalo, its radial distance from the cluster centre, and the fraction of other catalogues this subhalo exists in. Gray diamonds are subhaloes identified in all catalogues, black circles missing in all other catalogues, and coloured squares for subhaloes identified in some catalogues. 

\par 
If we pay particular attention to the subhaloes for which no counterpart is found, it is reassuring to know that most are composed of significantly less than $100$ particles. Most large subhaloes, those composed of $\gtrsim500$ particles, are present in all simulations, with a few interesting exceptions. The large subhalo identified in the \dmsim-\music\ catalogue composed of $\sim5\times10^3$ particles at a radius of $1500~\kpch$ is merging with the largest subhalo, also at $1500~\kpch$. Matches are identified in other catalogues but are not above the merit threshold used. This also applies for the object in the \radsim-\music\ catalogue. Similarly, the subhaloes in the \nrsim-\music\ catalogue located at very small radii have matches but these less than ideal matches are due to the difficulty of identifying subhaloes residing within the central regions of the halo hosts. Small differences in orbits will mean in some codes, different portions of the subhalo remain self-bound and are identified.  

\par 
Given that $\lesssim10\%$ of the subhalo population is ``missing'' in the three types of simulations, one-to-one comparison of well-resolved subhaloes is meaningful. From here on, we will restrain our comparison to subhaloes composed of $\geq100$ particles {\em in both catalogues}. This limits our comparison to $\approx105$, $\approx80$, and $\approx50$ objects in the \dmsim, \nrsim\ \& \radsim\ simulations respectively.

\subsection{Mass Proxies}
\label{sec:crosscomp:mass}
\begin{figure*}
    \centering
    \includegraphics[height=0.2\textheight,trim=0.0cm 0.5cm 2.6cm 2.cm, clip=true]{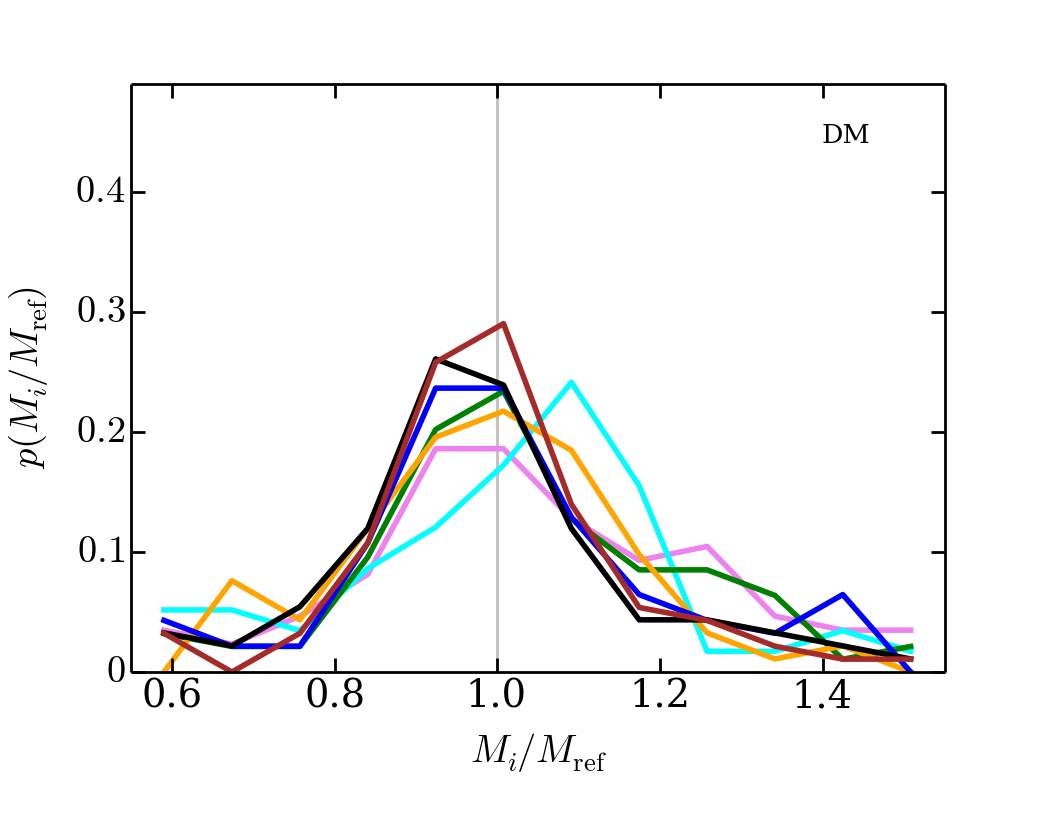}
    \hspace{-4.1pt}
    \includegraphics[height=0.2\textheight,trim=3.3cm 0.5cm 2.6cm 2.cm, clip=true]{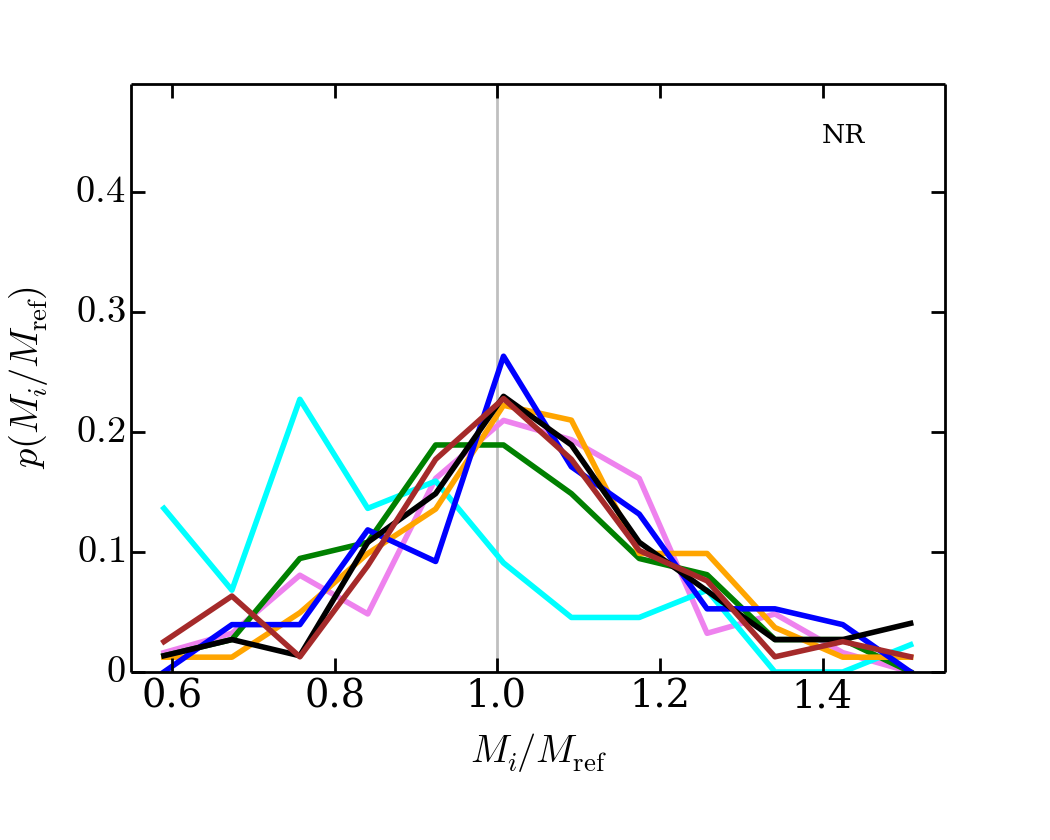}
    \hspace{-4.1pt}
    \includegraphics[height=0.2\textheight,trim=3.3cm 0.5cm 2.6cm 2.cm, clip=true]{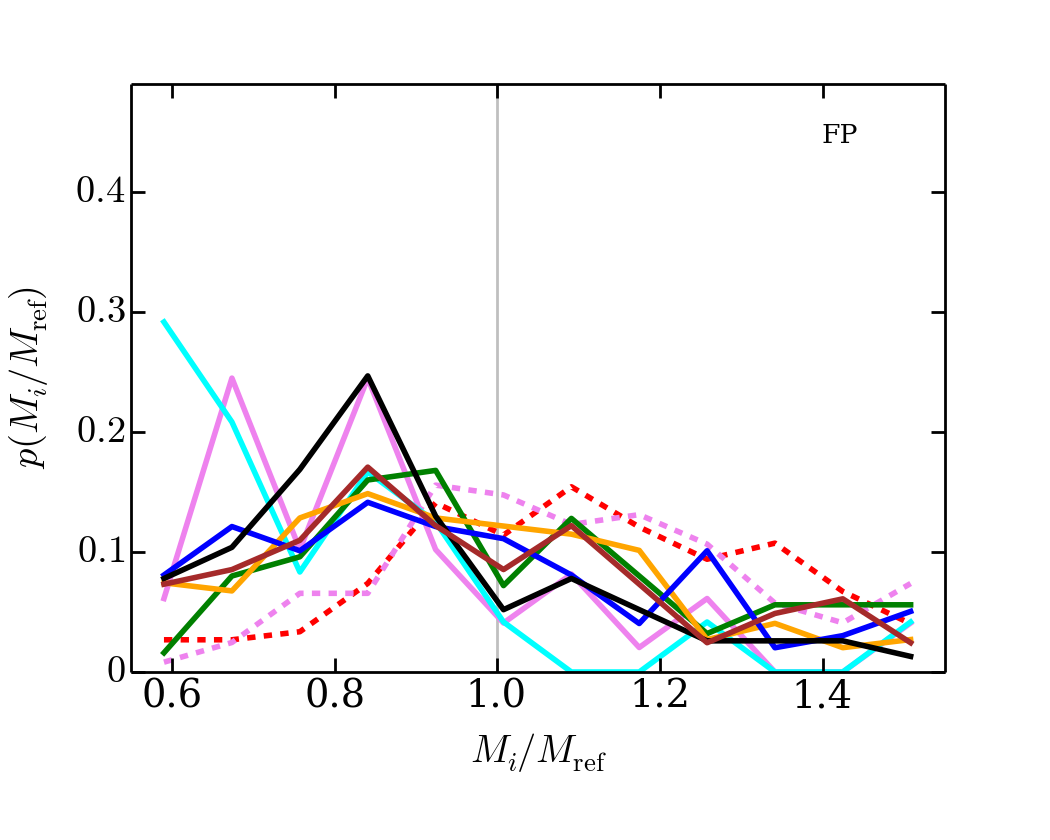}
    \\
    \includegraphics[height=0.2\textheight,trim=0.0cm 0.5cm 2.6cm 2.cm, clip=true]{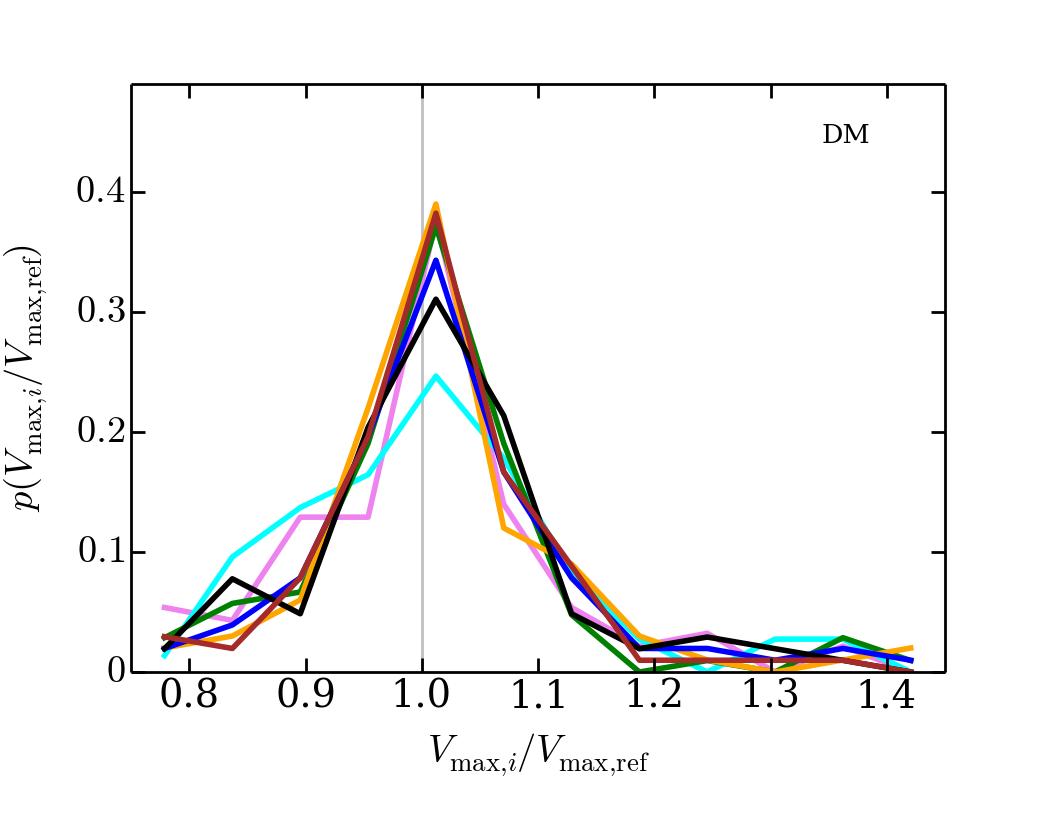}
    \hspace{-4.1pt}
    \includegraphics[height=0.2\textheight,trim=3.3cm 0.5cm 2.6cm 2.cm, clip=true]{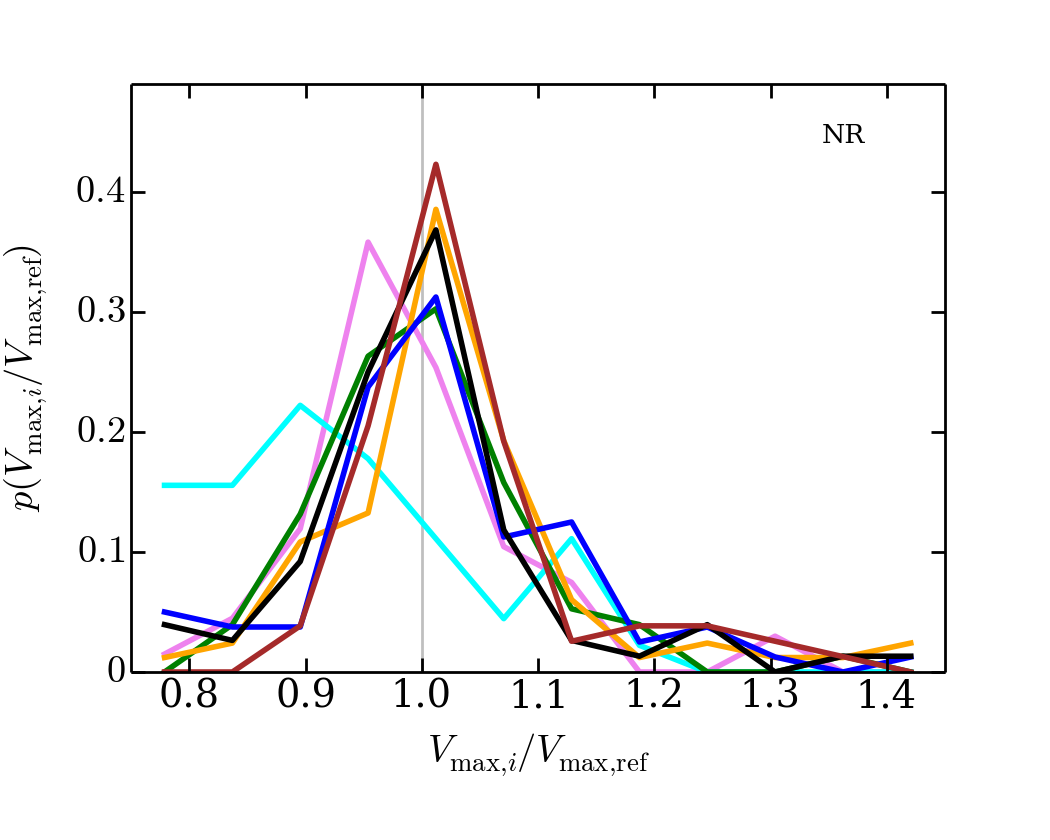}
    \hspace{-4.1pt}
    \includegraphics[height=0.2\textheight,trim=3.3cm 0.5cm 2.6cm 2.cm, clip=true]{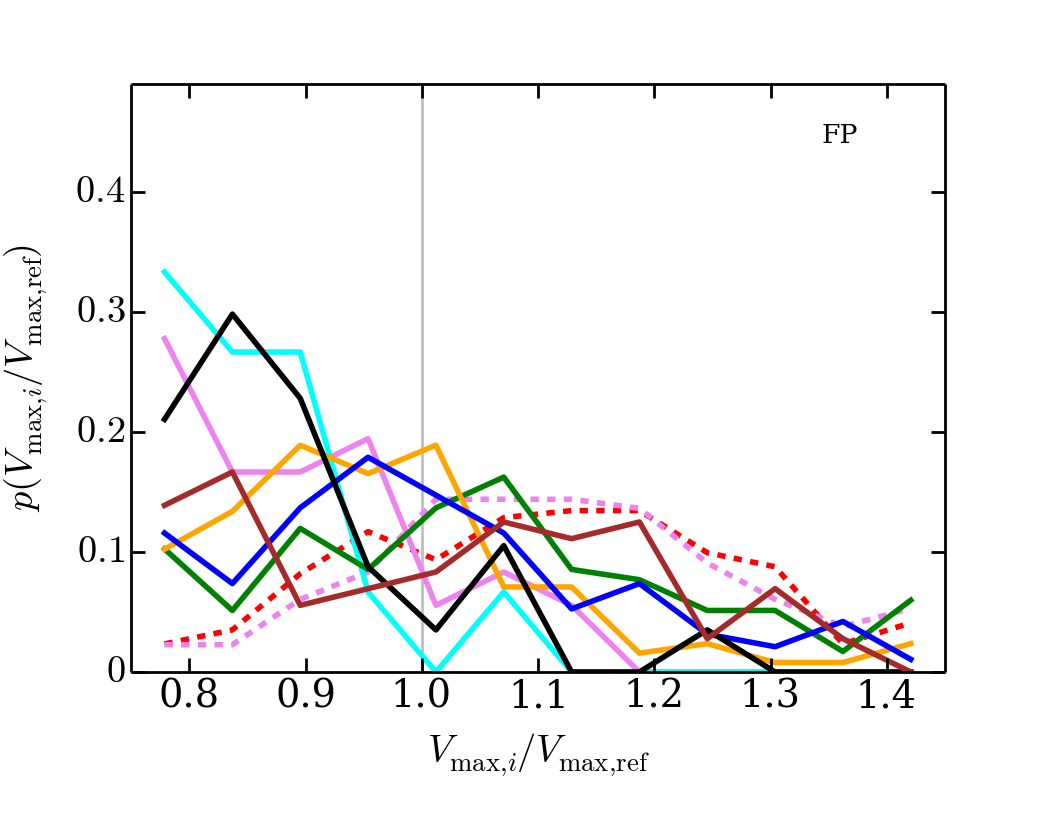}
    \caption{Ratio distribution of subhalo counter parts in simulation listed to those in the \music\ catalogue for mass (top) \& $\vmax$ (bottom) of the \dmsim\ (left), \nrsim\ (middle), \radsim\ (right) simulations respectively. Line styles and colours are the same as in \Figref{fig:massdistrib}. To guide the eye, we plot a solid gray vertical line at a ratio of one in both panels. For legend see \Figref{fig:stellarmassfunction}.}
    \label{fig:massvmaxcomp}
\end{figure*}

Figure \ref{fig:massvmaxcomp} shows the distribution of the ratio of a subhalo's bound mass and maximum circular velocity in one simulation to its counterpart in the \music\ catalogue for all well resolved ($N_p\gtrsim100$) subhaloes. For almost all codes, the \dmsim\ \& \nrsim\ runs have a ratio that follows a lognormal distribution. The typical variation is $\sim20\%$. The $\vmax$ distribution has a smaller scatter, $\approx10\%$, not surprising since the central region usually defining $\vmax$ is less effected by the tidal field of the cluster. The fact that all catalogues have similar variation suggests that this scatter is probably dominated by the differences in the exact orbits these subhaloes have taken in the highly nonlinear cluster environment, rather than different hydrodynamical implementations. The main outlier is \ramses, which primarily differs in the \nrsim\ simulations. \ramses\ produces smaller, less centrally concentrated subhaloes which are more susceptible to tidal disruption, hence the reason it has fewer subhaloes (see \Tableref{tab:nsubs}).

\par
Given the differences seen in \Figref{fig:vmaxdistrib} for the \radsim\ simulations, it is not surprising that even for subhaloes with a well defined counterpart in the \music\ simulation, the ratio of $\vmax$ shows systematic differences and vary greatly between codes. Feedback physics moves material out of cores of subhaloes, changing their circular velocity profiles significantly. What is somewhat unexpected is the variation in the mass. \music\ typically has more massive subhaloes than the other codes and there is a great deal of variation which is not that readily apparent from \Figref{fig:massdistrib}.

\subsection{Baryons}
\label{sec:crosscomp:baryons}
\begin{figure}
    \centering
    \includegraphics[width=0.38\textwidth,trim=0.0cm 6.75cm 2.5cm 2.65cm, clip=true]{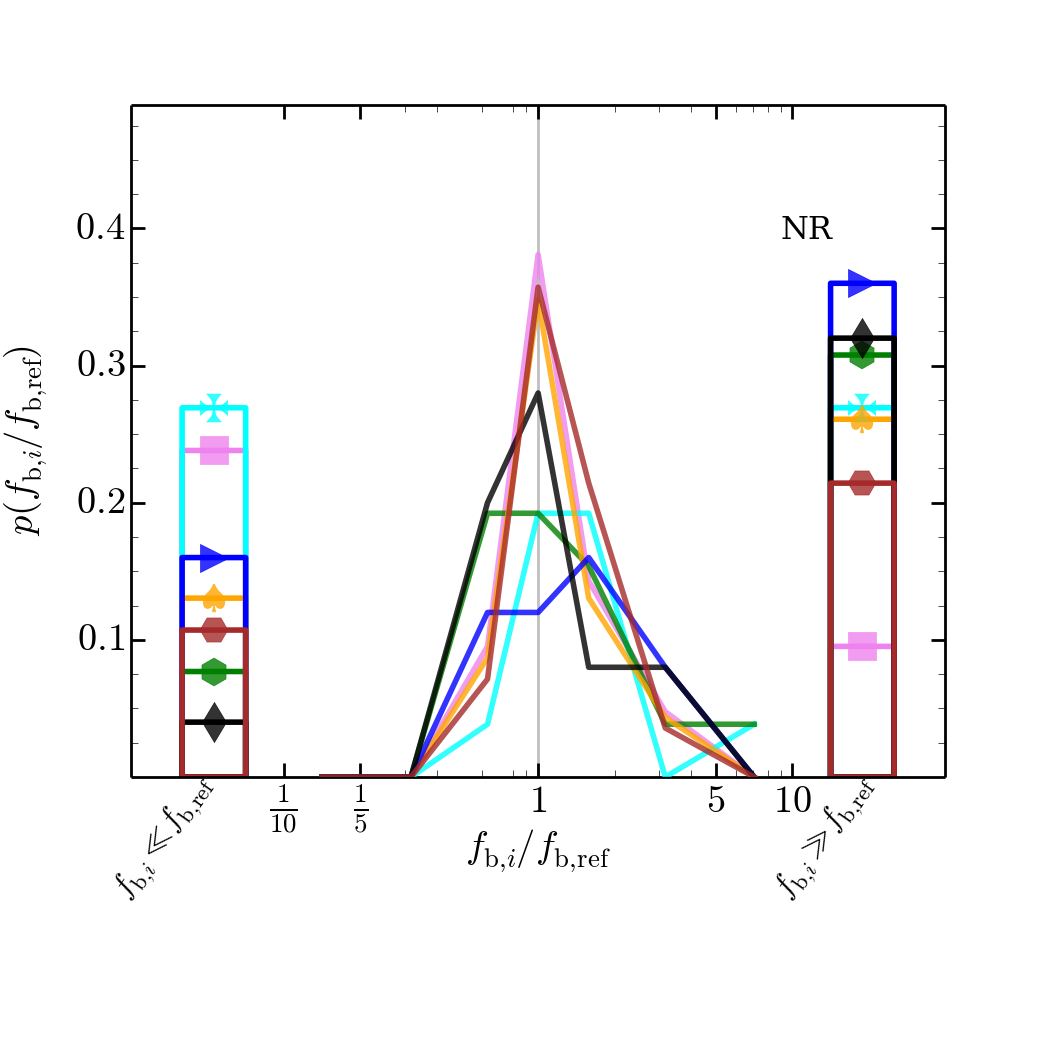}
    \\\vspace{-2.0pt}
    \includegraphics[width=0.38\textwidth,trim=0.0cm 3.25cm 2.5cm 2.65cm, clip=true]{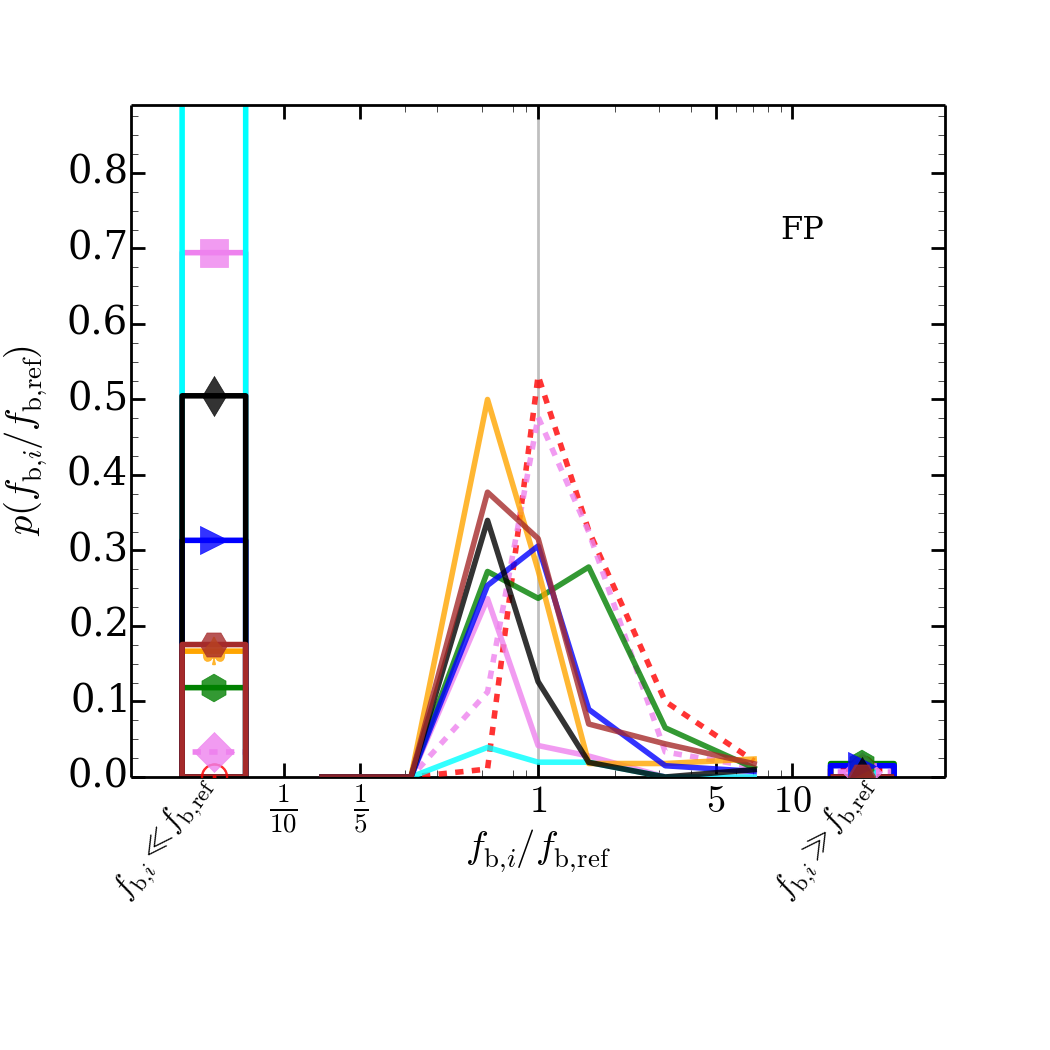}
    \caption{Baryon fraction comparison. Here we plot a histogram of the $f_{\rm b}$ ratio. We also plot two bins corresponding to subhaloes which contain negligible amounts baryons but have counterpart containing some, $f_{\rm b,i}\ll f_{\rm b,ref}$, and vice versa, $f_{\rm b,i}\gg f_{\rm b,ref}$. Line styles and colours are the same as in \Figref{fig:massdistrib}. For legend see \Figref{fig:gasfracmass} \& \ref{fig:stellarmassfunction}.}
    \label{fig:baryoncomp}
\end{figure}

We compare baryon fractions in \Figref{fig:baryoncomp}. When comparing the baryonic content of individual objects we must account for the possibility that either the subhalo or its counterpart has been completely stripped of baryonic material, resulting in a ratio $f_{{\rm b},i}/f_{{\rm b},{\rm ref}}$ that spans $(0,\infty)$. Therefore, we have binned objects where $f_{{\rm b},i}\leq 0.1 f_{{\rm b},{\rm ref}}$ and $f_{{\rm b},i}\geq 10 f_{{\rm b},{\rm ref}}$ separately in this figure. The non-radiative simulations have another issue: few objects contain non-negligible baryon fractions. For all the codes, $\approx70\%$ of the cross matched subhaloes have $f_{\rm b}\leq10^{-2}\Omega_b/\Omega_m$ in both catalogue. We ignore these stripped objects when comparing the ratio of the baryon fraction in \Figref{fig:baryoncomp}. 

\par
The first noticeable feature in the \nrsim\ simulations is that for most codes there are {\em two} significant populations, the largest centred at $f_{{\rm b},i}/f_{{\rm b},{\rm ref}}\approx1$. The major difference between codes lie which outlying bin contains a significant population. Subhaloes in \arepo\ are more likely to have been stripped of their gas relative to \music. Conversely, most other codes are systematically less stripped, with \gadgettwox\ and \gadgetxart, a classic SPH {\em and} modern SPH code, having the largest systematic offset. Interestingly, \ramses\ shows little bias in either direction. 

\par
In the \radsim\ runs, it appears that \music\ (and \musicpi) is the outlier, with subhaloes having higher baryon fractions. \arepo-SH is the only other code with counterparts having similar baryon fractions. \ramses, \arepo\ and \gadgettwox\ (both variants) have subhaloes biased to low $f_{\rm b}$. The question is whether \music's high baryon fractions are a consequence of it efficiently converting gas into stars, which are not subject to ram pressure or shocks, or whether these baryon rich objects have simply managed to retain gas in instances where other codes have been stripped. Or perhaps the counterpart is more massive and therefore able to better hold onto its baryons. A closer examination of these objects reveal that their \music\ counterparts typically have the same mass (within $20\%$) and contain {\em both} galaxies that have been stripped of or blown out all their gas and those that still have large reservoirs of fuel with which to form stars. Some of these galaxies even have gas fractions as high as $M_{\rm g}/M_{\rm b}\sim0.5$.

\par
\begin{figure}
    \centering
    \includegraphics[width=0.38\textwidth,trim=0.0cm 3.25cm 2.5cm 2.6cm, clip=true]{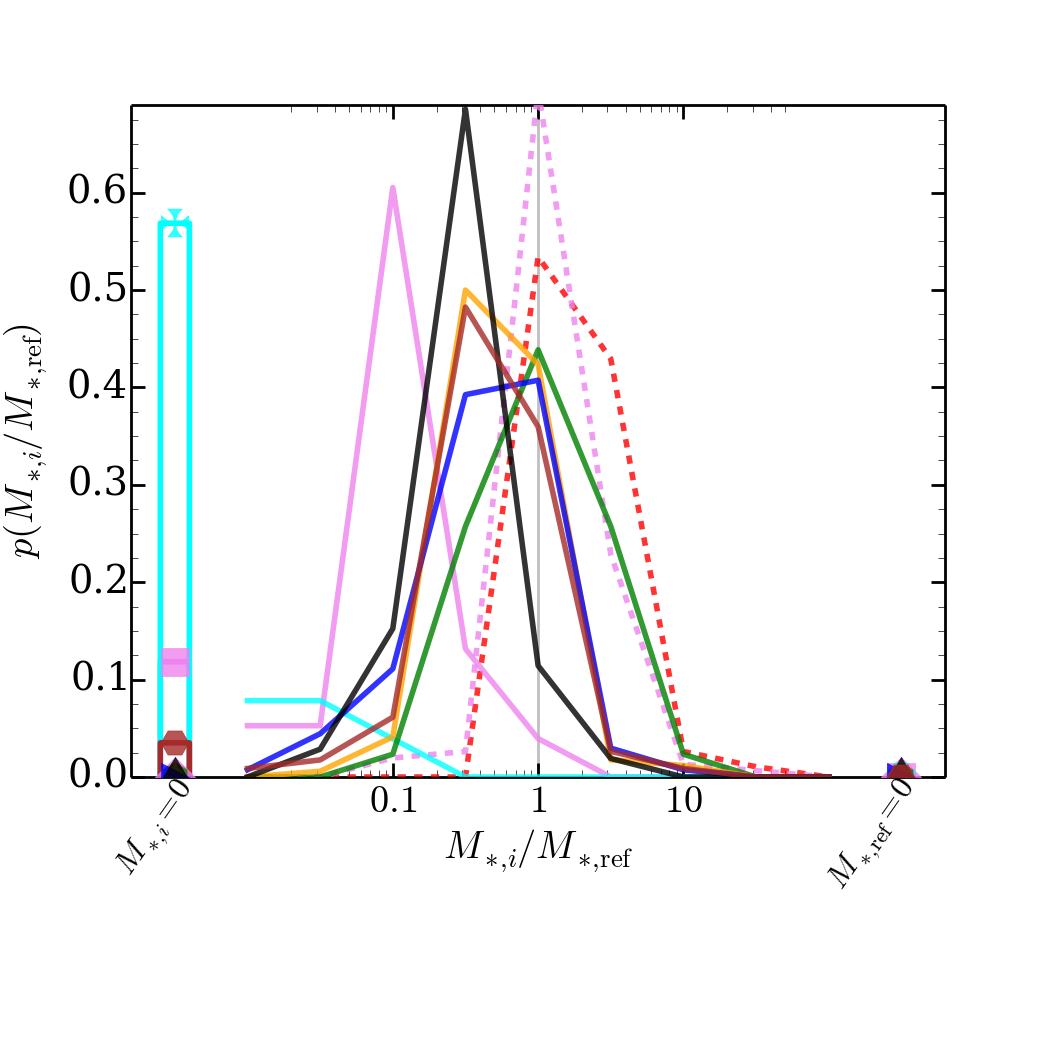}
    \caption{Stellar mass comparison similar to \Figref{fig:baryoncomp}. For legend see \Figref{fig:gasfracmass} \& \ref{fig:stellarmassfunction}.}
    \label{fig:stellarmasscomp}
\end{figure}

 
If we then focus on galaxies and their counterparts, we see in \Figref{fig:stellarmasscomp} significant systematic differences between codes in the stellar mass. \arepo\ typically not only has galaxies that are an order of magnitude less massive, it has a significant population of empty subhaloes whose \music\ counterparts do host a galaxy. \ramses\ is even more extreme. However, these are exceptions. Clearly for well resolved subhaloes composed of $\geq100$ particles, if a galaxy is present in one code, it is present in another. The difference lies in the size. Typically codes have less massive galaxies than \music, the exception being \arepo-SH, which lacks AGN, and \musicpi. AGN feedback is not the sole reason for the difference as \pesph\ (no AGN) has smaller galaxies than \owls, which does have AGN feedback.

\subsection{Galaxy/Subhalo Diversity}
\label{sec:crosscomp:summary}
\begin{figure}
    \centering
    \includegraphics[width=0.475\textwidth,trim=0.0cm 23.0cm 2.5cm 4.0cm, clip=true]{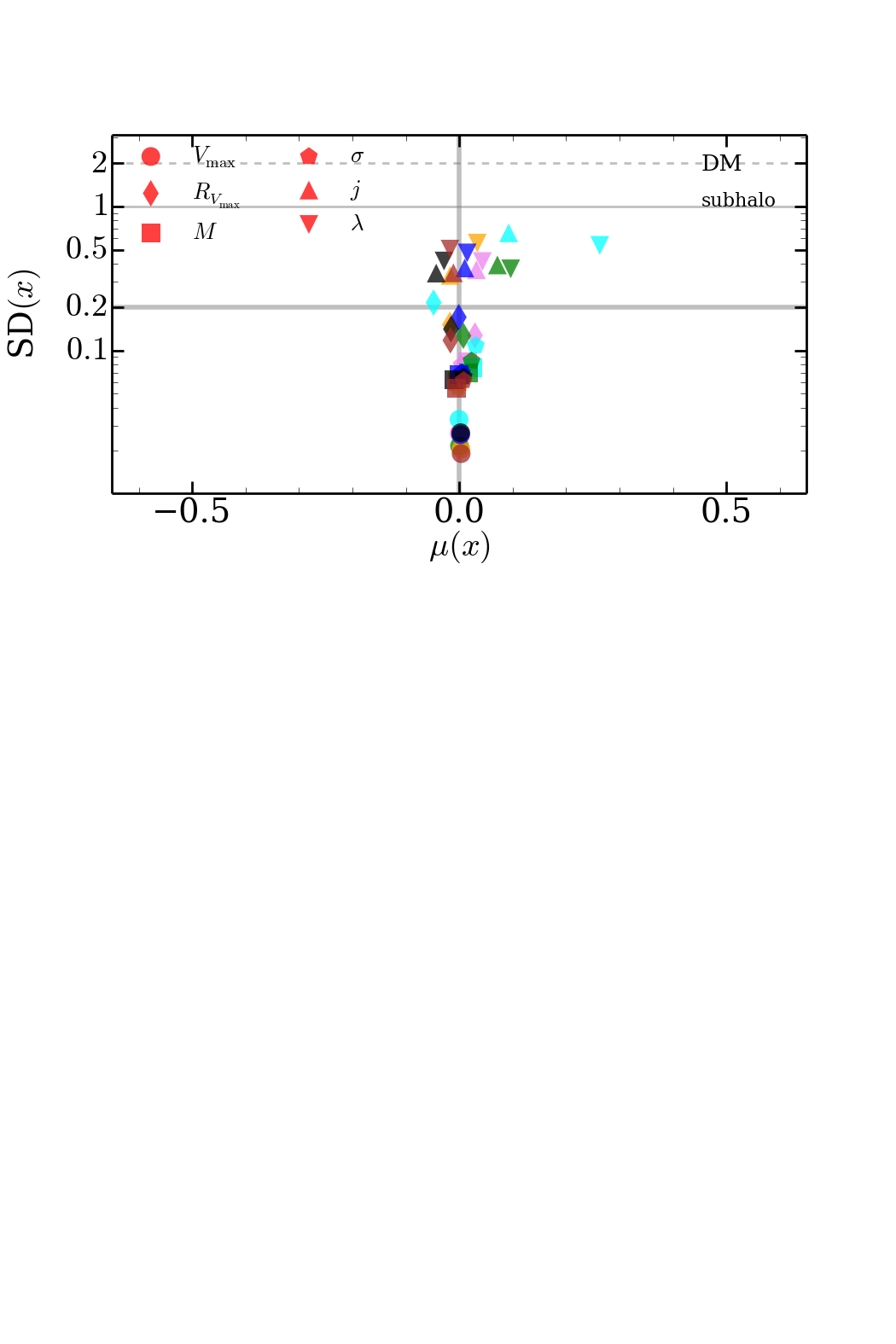}
    \hspace{-2.8pt}
    \includegraphics[width=0.475\textwidth,trim=0.0cm 12.0cm 2.5cm 4.0cm, clip=true]{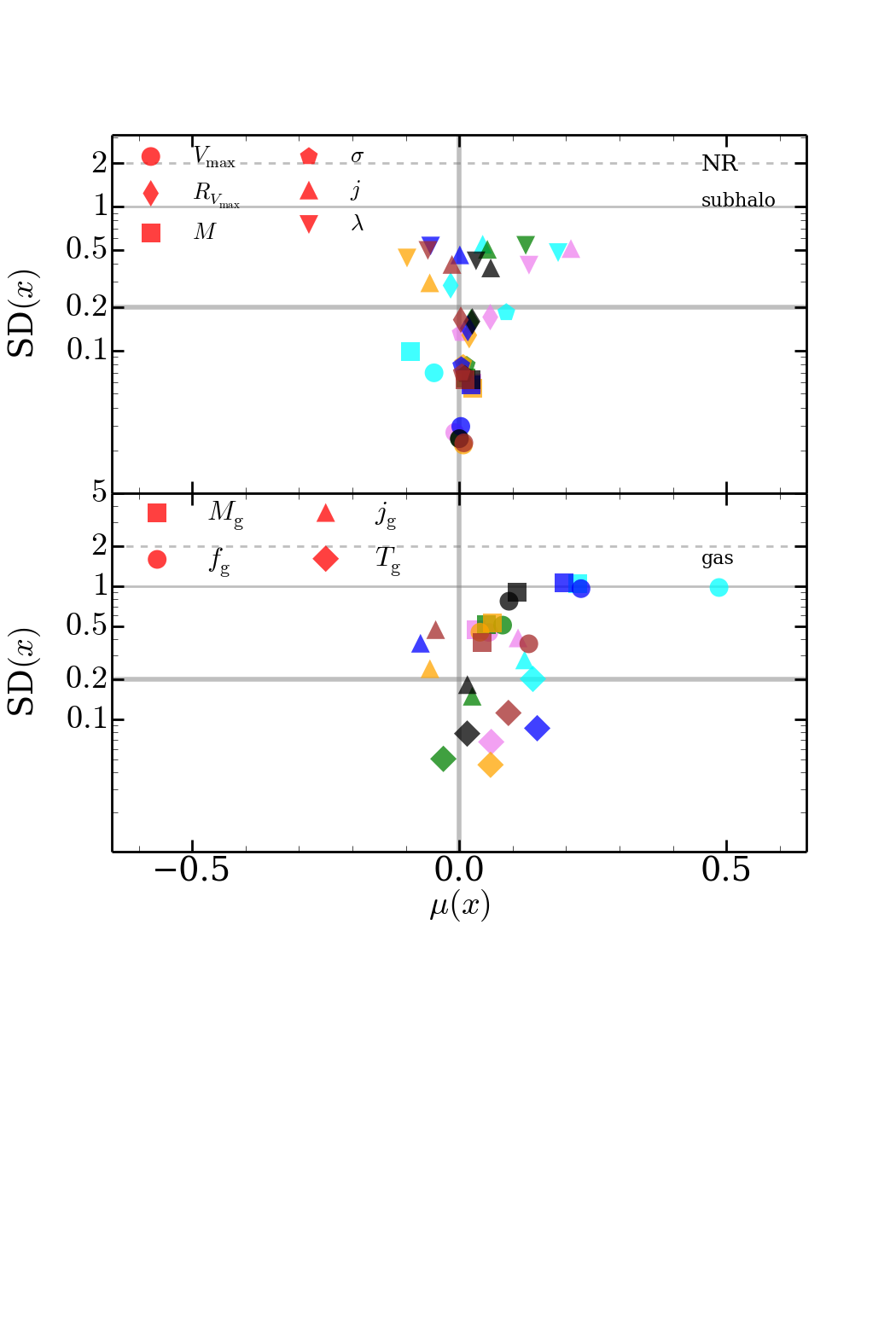}
    \hspace{-2.8pt}
    \caption{Properties comparison: we plot the median ($\mu$) \& standard deviation (SD) of the logarithmic ratio between the listed simulation \& \music. Subhalo properties are: $\vmax$, maximum circular velocity; $\rvmax$, radius of $\vmax$; $M$, virial mass; $\sigma$, velocity dispersion; $j$, specific angular momentum; $\lambda$, the spin parameter from Bullock et al. (2001). Gas: $M_{\rm g}$, gas mass; $f_{\rm g}$, gas fraction; $j_{\rm g}$, gas specific angular momentum; $T_{\rm g}$, average temperature. We show several lines to guide the eye: a thick gray line at $\mu=1$ and SD$=0.2$; and lines at SD~$=1$ \& $2$~dex. Marker {\em colours} are the same as in \Figref{fig:gasfracmass}, see legend in \Figref{fig:gasfracmass}.}
    \label{fig:summarycomp}
\end{figure}
\begin{figure}
    \centering
    \includegraphics[width=0.475\textwidth,trim=0.0cm 2.0cm 2.5cm 4.0cm, clip=true]{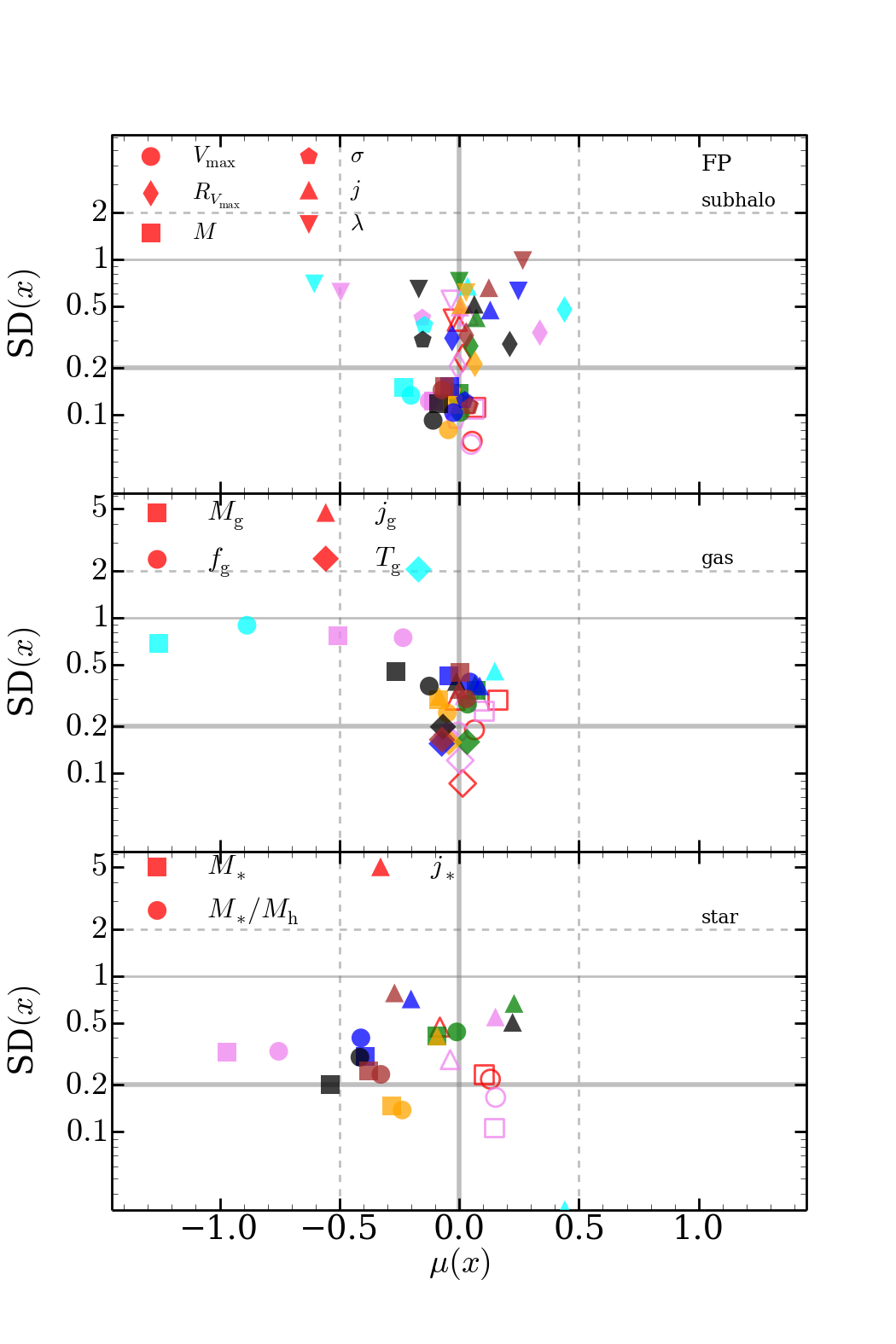}
    \hspace{-2.8pt}
    \caption{Similar to \Figref{fig:summarycomp} but for the \radsim\ simulations. Also includes galaxy properties: $M_{*}$, stellar mass, $M_*/M_{\rm h}$, stellar mass to halo mass; $j_*$, stellar specific angular momentum. We also extra lines at $\mu=-0.5,0.5$, the x-axis limits in \Figref{fig:summarycomp}. Code variants \musicpi\ \& \arepo-SH are plotted as open points. Note that \ramses\ only has a single well resolved galaxy with a match in \music\ where the angular momentum can be measured and hence has SD$=0$. To place this data point on the figure we set SD$=0.02$.}
    \label{fig:summarycomp2}
\end{figure}
We summarise the differences between subhaloes in a given simulation and their \music\ counterparts in \Figref{fig:summarycomp}. The logarithmic ratio, $\log(x_i/x_{\rm MUSIC})$, is typically well characterised by a normal distribution in the \dmsim\ \& \nrsim\ runs, although some distributions have significant tails or broad peaks in the \radsim\ runs (see \Figref{fig:massvmaxcomp}). Thus, we use the median, $\mu$, and calculate an {\em effective standard deviation}, SD, using the 0.32 \& 0.66 quantiles. Naturally, the median between a given catalogue and \music\ indicates whether systematic differences are present. Caution should be used in interpreting SD as it is the variation between \music\ and a given code, not the scatter between all codes. 
Note that here when comparing baryonic masses (and related quantities) we require that either the object or its reference counterpart have non-negligible amounts of gas/stars (depending on the comparison being made). For more complex properties such as spin, both must have non-negligible amounts. 

\par
First, examining the bulk subhalo properties in the \dmsim\ runs, we see here that the mass and $\vmax$ are well reproduced so long as star formation and feedback physics is {\em not} included. There is not significant systematic difference between codes and little scatter, with $\vmax$ varying by $\lesssim1\%$. The velocity dispersion and $R_{\vmax}$ are numerically converged for the non-full physics runs, with SD$\approx0.2$~dex. Angular momentum based quantities show large variations of up to $1~$dex, primarily as $j$ is affected by distant, marginally bound particles, and small differences in the exact position of a subhalo in one simulation to another will significantly contribute to the scatter. \ramses\ is the only code to show some systematic offset, having subhaloes with marginally high spins.

\par
We next present the \nrsim\ runs. The subhalo properties are almost as well converged as those in the \dmsim\ runs, with \ramses\ the only code with some systematic differences, producing smaller, less concentrated subhaloes. The \nrsim-gas panel of \Figref{fig:summarycomp} shows that the gas distribution is less numerically converged, particularly the amount of gas, with an average variation of $0.25$~dex. Some codes show greater code-to-code scatter of $\sim1$~dex (\gadgetxart,\gadgettwox,\owls). Most codes typically have more gas than \music. Interestingly, the gas temperature shows less scatter than the mass but the systematic differences between codes are more pronounced. The temperature bias does not appear to depend purely on numerical implementation as \ramses\ \& \gadgetxart, two very different codes have higher temperatures. However, we do find that both mesh codes have higher angular momentum gas than SPH codes.

\par 
However, it is important to recall that the number of subhaloes with gas is small, so the $\mu$ and SD estimators suffer from small number statistics. Additionally, the $M_{\rm g}$ \& $f_{\rm g}$ ratios have a bimodal distribution since a subhalo can retain gas in one code but have been completely stripped in another. As we have used quantiles to estimate the mean and standard deviation, these subhaloes do not drastically skew these estimates (we treat them as containing one gas particle for the purposes of mass comparisons) and excluded when comparing other properties. Generally, $\approx20-30\%$ of subhaloes fall into this category, therefore the systematic differences and variance presented here are {\em underestimates} but the general features will not change. 

\par
In the full physics runs seen in \Figref{fig:summarycomp2}, the scatter in the bulk properties of the galaxy/(sub)halo host have increased by $\sim0.1$~dex for $\vmax$ and $M$ respectively. However, systematic differences are becoming noticeable, with $\vmax$ in \ramses\ \& \arepo\ being lower by $\sim0.2$~dex. The amount of gas has similarly increased scatter along with systematic differences. Both mesh codes differ significantly from the SPH results, being more gas poor by a factor of 2 for \arepo\ and up to an order of magnitude for \ramses. The scatter is also very high at $\sim1$~dex, whereas the SPH codes show $\sim0.3$~dex scatter.

\par
The galaxies stellar content shows a minimum of $\sim0.15$~dex scatter. More importantly, codes typically produce less massive galaxies than \music, with the two mesh codes with AGN being the most extreme. \arepo's galaxies are $1$~dex smaller. \ramses's galaxies are $2.4$ dex smaller (this lies off the figure) and has SD$=0.43$. Only \owls, \arepo-SH and the \music\ variant itself produce similar galaxies to \music. The stellar angular momentum shows large code-to-code variation and scatter. For example \arepo, \owls\ and \gadgettwox\ galaxies relative to \music\ are more rotationally supported. The sole well resolved \ramses\ galaxy is also biased high. The other codes show that the overall picture is mixed as codes with/without AGN and modern and classic SPH codes have stellar distributions with similar angular momentum as \music.

\par
We also calculate the effective standard deviation based on comparing objects to the median object. Recall that this median is calculated based on possibly incomplete list of catalogues as an object may not be present in all codes. Nevertheless this comparison, although it hides some systematic differences, is informative in estimating the code-to-code scatter. We find that for \dmsim\ \& \nrsim\ runs the scatter for subhalo properties save angular momentum is similar to that seen in \Figref{fig:summarycomp} at around $\lesssim0.1~$dex. The angular momentum related properties have higher scatter $0.2$~dex scatter (lower than that calculated using \music\ as a reference). In the \nrsim\ simulations, gas properties typically vary by $0.1-0.2$~dex. Including star formation and feedback physics increases the scatter for all properties, with subhalo quantities like mass varying by $\sim0.1$~dex, gas properties varying by $0.2$~dex and stellar properties vary by $0.2-0.4$~dex.

\section{Discussion \& Conclusion}
\label{sec:discussion} 
Hydrodynamical codes, regardless of specific numerical approach used, attempt to model (some of) the complex processes involved in forming a galaxy. In this paper, we have assessed how well hydrodynamic codes reproduce the same subhaloes and galaxies in a cluster environment using the nIFTy cluster data set. To address this goal, we have compared both the overall distribution of subhaloes and galaxies and compared individual objects. 

\par
We find that in DM only and non-radiative simulations, codes show $5-10\%$ scatter in the dark matter {\em subhalo} population and even on an individual object basis the scatter is only $0.1$~dex for properties like $\vmax$ and mass\footnote{Note that we only used codes that have full physics modules, mostly limiting our analysis to codes with similar Tree-PM gravity back-ends, the exception being \ramses.}. This is unsurprising considering the small amount of scatter in the dark matter distribution observed in \cite{nifty1}. 

\par 
The differences lie in the baryonic component. In \cite{nifty1} we found that even in \nrsim\ runs, the gas entropy and density profiles of the cluster differed significantly from code-to-code, with mesh and modern SPH codes producing entropy cores whereas classic SPH codes produced ever falling entropy profiles. Here we find that individual subhaloes show large variation in the baryonic fraction depending on the code used. The code-to-code scatter is $0.2-0.4$~dex despite the overall similarity between codes in the likelihood of a subhalo being baryon poor. However, subhaloes do not show a strong separation between classic SPH and other codes. 

\par 
The key result of this paper is that codes produce different galaxy populations and that the diversity is significant, despite all codes approaching galaxy formation in a similar fashion. Codes convert gas particles or cells into a ``star'' particle when some criterion is satisfied, typically if a converging flow of gaseous material has high enough local densities and able to cool. This newly formed particle represents a star cluster, the basic galaxy building block. Star particles feed energy and metals back into the local environment. The issue is that these processes occur at unresolved scales, thus each code uses their own subgrid modules to model this complex process. Add to this mix, supermassive black holes, their growth by accretion and the associated injection of energy via AGN. Some codes include AGN feedback, some do not. Considering the variety of subgrid physics, some diversity is to be expected but perhaps not to the extent seen here. Even the bulk gas and stellar fractions of the entire cluster show marked differences, with gas and stellar fractions ranging from $\sim0.12-0.18$ \& $\sim0.01-0.05$ respectively \cite[][]{nifty2}.

\par
We find that the number of galaxies of a given stellar mass can vary by a factor of 4 in the cluster environment. The exception is \ramses, which severely suppresses galaxy formation inside clusters, producing a paltry number of galaxies despite having no supernova feedback. Among the other codes, \arepo\ produces the fewest, followed by \magneticum, whereas \music\ \& \owls\ produce the most. 

\par 
Not only do the number of galaxies differ, codes do not produce the same stellar mass to halo mass relation. Codes with AGN physics have massive galaxies with much lower $\MstarMh$, yet some have higher $\MstarMh$ values than \music\ for the lowest mass galaxies resolved here. Despite all this variety, codes generally produce the same effective baryonic Tully-Fisher (Faber-Jackson) relation, i.e., $M_*$-$\vmax$ relation, indicating that observations like those of \cite{bell2001,reyes2011} has limited use in pinning subgrid physics.

\par 
By comparing well resolved individual objects between codes, we find that if a subhalo hosts a galaxy in one code, generally it will host a galaxy in other codes. The exceptions are the two mesh codes that include AGN, \ramses\ \& \arepo, which have the lowest star formation efficiencies. Of greater importance is that this synthetic galaxy will not have the same stellar mass across codes, despite having a similar merger and orbit history. First we note that galaxies show large scatter of $\sim0.2-0.5$~dex in stellar mass, $\MstarMh$ and stellar angular momentum. Second, there are significant systematic differences between codes. For example, galaxies in \arepo\ are $\sim1$~dex less massive than those in \music. 

\par 
The variety in synthetic galaxies and input subgrid physics is telling. Some codes with similar subgrid schemes, such as \gadgettwox\ \& \owls, which have similar SF and AGN but different IMFs and SN feedback and significantly different cooling curves (\gadgettwox\ assumes solar metallacity), produce different numbers of galaxies. The number of galaxies here differ by $60\%$, and distributions like gas fractions and luminosity functions differ in shape. Changes in the cooling curve might account for some of these differences. Another example of similar codes is \gadgetxart \& \magneticum. These two modern SPH codes have the same SF, IMF, similar AGN and differ in the SPH conduction scheme and significantly in the SN feedback scheme (\gadgetxart\ has kinetic SN, \magneticum\ has both thermal and kinetic). Here the differences are more subtle: the GSMFs have similar shapes but \magneticum\ has fewer low stellar mass galaxies likely due to stronger quenching from the addition of thermal SN feedback, resulting in \gadgetxart\ having  $50\%$ more galaxies with $M_*>10^{9}\Msunh$. 

\par 
Mesh codes at first glance are far less efficient than similar SPH codes at producing galaxies. \arepo\ has similar subgrid schemes to the modern SPH code \gadgetxart, yet has only $30\%$ of the galaxies that \gadgetxart\ has. The galaxies in the \arepo\ cluster are more likely to be stripped of gas, have lower stellar masses and do not follow the same GSMF, although they have a similar mass BCG (including intracluster stars). \arepo\ also has galaxies with higher angular momentum than \gadgetxart. This higher angular momentum difference appears to hinge on the AGN feedback scheme. \arepo-SH, lacking AGN feedback, produces numbers much closer to that of \music, the only code lacking AGN feedback. Moreover, the distributions and even individual galaxies themselves are similar, although \music\ tends to produce a larger number of low stellar mass galaxies. The dependence of subgrid physics on the method used to evolve gas has been noted for the subgrid physics implemented \cite[in for instance the EAGLE simulations][]{schaye2015a,schaller2015a}. 

\par
Many numerical studies show AGN feedback can play an important role (e.g.~\citealp{puchwein2010,mccarthy2010a,teyssier2011,cui2014a}, although observational evidence may not be as clear cut, see \citealp{schawinski2014a}). However, the galaxy diversity seen between our suite of codes tells us that differences {\em do not solely arise from the inclusion of AGN feedback}. \owls, which has AGN, has similar mass galaxies to \music. Conversely, \pesph\ produces systematically lower mass galaxies than \owls\ yet it does not include AGN, although the use of a quenching model for massive galaxies in \pesph\ might mimic the statistical suppression of star formation that AGN have.

\par 
In general, codes that reproduce the observed galaxy population in some respects, such as the luminosity function, in certain environments will need to be adjusted to reproduce galaxies in another environment. Therefore, subgrid physics {\em as it stands} is fine-tuned. The fact that subgrid physics requires tuning has been noted before \cite[e.g.][]{haas2013a,haas2013b,lebrun2014a, schaye2015a,crain2015a}. However, the {\em similarity of galaxies} produced by codes with {\em different subgrid physics} and {\em differences} between codes with {\em similar schemes} implies the diversity and similarity is not solely a matter of fine-tuning a particular subgrid scheme. Rather current subgrid physics schemes does not fully capture the real processes governing galaxies. 

\par 
In conclusion, our comparison suggests that the properties of any individual synthetic galaxy should be treated with errors bars of at least $\sim0.2-0.4$~dex. 
\section*{Acknowledgements}
The authors thank Joop Schaye for insightful comments. The authors also express special thanks to the Instituto de Fisica Teorica (IFT-UAM/CSIC in Madrid) for its hospitality and support, via the Centro de Excelencia Severo Ochoa Program under Grant No. SEV-2012-0249, during the three week workshop ``nIFTy Cosmology'' where this work developed. We further acknowledge the financial support of the University of Western 2014 Australia Research Collaboration Award for ”Fast Approximate Synthetic Universes for the SKA”, the ARC Centre of Excellence for All Sky Astrophysics (CAASTRO) grant number CE110001020, and the two ARC Discovery Projects DP130100117 and DP140100198. We also recognize support from the Universidad Autonoma de Madrid (UAM) for the workshop infrastructure.

\par
PJE is supported by the SSimPL programme and the Sydney Institute for Astronomy (SIfA), and {\it Australian Research Council} (ARC) grants DP130100117 and DP140100198. STK acknowledges support from STFC through grant ST/L000768/1. AK is supported by the {\it Ministerio de Econom\'ia y Competitividad} (MINECO) in Spain through grant AYA2012-31101 as well as the Consolider-Ingenio 2010 Programme of the {\it Spanish Ministerio de Ciencia e Innovaci\'on} (MICINN) under grant MultiDark CSD2009-00064. He also acknowledges support from the {\it Australian Research Council} (ARC) grants DP130100117 and DP140100198. He further thanks The Feelies for crazy rhythms. CP acknowledges support of the Australian Research Council (ARC) through Future Fellowship FT130100041 and Discovery Project DP140100198. WC and CP acknowledge support of ARC DP130100117. GY and FS acknowledge support from MINECO (Spain) through the grant AYA 2012-31101. GY thanks also the  Red Espa\~{n}ola de Supercomputacion for granting the computing time in the Marenostrum Supercomputer at BSC, where all the MUSIC simulations have been performed. AMB is supported by the DFG Research Unit 1254 ``Magnetisation of interstellar and intergalactic media'' and by the DFG Cluster of Excellence ``Universe''. SB \& GM acknowledge support from the PRIN-MIUR 2012 Grant "The Evolution of Cosmic Baryons" funded by the Italian Minister of University and Research, by the PRIN-INAF 2012 Grant "Multi-scale Simulations of Cosmic Structures", by the INFN INDARK Grant and by the "Consorzio per la Fisica di Trieste". IGM acknowledges support from a STFC Advanced Fellowship. EP acknowledges support by the ERC grant ``The Emergence of Structure during the epoch of Reionization''. JS acknowledges support from the European Research Council under the European Unions Seventh Framework Programme (FP7/2007-2013)/ERC grant agreement 278594-GasAroundGalaxies.

\par
The authors contributed to this paper in the following ways: PJE organized and analysed the data, made the plots and wrote the paper. AK, GY \& FRP organized the nIFTy workshop at which this program was completed. GY supplied the initial conditions. All the other authors, as listed in Section 2 performed the simulations using their codes. All authors have read and commented on the paper.

\par 
The simulations used for this paper have been run on a variety of supercomputers and are publicly available at the MUSIC website, \href{http://www.music.ft.uam.es}{\url{http://www.music.ft.uam.es}}. MUSIC simulations were carried out on Marenostrum. AREPO simulations were performed with resources awarded through STFCs DiRAC initiative. The authors thank Volker Springel for helpful discussions and for making AREPO and the original GADGET version available for this project. G3-PESPH Simulations were carried out using resources at the Center for High Performance Computing in Cape Town, South Africa.

\par 
This research has made use of NASA's Astrophysics Data System (ADS) and the arXiv preprint server.

\pdfbookmark[1]{References}{sec:ref}
\bibliographystyle{mn2e}
\bibliography{substructure.bbl}

\appendix
\section{Codes} \label{sec:app:codes}
\subsection{Mesh-based Codes}
\subsubsection{AMR}
\paragraph*{{\sc\bf RAMSES} (Teyssier, Perret)}\label{code:ramses}
\ramses\ is an adaptive mesh refinement code that uses a directionally unsplit, second order Godunov scheme with the HLLC Riemann solver to solve hydrodynamics and an adaptive particle mesh code to solve the Poisson equation. The grid is adaptively refined on a cell-by-cell basis, following a quasi-Lagrangian refinement strategy whereby a cell is refined into 8 smaller new cells if its dark matter or baryonic mass grows by more than a factor of eight. Time integration is performed using an adaptive, level-by-level, time stepping strategy. 

\medskip

\noindent \emph{Cooling \& Heating:} Gas cooling and heating is performed assuming coronal equilibrium with a modification of the \cite{haardt1996} UV background and a self-shielding recipe based on \cite{aubert2010}. Hydrogen and Helium cooling and heating processes are included following \cite{katz1996}, metal cooling follows \cite{sutherland1993}. Here, the code also uses a temperature floor of $10^4$~K to prevent spurious fragmentation of relatively poorly resolved galactic discs.

\medskip

\noindent \emph{Star Formation:} Star formation is implemented as a stochastic process using a local Schmidt law as in \cite{rasera2006}. The density threshold for star formation was set to $n_*=0.1 H/cc$, and the local star formation efficiency per gas free fall time was set to $5\%$.

\medskip

\noindent \emph{Stellar Population Properties \& Chemistry:} Each star particle is treated as a single stellar population (SSP) with a \cite{chabrier2003} IMF. Mass and metal return to the gas phase by core collapse supernovae only. A single average metallacity is followed during this process and advected in the gas as a passive scalar, to be used as an indicator of the gas metallicity in the cooling function.

\medskip

\noindent \emph{Stellar Feedback:} In this project, no feedback processes related to the stellar population are used.

\medskip

\noindent \emph{SMBH Growth \& AGN Feedback:} SMBH particles are represented by sink particles \cite[][]{teyssier2011}. The SMBH accretion follows Bondi accretion with the rate constrained by the instantaneous Eddington limit. When the gas density is larger than the star formation density threshold the Bondi accretion rate is boosted \cite[][]{boothschaye2009}. SMBH particles are evolved using a direct gravity solver, to obtain a more accurate treatment of their orbital evolution. SMBH particles more massive than $10^8 M_\odot$ are
allowed to merge if their relative velocity is smaller than their pair-wise scale velocity. Less massive SMBH particles, on the other
hand, are merged as soon as they fall within 4 cells from another SMBH particle. The AGN feedback used is a simple thermal energy
dump with $0.1c^2$ of specific energy, multiplied by the instantaneous SMBH accretion rate.


\subsubsection{Moving Mesh}

\paragraph*{{\sc\bf Arepo} (Puchwein)}\label{code:arepo}
\noindent \arepo\ uses a Godunov scheme on an unstructured moving Voronoi mesh; mesh cells move (roughly) with the fluid. The main difference between \arepo\ and traditional Eulerian AMR codes (such as {\sc{ART}}) is that \arepo\ is almost Lagrangian and Galilean invariant by construction. The main difference between \arepo\ and SPH codes (see next subsection) is that the hydrodynamic equations are solved with a finite-volume Godunov scheme. The version of \arepo\ used in this study conserves total energy in the Godunov scheme, rather than the entropy-energy formalism described in \cite{arepo}. Detailed descriptions of the galaxy formation models implemented in \arepo\ can be found in \cite{vogelsberger2013} and \cite{vogelsberger2014a}, but the key features can be summarised as follows.

\medskip

\noindent \emph{Cooling \& Heating:} Gas cooling takes the metal abundance into account. The metal cooling rate is computed for solar composition gas and scaled to the total metallicity of the cell. Photoionization and photoheating are followed based on the homogeneous UV background model of \cite{fauchergiguere2009} and the self-shielding prescription of \cite{rahmati2013}. In addition to the homogeneous UV background, the ionizing UV emission of nearby AGN is taken into account.

\medskip

\noindent \emph{Star Formation:} The formation of stars is followed with a multiphase model of the interstellar medium which is based on \cite[][hereafter SH03]{springel2003} but includes a modified effective equation of state above the star formation threshold, i.e. above a hydrogen number density of $ 0.13 \, \mathrm{cm}^{-3}$.

\medskip

\noindent \emph{Stellar Population Properties \& Chemistry:} Each star particle is treated as a single stellar population (SSP) with a \citet{chabrier2003} IMF. Mass and metal return to the gas phase by AGB stars, core collapse supernovae and Type Ia supernovae is taken into account. Nine elements are followed during this process (H, He, C, N, O, Ne, Mg, Si, Fe).

\medskip

\noindent \emph{Stellar Feedback:} Feedback by core collapse supernovae is implicitly invoked by the multiphase star formation model. In addition, we include a kinetic wind model in which the wind velocity scales with the local dark matter velocity dispersion ($v_{\mathrm{wind}} \sim 3.7 \sigma_{\mathrm{DM,1D}}$). The mass-loading is determined by the available energy which is assumed to be $1.09\times10^{51}\,\mathrm{erg}$ per core collapse supernova. Wind particles are decoupled from the hydrodynamics until they fall below a specific density threshold or exceed a maximum travel time. This ensures that they can escape form the dense interstellar medium.

\medskip

\noindent \emph{SMBH Growth \& AGN Feedback:} SMBHs are treated as collisionless sink particles. $10^5 M_\odot / h$ BHs are seeded into haloes once they exceed a mass of $5 \times 10^{10} M_\odot / h$. The BHs subsequently grow by Bondi-Hoyle accretion with a boost factor of $\alpha = 100$. The Eddington limit on the accretion rate is enforced in addition. AGN are assumed to be in the quasar mode for accretion rates larger than 5\% of the Eddington rate. In this case $1\%$ of the accreted rest mass energy is thermally injected into nearby gas. For accretion rates smaller than $5\%$ of the Eddington rate, AGN are in the radio mode in which $7\%$ of the accreted rest mass energy is thermally injected into spherical bubbles \cite[similar to][]{sijacki2007}. Full details of the black hole model are given in \cite{sijacki2014}.

\medskip

In addition to the main run, we have performed a simulation with simplified galaxy formation physics which allows a direct comparison to \gadget\ simulations with the same baryonic physics. In this simulation, we account only for primordial cooling, photoheating by the UV background, star formation with the \citetalias{springel2003} model, and kinetic wind feedback with a mass-loading of two times the star formation rate and a wind velocity of $\sim 342 \, \mathrm{km}/\mathrm{s}$, essentially the subgrid physics of \music.


\subsection{SPH Codes}
\subsubsection{Classic}
\paragraph*{{\sc\bf Gadget3-MUSIC} (Yepes, Sembolini)}\label{code:music} This is modified version of the \gadget3\ Tree-PM code that uses classic entropy-conserving SPH formulation with a 40 neighbour M3 kernel. The basic \citetalias{springel2003} model was used. The variant, \musicpi, uses the same SPH formulation but different feedback \cite[there are differences in how SN energy is distributed to surrounding SPH particles, the cooling function is metal dependant, it traces different metal species from Type IA and SN-II separately and it switches off cooling around SN explosions; see][]{piontek2011}.

\medskip

\noindent \emph{Cooling \& Heating:} Radiative cooling is assumed for a gas of primordial composition, with no metallicity dependence, and the effects of a background homogeneous UV ionising field is assumed, following \cite{haardtmadau2001}.

\medskip

\noindent \emph{Star Formation:} The \citetalias{springel2003} model is implemented.

\medskip

\noindent \emph{Stellar Population Properties \& Chemistry:} A \cite{salpeter1955} IMF is assumed, with a slope of -1.35 and upper and lower mass limits of $40 \rm M_{\odot}$ and $0.1 \rm M_{\odot}$ respectively.

\medskip

\noindent \emph{Stellar Feedback:} This has both a thermal and a kinetic mode; thermal feedback evaporates the cold phase within SPH particles and increases the temperature of the hot phase, while kinetic feedback is modelled as a stochastic wind (as in \citetalias{springel2003}) -- gas mass is lost due to galactic winds at a rate $\dot{M}_w$, which is proportional to the star formation rate $\dot{M}_{\ast}$, such that $\dot{M}_w = \eta\dot{M}_{\ast}$, with $\eta=2$. SPH particles near the star forming region will be subjected to enter in the wind in an stochastic way. Those particles impacted upon by the wind will be given an isotropic velocity kick of $v_w=400\rm kms^{-1}$ and will freely travel without feeling pressure forces up to 20 kpc distance from their original positions

\medskip

\noindent \emph{SMBH Growth \& AGN Feedback:} These processes are not included.

\medskip


\paragraph*{{\sc\bf Gadget3-OWLS} (McCarthy, Schaye)}\label{code:owls} The is a heavily modified version of \gadget3\, using a classic entropy-conserving SPH formulation with a 40 neighbour M3 kernel. 

\medskip

\noindent \emph{Cooling \& Heating:} Radiative cooling rates for the gas are computed on an element-by-element basis by interpolating within pre-computed tables \cite[generated with the {\small{CLOUDY}} code; cf.][]{cloudy} that contain cooling rates as a function of density, temperature and redshift calculated in the presence of the cosmic microwave background and photoionization from a \cite{haardtmadau2001} ionising UV/X-ray background \cite[further details in ][]{wiersma2009a}.

\medskip

\noindent \emph{Star Formation:} Star formation follows the prescription of \citetalias{schaye2008} -- gas with densities exceeding the critical density for the onset of the thermogravitational instability is expected to be multiphase and to form stars \cite[][]{schaye2004}. Because the simulations lack both the physics and numerical resolution to model the cold interstellar gas phase, an effective equation of state (EOS) is imposed with pressure $P\propto\rho^{4/3}$ for densities $n_{\rm H}>n_{\ast}$ where $n_{\ast}=0.1{\rm cm}^{-3}$. Gas on the effective EOS is allowed to form stars at a pressure-dependent rate that reproduces the observed Kennicutt-Schmidt law \cite[][]{schmidt1959,kennicutt1998} by construction.

\medskip

\noindent \emph{Stellar Population Properties \& Chemistry:} The ejection of metals by massive- (SNeII and stellar winds) and intermediate-mass stars (SNeIa, AGB stars) is included following the prescription of \cite{wiersma2009b}. A set of 11 individual elements are followed (H, He, C, Ca, N, O, Ne, Mg, S, Si and Fe), which represent all the important species for computing radiative cooling rates.

\medskip

\noindent \emph{Stellar Feedback:} Feedback is modelled as a kinetic wind \cite[][]{dallavecchia2008} with a wind velocity $v_w=600\rm km\,s^{-1}$ and a mass loading $\eta=2$, which corresponds to using approximately 40 per cent of the total energy available from SNe for the adopted \cite{chabrier2003} IMF. This choice of parameters results in a good match to the peak of the SFR history of the universe \cite[][]{schaye2010}.

\medskip
 
\noindent \emph{SMBH Growth \& AGN Feedback:} Each black hole can grow either via mergers with other black holes within the softening length or via Eddington-limited gas accretion, the rate of which is calculated using the Bondi-Hoyle formula with a modified efficiency, setting $\beta=2$ as in \cite{boothschaye2009}. The black hole is forced to sit on the local potential minimum to suppress spurious gravitational scattering \cite[][]{springel2005}. Feedback is done by storing up the accretion energy (assuming $\epsilon_r=0.1$, $\epsilon_f=0.15$) until at least one particle can be heated to a fixed temperature of $T_{\rm AGN}=10^8 \rm K$ \cite[][]{boothschaye2009}.

\medskip


\paragraph*{{\sc\bf Gadget2-X} (Kay)}\label{code:gadget2x} This is a modified version of the original \gadget2\ Tree-PM code that uses the classic entropy-conserving SPH formulation with a 40 neighbour M3 kernel. A detailed description of the code can be found in \cite{pike2014}, but its key features can be summarised as follows.

\medskip

\noindent \emph{Cooling \& Heating:} Cooling follows the prescription of \cite{thomascouchman1992} -- a gas particle is assumed to radiate isochorically
over the duration of its timestep. Collisional ionisation equilibrium is assumed and the cooling functions of \cite{sutherland1993} are used, with the metallicity $Z$=0 to ignore the increase in cooling rate due to heavy elements. Photo-heating rates are not included but the gas is heated to a minimum $T=10^4\rm K$ at $z<10$ and $n_{\rm H}<0.1 {\rm cm}^{-3}$.

\medskip

\noindent \emph{Star Formation:} Star formation follows the method of \citetalias{schaye2008}; it assumes an equation of state for the gas with $n_{\rm H}>0.1\,{\rm cm}^{-3}$, with an effective adiabatic index of $\gamma_{\rm eff}=\nicefrac{4}{3}$ for constant Jeans mass. Gas is converted to stars at a rate given by the Kennicutt-Schmidt relation \cite[][]{schmidt1959,kennicutt1998}, assuming a disc mass fraction $f_g$=1. The conversion is done stochastically on a particle-by-particle basis so the gas and star particles have the same mass.

\medskip

\noindent \emph{Stellar Population Properties \& Chemistry:} Each star particle is assumed to be a single stellar population with a \cite{salpeter1955} IMF.

\medskip

\noindent \emph{Stellar Feedback:} A prompt thermal Type II SNe feedback model is used. This assumes that a fixed number, $N_SN$, of gas particles are heated to a fixed temperature, $T_SN$, with values of $N_SN=3$ and $T_SN=1e7K$ chosen to match observed hot gas and star fractions \cite[cf.][]{pike2014}. Heated gas is allowed to interact hydrodynamically with its surroundings and radiate.

\medskip

\noindent \emph{SMBH Growth \& AGN Feedback:} A variation on the \cite{boothschaye2009} model is used. Black holes are seeded in friends-of-friends (FOF) haloes with more than 50 particles at $z$=5, at the position of the most bound star or gas particle, which is replaced with a black hole particle. The gravitational mass of the replaced particle is unchanged but an \emph{internal} mass of $10^6 h^{-1} {\rm M}_{\odot}$ is adopted for the calculation of feedback. Each black hole can grow either via mergers with other black holes within the softening length or via Eddington-limited gas accretion, the rate of which is calculated using the Bondi-Hoyle formula with a modified efficiency, setting $\beta$=2 as in \cite{boothschaye2009}. The black hole is forced to sit on the local potential minimum, to suppress spurious gravitational scattering. Feedback is done by storing up the accretion energy (assuming $\epsilon_r=0.1$, $\epsilon_f=0.15$) until at least one particle can be heated to a fixed temperature of $T_{\rm AGN}=3 \times 10^8 \rm K$. This high temperature was chosen for high-mass clusters to match their observed pressure profiles -- a lower temperature causes too much gas to accumulate in cluster cores because there is insufficient entropy to escape to larger radius.


\subsubsection{Modern}

\paragraph*{{\sc\bf Gadget3-X} (Murante, Beck)}\label{code:gadget3x}
\noindent This is a modified version of the non-public \gadget3 that includes: an artificial conduction term that largely improves the SPH capability of following gas-dynamical instabilities and mixing processes; a higher-order Wendland C4 kernel \cite[][]{dehnen2012} to better describe discontinuities and reduce clumpiness instability; and a time-dependent artificial viscosity term to minimize viscosity away from shock regions. Pure hydrodynamical and hydro/gravitational tests on the performance of modified SPH scheme are presented in \cite{beck2016a}.

\medskip

\noindent \emph{Cooling \& Heating:} Gas cooling is computed for an optically thin gas and takes into account the contribution of metals, using the procedure of \cite{wiersma2009a}, while a uniform UV background is included following the procedure of \cite{haardtmadau2001}.

\medskip

\noindent \emph{Star Formation:} Star formation is implemented as in \cite{tornatore2007}, and follows the star formation algorithm is that of
\citetalias{springel2003} -- gas particles above a given density threshold are treated as multi-phase. The effective model of \citetalias{springel2003} describes a self-regulated, equilibrium inter-stellar medium and provides a star formation rate that depends upon the gas density only, given the model parameters.

\medskip

\noindent \emph{Stellar Population Properties \& Chemistry:} Each star particle is considered to be a single stellar population (SSP). We follow the evolution of each SSP, according to the \cite{chabrier2003} IMF. We account for metals produced in the SNeIa, SNeII and AGB phases, and follow 16 chemical species. Star particles are stochastically spawned from parent gas particles as in \citetalias{springel2003}, and get their chemical composition of their parent gas. Stellar lifetimes are from \cite{padovani1993}; metal yields from \cite{woosley1995} for SNeII, \cite{thielemann2003} for SNeIa, and \cite{vandehoek1997} for AGB stars. 

\medskip

\noindent \emph{Stellar Feedback:} SNeII release energy into their surroundings, but this only sets the hot gas phase temperature and, as a consequence, the average SPH temperature of gas particles. Supernova feedback is therefore modelled as kinetic and the prescription of \citetalias{springel2003} is followed (i.e. energy-driven scheme with a fixed wind velocity of $350 {\rm km} {\rm s}^{-1}$, wind particles decoupled from surrounding gas for a period of 30 Myr or until ambient gas density drops below 0.5 times the multiphase density threshold).

\medskip

\noindent \emph{SMBH Growth \& AGN Feedback:} AGN feedback follows \cite{steinborn2015a}. In the aforementioned model, SMBHs grow via Bondi-Hoyle like gas accretion (Eddington limited) with the model distinguishing between cold and hot component (see their Eq. 19). Here only cold accretion is considered, using a fudge-factor $\alpha_{\rm cold}=100$ in the Bondi-Hoyle formula (i.e., $\alpha_{\rm hot}=0$). The radiative efficiency is variable, and it is evaluated using the model of \cite{churazov2005}. Such a model outputs separately the AGN mechanical and radiative power as a function of the SMBH mass and the accretion rate. Here these are summed to give the resulting energy thermally to the surrounding gas with an AGN feedback/gas coupling efficiency of $\epsilon_{\rm fb}=0.5$. The parameters of the hydro model were tuned using the tests presented in \cite{beck2016a} and those of the AGN model for reproducing  observational  scaling  relations  between  SMBH mass  and  stellar  mass  of  the  host  galaxies. No attempt was made to reproduce any of the observational properties of the ICM. First results on the application of this code to simulations of galaxy clusters, including the reproduction of the Cool Core/Non-Cool Core dichotomy, can be found in \cite{rasia2015a}.


\paragraph*{{\sc\bf Gadget3-PESPH} (February, Dav\'e, Huang, Katz)} \label{code:pesph}
This version of \gadget\ uses the pressure-entropy SPH formulation of \cite{hopkins2013} with a 128 neighbour HOCTS(n=5) kernel and the time-dependent artificial viscosity scheme of \cite{morris1997}.

\medskip

\noindent \emph{Cooling \& Heating:} Radiative cooling using primordial abundances is modelled as described in \cite{katz1996}, with additional cooling from metal lines assuming photo-ionisation equilibrium follows \cite{wiersma2009a}. A \cite{haardtmadau2001} uniform ionising UV background is assumed.

\medskip

\noindent \emph{Star Formation:} Star formation follows the approach set out in \citetalias{springel2003}, where a gas particle above a density threshold of $n_{\rm H} = 0.13$ cm$^{−3}$ is modelled as a fraction of cold clouds embedded in a warm ionised medium, following \cite{mckee1977}. The star formation rate obeys the \cite{schmidt1959} law and is proportional to $n_{\rm H}^{1.5}$, with the star formation timescale scaled to match the $z$=0 \cite{kennicutt1998} relation. In addition, the heuristic model of \cite{rafieferantsoa2015}, tuned to reproduce the exponential truncation of the stellar mass function, is used to quench star formation in massive galaxies. A quenching probability $P_Q$, which depends on the velocity dispersion of the galaxy, determines whether or not star formation is stopped in a given galaxy; if it is stopped, each gas particle eligible for star formation first has its quenching probability assessed, and if it is selected for quenching then it is heated to 50 times the galaxy’s virial temperature, which unbinds it from the galaxy.

\medskip
  
\noindent \emph{Stellar Population Properties \& Chemistry:} Each star particle is treated as a single stellar population with a \cite{chabrier2003} IMF throughout. Metal enrichment from SNeIa, SNeII and AGB stars are tracked, while 4 elements -- C, O, Si and Fe -- are also tracked individually, as described by \cite{oppenheimer2008}.

\medskip

\noindent \emph{Stellar Feedback:}  Supernova feedback is assumed to drive galactic outflows, which are implemented using a Monte Carlo approach analogous to that used in the star formation prescription. Outflows are directly tied to the star formation rate, using the relation $\dot{M}_{wind} = \eta \times$SFR, where $\eta$ is the outflow mass loading factor. The probability for a gas particle to spawn a star particle is calculated from the subgrid model described above, and the probability to be launched in a wind is $\eta$ times the star formation probability. If the particle is selected to be launched, it is given a velocity boost of $v_w$ in the direction of $v\times a$, where $v$ and $a$ are the particle’s instantaneous velocity and acceleration, respectively.

\medskip

\noindent This is a highly constrained heuristic model for galactic outflows, described in detail in \cite{dave2013}, which utilises outflows scalings expected for momentum-driven winds in sizeable galaxies ($\sigma > 75 $km s$^{−1}$), and energy-driven scalings in dwarf galaxies. In particular, the mass loading factor (i.e. the mass outflow rate in units of the star formation rate) is $\eta = 150 \text{km s}^{-1} /\sigma$ for galaxies with velocity dispersion $\sigma > 75 $km s$^{−1}$ , and $\eta = 150 \text{km s}^{-1} /\sigma^2$ for $\sigma < 75 $km s$^{−1}$.

\medskip

\noindent \emph{SMBH Growth \& AGN Feedback:} These processes are not included.

\medskip


\paragraph*{{\sc\bf Gadget3-Magneticum} (Saro)}\label{code:magneticum}
\magneticum\ is a modified version of \gadget3 using a kernel based on the bias-corrected, sixth-order Wendland kernel \cite[][]{dehnen2012} with 295 neighbours. The code also incorporates a low viscosity scheme to track turbulence \cite[][]{dolag2005a,beck2016a}, gradients computed with high-order scheme \cite[][]{price2012a}, thermal conduction is modelled isotropically at 1/20th of the Spitzer rate \cite[][]{dolag2004b}, and a time-step limiting particle wake-up algorithm \cite[][]{pakmor2012}.

\medskip

\noindent \emph{Cooling \& Heating:} Cooling follows the prescription of \cite{wiersma2009a} and photoionization from a \cite{haardtmadau2001} ionising UV/X-ray background. Radiative cooling rates for 11 elements (H, He, C, N, O, Ne, Mg, Si, S, Ca, Fe) are computed by interpolating within pre-computed tables \cite[generated with the {\small{CLOUDY}} code; cf.][]{cloudy}

\medskip

\noindent \emph{Star Formation:} The \citetalias{springel2003} model is implemented.

\medskip

\noindent \emph{Stellar Population Properties \& Chemistry:} Stars follow a Chabrier IMF. Chemical evolution follows \cite{tornatore2007}: metals are produced by SNII, by supernovae type Ia (SNIa) and by intermediate and low-mass stars in the asymptotic giant branch (AGB). Metals and energy are released  accounting for mass-dependent life-times with lifetimes according to \cite{padovani1993}, metallicity-dependent stellar yields according to \cite{woosley1995} for SNII, \cite{vandehoek1997} for AGB stars, and \cite{thielemann2003} for SNIa.

\medskip

\noindent \emph{Stellar Feedback:} The hot gas within the multiphase ISM model is heated by supernovae and can evaporate the cold clouds. A certain fraction of massive stars ($10\%$) is assumed to explode as SNII triggering galactic winds with a mass loading rate proportional to the SFR and a wind velocity of 350 km/s.

\medskip

\noindent \emph{SMBH Growth \& AGN Feedback:} SMBH and AGN feedback are based on \cite{springel2005,dimatteo2005} and modifications of \cite{sijacki2007,fabjan2010,hirschmann2014a,dolag2015a}. SMBH's grow via Bondi-Hoyle accretion of gas or mergers. The accretion rate is limited to the Eddington rate and a characteristic boost factor of 100 is applied as only the accretion to large scale is captured. Unlike \cite{springel2005} in which entire gas particles are accreted, here 1/4 of a gas particle's mass can be captured in an accretion event. During accretion events, $10\%$ of the accreted mass is converted into energy, $10\%$ of which is thermally coupled with gas within the smoothing length of the SMBH, weighted using the hydrodynamics SPH kernel. When the accretion rate drops below a threshold, it is assumed that there is a transition from a quasar mode to a radio mode of AGN feedback, and the feedback efficiency is enhanced by a factor of 4. 

\end{document}